\documentstyle[12pt,psfig]{article}
\def\ni{\noindent}

\begin{document}
\thispagestyle{empty}
\begin{tabular}{p{12.4cm}r}
& {\bf CHAPTER 5}\\[3.5cm]
\end{tabular}
\hspace*{1.5cm} {\LARGE \bf Theory of Adsorption on Metal Substrates}\\[1.5cm]

\begin{tabular}{p{6.4cm}r}
& {\bf M. Scheffler and C. Stampfl}\\[0.5cm]
& Fritz-Haber-Institut der Max-Planck-Gesellschaft\\
& Faradayweg 4-6\\
& D-14195 Berlin, Germany\\[10cm]

& {\scriptsize Handbook of Surface Science}\\
 & {\scriptsize Volume 2, edited by K. Horn and M. Scheffler}\\[2cm]

\end{tabular}

\newpage

{\Large \bf Contents}
\begin{enumerate}

\item[5.1] Introduction
%(sec:intro)

    \begin{enumerate}
    \item[5.1.1] The nature of the surface chemical bond: indeterminate
               concepts,\\
               yet  useful ...
    \item[5.1.2]  What will be discussed and why, and what is missing
    \end{enumerate}

\item[5.2] Concepts and definitions

\begin{enumerate}
    \item[5.2.1] Density of states
    \item[5.2.2] Energies
   \item[5.2.3] Binding energy at kink sites
    \item[5.2.4] The surface energy barrier

\end{enumerate}

\item[5.3] The tight-binding picture of bonding
\begin{enumerate}
    \item[5.3.1] Adsorbate-substrate interaction
    \item[5.3.2] Adsorbate band structure
  \end{enumerate}

\item[5.4] Adsorption of isolated adatoms

\begin{enumerate}
\item[5.4.1] Geometry
\item[5.4.2] Density of states $\Delta N(\epsilon)$
\item[5.4.3] Electron density: $n({\bf r})$,
$\Delta n({\bf r})$, and $n^\Delta ({\bf r})$
\item[5.4.4] Surface dipole moments
\end{enumerate}

\item[5.5] Alkali-metal adsorption: the traditional picture of 
           {\em on-surface}  adsorption

    \begin{enumerate}
    \item[5.5.1] The Langmuir-Gurney picture
    \item[5.5.2] Coverage dependence of the work function
    \item[5.5.3] Ionization of the adsorbate and screening by the
                  substrate
                  electrons
    \item[5.5.4] Surface band structure
    \end{enumerate}

\item[5.6] Substitutional adsorption and formation of surface alloys

    \begin{enumerate}
    \item[5.6.1] Na on Al(001)
    \item[5.6.2] Na on Al(111)
    \item[5.6.3] Co on Cu(001)
    \end{enumerate}

\item[5.7] Adsorption of CO on transition-metal surfaces -- a
      model system for a simple molecular adsorbate

\item[5.8] Co-adsorption [the example CO plus O on Ru(0001)]

\item[5.9] Chemical reactions at metal surfaces

    \begin{enumerate}
    \item[5.9.1] The problems with ``the'' transition state
    \item[5.9.2] Dissociative adsorption and associative desorption
                   of H$_2$ at transition metals
\begin{enumerate}
    \item[5.9.2.1] The potential-energy surface of H$_2$ at transition-metal surfaces
    \item[5.9.2.2] The dynamics of H$_2$ dissociation at transition-metal
             surfaces
 \end{enumerate}
\end{enumerate}

    \item[5.10] The catalytic oxidation of CO

\item[5.11] Summary outline of main points
\end{enumerate}

\ni References

\newpage
\setcounter{section}{5}

\subsection{Introduction}
\label{sec:intro}

The theory of adsorption has reached a level where it is possible
to calculate free energies, as well as the electronic and atomic
structure,
of
medium-sized systems with predictive accuracy. Such {\em ab initio}
calculations (i.e., starting from the electronic structure) typically
employ
complicated methods and significant computational resources. Clearly,
the methodological developments of recent years have been impressive,
although further developments, enhancements, and speed-ups of such
methods are still necessary. Acknowledging the predictive
power of density-functional theory calculations,
however, we also note that the need remains for finding explanations and
developing simple concepts. Today's concepts are largely based on
experience from gas-phase chemistry, as for example the concepts of
electronegativity, HOMOs (highest occupied molecular orbitals), LUMOs
(lowest unoccupied molecular orbitals),
and the reactivity of open-shell systems. However,
it is also clear that some concepts, which are powerful in the gas-phase
chemistry of molecules, can be quite misleading when it comes to
surfaces. One example is that of ``the'' transition state of a chemical
reaction: For molecular chemistry this concept is typically useful, but
for
molecules at surfaces it turns out that the dimension of phase space
is so high that not just one, but many transition states exist,
and all of them may play a role (see Section \ref{sec:reactions}).

What is needed now, and in the years to come, is to perform more
predictive
simulations of surface chemical reactions; but also the next step
can and should be done, which is the development of {\em explanations}
and {\em understanding}. For example, at this point we are not able to
rationalize the factors that determine at which transition-metal
surface a rotationally excited molecule will dissociate more easily
than a molecule that is not rotationally excited
(see Section \ref{sec:H2}). Also we are just starting to
develop an understanding of why some adsorbates occupy a substitutional
site, and not just simply adsorb {\em on} the surface, and why this
geometry
may change with coverage. Note that ten years ago substitutional
adsorption of single adatoms was unheard of, but now, it is a well known
and common phenomenon (see Section \ref{sec:substitutional}).

The discussion presented in this chapter is based on results from
density-functional theory calculations. Rather than discussing
details of these calculations, we simply use the results to show where
we are on
the way to attaining a  rationalization of the factors which actuate the
surface chemistry for different systems.
This goal does   not (and cannot) aim at a quantitative description, and
it is clear that concepts, e.g., that of electronegativity, often do
not withstand a quantitative analysis. However, describing the results
in words and in a chemical language, is the basis on which
``understanding''
is built. Clearly, the quality (i.e., the nature or character)
of an effect is linked to the quantitative
numbers, and sometimes a small change in the quantity changes the
quality. In this sense our goal is insecure. But we strongly believe
that capturing the nature of a situation and understanding trends
is the essence of ``understanding''. As we said above, theoretical
surface science is now able to do this, and some examples along this
route already exist, some of which will be discussed in this chapter.

\subsubsection{The nature of the surface chemical bond: indeterminate
concepts, yet  useful ...}
\label{sec:intro-1}

A crude classification of adsorption distinguishes two classes,  
namely that
of a very weak (van der Waals type) interaction
between  adsorbate and substrate (phy\-si\-sorption), where the
adsorption energy is typically less than 0.3 eV per adsorbed particle
%%%(29kJ mol$^{-1}$),
(6.9 kcal mol$^{-1}$),
and that of chemisorption, where the adsorption energy
is larger.

For chemisorption systems there is a further classification of
the nature of bonding which is frequently applied, although it is
neither unique nor general. Nevertheless, it has some natural
advantages and will also be applied in this chapter. It is based on a
survey of electronic, electrical, vibrational, and thermal
properties. Thus, altogether we will distinguish four different
types of bonding:
\begin{itemize}
   \item [1.] van der Waals,
   \item [2.] covalent,
   \item [3.] metallic,
   \item [4.] ionic.
\end{itemize}
As little the nature of bonds of types 2, 3, and 4 is well defined,
that of van der Waals bonding is equally so. The concept of the
van der Waals interaction is valid for large distances  where
orbitals do not overlap. It is due to electron-density fluctuations 
at the
different atoms and the polarization that the fluctuations induce at
the other atoms.
However, at the equilibrium geometry of an adsorbate on a surface,
the direct
interaction of adsorbate and substrate orbitals  is significant. This
is indeed plausible as equilibrium geometries are determined by the
interplay of attractive interactions and Pauli repulsion. As a 
consequence,
at the bonding geometry of adsorbed noble-gas atoms which are regarded
as
exhibiting a van der Waals like bonding,
and even more so at the
turning point of noble-gas atom scattering at surfaces, the physics is
largely
determined by the interaction of orbitals and {\em static} polarization.
Thus, {\em physisorption}, although implying a weak
interaction strength, can induce a static dipole moment
at the adsorbate (in particular for larger atoms,  e.g., Xe),
and the electrostatic interaction of this adsorbate dipole
with the substrate contributes noticeably to the bond strength.
Indeed, density-functional theory (DFT) calculations performed using
either the local-density approximation (LDA) or the generalized gradient
approximation (GGA) for the exchange-correlation interaction,
which both lack a description of the nature of the van der
Waals interaction, as this is intrinsically non-local,
seem to give a reasonable description of adsorption
of noble-gas atoms at surfaces (see, e.g., Brivio and Trioni, 1999).
Nevertheless, it is not clear if the acting ``overlap-modified
dispersion forces'' are described by the present exchange-correlation functionals with sufficient accuracy. In any case, what we like to emphasize is that even weak
bonding can give rise to noticeable changes in the electrostatic field
at the surface and therefore to changes in surface properties.
We elaborate here on this discussion more than usual because the
use of the above noted four bonding types
has often caused confusion when they were interpreted literally.
Knowing about the danger
of over-interpreting the relevance of these concepts we
still feel that  the advantage of categorizing systems
is important for identifying trends, noting unusual effects, and for
{\em scientific understanding} in general.
In Sections \ref{sec:tight}, \ref{sec:nature}, and \ref{sec:LT-alkali} we
will continue this discussion with respect to ionic and covalent bonding.

Metallic bonding, also noted in the above list, is a special case of
covalent interaction where the electron density is more delocalized and
not (strongly) peaked between the atoms. In principle, the attractive
interaction is due to the fact that the electrons act like a
structure-less ``glue'' between the nuclei. This attraction is stabilized
by a repulsive term due to the kinetic energy of the electrons, because
electrons do not like to become localized, but instead, as
a consequence of the
Heisenberg uncertainty principle, they like to spread out.
For {\em isolated} adatoms, the concept of metallic bonding is not very
useful because metallic bonding, and the concept of
delocalized electrons, requires a high coordination.
Therefore we will not allude to this concept in Section \ref{sec:nature}
where
isolated atoms are discussed, but we will in Sections
\ref{sec:LT-bands} and
\ref{sec:substitutional}
when we compare adlayers of different density
and discuss the formation of a surface electronic band structure.

\subsubsection{What will be discussed and why,
and what is missing}
\label{sec:intro-2}

This chapter deals with various adsorbates on metal surfaces, where
we focus on the electronic properties and the nature of the chemical
bond. To some extent this also necessitates a discussion of
the atomic geometries because the distances and directions
between the atoms and their neighbors, determine which, and how,
orbitals will hybridize. Just as one example we mention
aluminum which is a nearly-free-electron metal. However, when an Al atom 
is
placed in a certain local coordination, its $s$- and $p$-electrons can
hybridize and form directional bonds. This is in fact well known even
for bulk systems, as for example AlAs. This property of Al, namely
being close to a covalent material, contributes to the low formation
energy
of surface vacancies at Al(111) (see Neugebauer and Scheffler, 1992) and
is instrumental in the theoretical result that self diffusion at Al(001)
proceeds via the exchange mechanism (see Feibelman, 1990; Yu and
Scheffler,
1997).

We will keep the discussion in this chapter simple and accentuate
the {\em qualitative} nature of the various mechanisms. Nevertheless, we
emphasize that
all results discussed below are based on quantitative calculations 
performed
using density-functional theory (DFT) (Dreizler and Gross, 1990).
Indeed, the dialectic relation of quantity and quality will become
obvious through several examples discussed below:
Often a small quantitative difference in certain values will cause a
significantly changed electronic and/or geometric structure and as a
consequence
a different bonding quality or nature.

The exchange-correlation functional employed
in many of the studies discussed below is the local-density
approximation (LDA), which gives a reliable description of geometries
and the nature of the bond. In the more recent studies, which are
presented in Sections \ref{sec:CO} -- \ref{sec:catalytic},
the generalized gradient approximation (GGA) is employed.
For equilibrium geometries the GGA gives results similar to the LDA.
For the description of chemical reactions, in particular of
transition states, where the breaking of old bonds and making
of new bonds occurs, the GGA is in fact mandatory,  i.e.,
the LDA often gives even qualitatively incorrect results
(see for example Hammer et al., 1994, and references therein).

For simplicity's sake, we restrict ourselves to close-packed
substrate surafces [mainly fcc (111) and fcc (001)]. The more open surfaces
are typically close to a structural instability and therefore
sometimes already reconstruct when clean, or when adsorbates are added.
We also limit the number of substrate materials with the view
that this will ease the readability of the chapter. However, essentially 
all
important mechanisms that we would like to discuss are represented by
the systems presented.

Because of their large dipole moments and special role in various
industrial
applications,
the discussion of alkali-metal adsorbates is addressed in particular
detail.
This presentation also focuses on the general
properties;  for more complete discussions we refer to some
recent review papers, e.g., Stampfl and Scheffler (1995) and
Adams (1996).
Noble-gas adatoms\footnote{For a discussion
of recent work on noble-gas atom adsorption see
Bruch et al., 1997; Brivio and Trioni, 1999; Seyller et al., 1998;
Petersen et al., 1996, 2000.}
and $f$-electron
systems are not discussed, and also for metallic substrates not
all ``classes'' are covered: We concentrate on elemental substrates
(only a few words are said about alloys) and consider
low coverages,  i.e., from a single adatom up to a full
monolayer.\footnote{The coverage is defined in this chapter such that
                   for $\Theta=1$, the number of adatoms is the same
                   as the number of atoms in the clean, unreconstructed
                   surface. \label{coverage}}

As the properties of adsorbates on metals are manifold,
we are unable to address all features which may be
relevant in one or another situation. For example,
core-levels and surface core-level shifts (with their
interesting initial state and final state effects) are not mentioned
(see for example Andersen et al., 1994; Methfessel et al., 1995;
Hennig et
al.,
1996; Ganduglia-Pirovano et al., 1997,
and references therein).
The same is true for surface electric resistance, magnetism, and more.

Otherwise, we trust that the selection of systems is representative.
It covers a small fraction of work done by us over the last several years
on adsorption, co-adsorption, and chemical reactions.
We note that similar work has also been published by other groups
(see, e.g., Hammer and N{\o}rskov, 1997, and  references therein).

\subsection{Concepts and definitions}
\label{sec:definitions}

This section starts with some general remarks and then collects
definitions of important quantities which are typically
calculated in theoretical work on adsorption and will be used later
in this chapter.

Accurate knowledge of the geometry of the adsorbate and
substrate atoms is a prerequisite for any additional analysis of the
adsorbate properties, as for example the surface electronic structure,
adsorbate-induced work function changes, and the chemical
reactivity. The apparent hierarchy expressed in this sentence reflects
the fact that
the atomic structure is somewhat more directly accessible for 
experimental
studies. Nevertheless, we emphasize that electronic and atomic structure
are closely interconnected and do not consider it very useful to analyze
``the chicken and the egg problem'',  i.e., to discuss
whether the geometry creates the electronic structure or vice versa.
In fact, we will stress in this chapter (particularly in Sections
\ref{sec:LT-alkali} and \ref{sec:substitutional})
that the same adsorbate
can exhibit a different bonding character, depending on the adsorbate
coverage and the local adsorbate geometry. We will discuss these aspects
for special examples but note that these are  {\em examples} which
represent many systems; they are selected because they demonstrate
the effects most clearly.

\subsubsection{Density of states}
\label{sec:DOS}

\begin{figure}[b]
\unitlength1cm
   \begin{picture}(0,5.0)
\centerline{
       \psfig{file=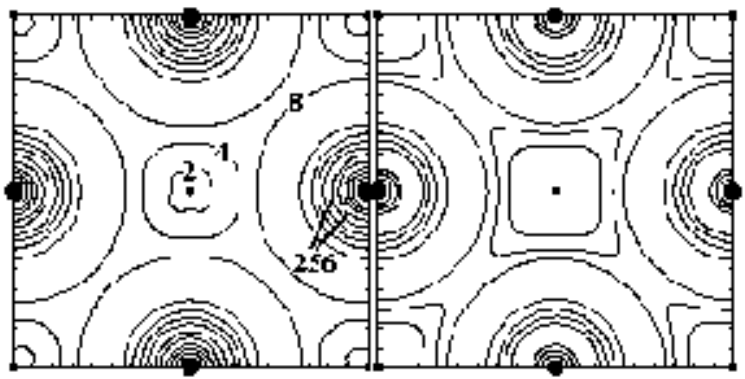,width=10.0cm}
}
   \end{picture}
\vspace{-0.3cm}
   \caption{\small  Lines of constant electron density (valence only) for a
NaCl crystal, the prototype of ionic bonding.
Left: Result obtained by a superposition of the electron densities
of {\em neutral} Na and Cl atoms.
Right: Result from a self-consistent DFT-LDA calculation.
The units are $10^{-3}$ bohr$^{-3}$. Large dots
mark the positions of the Cl atoms, and small dots mark the Na atoms.
Adjacent contour lines differ by a factor of 2. The maximum density 
for the
non-self-consistent calculation (left) is $265 \times 10^{-3}$ 
bohr$^{-3} $
and for the self-consistent calculation (right) the highest density
is $253 \times 10^{-3}$ bohr$^{-3}$ (from Bormet et al., 1994a).}
\label{nacl}
\end{figure}

An important quantity accessible in calculations, though not in 
experiments,
is the local density of states
\begin{equation}
n({\bf r}, \epsilon) = \sum_{i=1}^\infty |\varphi_i({\bf r})|^2
\delta(\epsilon - \epsilon_i),
\label{eq:l-dos}
\end{equation}
where $\varphi_i({\bf r})$ are the single-particle eigenfunctions
of the Kohn-Sham Hamiltonian.
For those who like Green functions this is written as
\begin{equation}
n({\bf r}, \epsilon) = -\frac{2}{\pi} {\rm Im}
\cal{G}({\bf r},{\bf r}, \epsilon),
\end{equation}
where $\cal{G}(\epsilon)$ is the retarded Green-function  operator
of the adsorbate system.
The electron density is
\begin{equation}
n({\bf r}) =
\int \limits_{-\infty}\limits^\infty f(\epsilon, T) n({\bf r},
\epsilon) d \epsilon =
\sum_{i= 1}^{ \infty } f(\epsilon_i, T) |\varphi_i({\bf r })|^2,
\label{eq:n(r)}
\end{equation}
with the Fermi distribution $f(\epsilon, T)$ at temperature $T$.
Often, when the nature of a chemical bond is analyzed, electron density
plots are shown, sometimes complemented with a discussion of charge
transfer from atom A to atom B.
We do not consider this approach so useful and emphasize
that a small difference in the electron density sometimes implies
a significantly different physical-chemical interaction. This
is demonstrated in Fig. \ref{nacl}
which shows the electron density of
bulk  NaCl, that is a system, which is well accepted to have ionic bond
character. However, even in this example the self-consistently calculated
electron density  (right) and that
constructed by a superposition of electron densities of the {\em neutral}
Na and {\em neutral} Cl atoms (left) differ only slightly, and just on the
basis of inspecting $n({\bf r})$
charge transfer can hardly be identified. The small changes brought about
by the self-consistent rearrangement of electron density are that
around the Na nucleus the charge density is very slightly decreased; but
also at the Cl atom we see that the maximum of the electron density,
which
is rather close to the nucleus, becomes slightly lower.
Thus, when one only inspects the electron density, even NaCl does not
present a clear case for a system with ionic bonding. We emphasize that
for
low-symmetry situations, as for example a surface, this problem is even
more
pronounced. However, an analysis of the density of states (see, e.g.,
also Section \ref{sec:Delta-N}) shows that it is
indeed appropriate
to  describe the upper valence band in terms of Cl $3p$ and the lower
conduction band in terms of Na $3s$ orbitals.

The example of Fig. \ref{nacl} shows that despite the fact that the
electron density can be considered the ruling quantity in
density-functional theory, its inspection can be  misleading. Sometimes,
though not always, plots of density {\em differences}
\begin{equation}
\Delta n({\bf r}) = n({\bf r}) - n^0({\bf r})
\label{eq:Deltan}
\end{equation}
give a better impression (see Sections \ref{sec:nature} and
\ref{sec:LT-screening}). Here $n^0({\bf r})$ is the electron density
of the
clean substrate where the geometry is (typically) chosen to be that of
the adsorbate system.
Also the {\em difference density}
\begin{equation}
\label{n-Delta}
   n^\Delta({\bf r}) = n({\bf r}) - n^{\rm 0}({\bf r})
                      -n^{{\rm Na},f_{3 s}}({\bf r})   \quad
\end{equation}
is often a helpful quantity.
Here $n^{{\rm Na},f_{3 s}}({\bf r})$ is the (spherical) electron
density of the partially ionized atom to be adsorbed, where the
occupation
of the valence level is given by the parameter $f_{3s}$.
Just as an example, our indices here refer to a Na atom, where the
valence level is $3s$.

A typically more sensitive quantity is the density of states
(DOS),
\begin{equation}
N(\epsilon) = \int n({\bf r}, \epsilon) d {\bf r} = \sum_{i= 1}^{ \infty }
\delta(\epsilon -\epsilon_i),
\label{eq:DOS}
\end{equation}
where the sum goes over all eigenstates of the Kohn-Sham Hamiltonian.
The DOS gives a noticeable contribution to the electrons' kinetic energy
\begin{equation}
T_s[n({\bf r})] = \int  \limits_{-\infty}\limits^\infty
f(T, \epsilon) N(\epsilon)\, \epsilon\, d \epsilon
- \int  V^{\rm eff}({\bf r}) n({\bf r}) d{\bf r},
\label{eq:T_s}
\end{equation}
where $V^{\rm eff}$ is the effective potential of the Kohn-Sham
Hamiltonian (see Chapter 1 of this book).
We also note that a one-to-one correspondence exists between $N(\epsilon)
$
and the electron density,\footnote{Obviously, as $N(\epsilon)$
defines the Kohn-Sham Hamiltonian,
it also defines all ground-state and all excited-state properties.}
which (in addition to the argument of practicability and clarity)
gives a premise to its use.

Throughout this chapter we will follow the thinking that the nature
of bonding is determined by the interaction of the orbitals at the
different atoms involved, and thus, it is more clearly identified by
an inspection of the DOS and/or by the adsorbate-induced change
($\Delta$DOS)
\begin{equation}
\Delta N(\epsilon) = N(\epsilon) - N^0(\epsilon),
\label{eq:Delta-DOS}
\end{equation}
where $N^0(\epsilon)$ is the DOS of the clean substrate where the geometry is typically chosen to be that of the adsorbate system.
Furthermore, we note the state-resolved DOS, also called the
projected DOS, as a very useful quantity:
\begin{equation}
N_{\alpha}(\epsilon) = \sum_{i= 1}^{\infty} |\langle \phi_\alpha |
\varphi_i
\rangle|^2 \delta(\epsilon -\epsilon_i),
\label{eq:DOS_a}
\end{equation}
where $\phi_\alpha$ is a properly chosen localized function.
The spatial distribution of the electron density is viewed as a 
consequence
(not the origin) of the hybridization of different orbitals.
The critic may argue that this is just (another) chicken-egg problem,
and we agree with this assessment,  i.e., our approach is
simply taken because it is useful.
We also note that information about $N_{\alpha}(\epsilon)$ can be
obtained
(though in a somewhat distorted way) from angle-resolved
photoemission, inverse photoemission, and scanning tunneling 
spectroscopy.

Unlike the bond formation in molecular
chemistry, the case of adsorption is one between very different partners:
The adatom comes with discrete energy levels and few occupied
states, but  the substrate has a near-infinite number ($\sim$$10^{23}$) of
electrons.
Thus, the valence level of the adparticle will interact with a
semi-infinite continuum of levels $\epsilon \geq \epsilon_{0}$, where
$\epsilon_{0}$ is the bottom of the
substrate valence band. The location of the highest occupied level, the
Fermi energy, is determined by the substrate. On a surface, a
description of
the
chemical bonding in terms of an elementary discrete level scheme has
therefore
to be extended: Levels lying in the band region of the substrate 
receive a
finite width. These broad levels
are called adsorbate-induced resonances
and they get filled up to the substrate Fermi level.
The wave functions of these resonances can be understood to arise from
the adsorbate orbitals. In an alternative view, the states of the clean
surface are described as standing waves, incident from the bulk and
reflected at the potential-energy barrier of the clean surface, with a node
at the surface. An
adsorbed particle modifies the reflection properties, which can be
described by introducing an energy-dependent phase shift,
$\delta_{\alpha}(\epsilon)$,
describing the scattering of bulk states of a certain representation
$\alpha$ by the
adparticle. This phase shift is defined by (Callaway, 1964, 1967)
\begin{eqnarray}
{\rm tan} \delta_{\alpha}(\epsilon) = \frac{-{\rm Im} D_{\alpha}
(\epsilon)}{{\rm Re} D_{\alpha}(\epsilon)},
\end{eqnarray}
where $D_{\alpha}$ is the determinant
\begin{eqnarray}
 D_{\alpha}(\epsilon) = {\rm det}\{1 - {\cal G}^0 (\epsilon)
\Delta V \}_{\alpha}.
\end{eqnarray}
${\cal G}^0$ is the Green's function of the bare substrate, and
$\Delta V$ is the change in the potential due to the adsorbate
\begin{eqnarray}
\Delta V = V^{\rm eff}[n] - V^{\rm eff}[n^0].
\end{eqnarray}
The density of states, induced by the adparticle, is given by the
derivative of the phase shift
\begin{eqnarray}
\Delta N(\epsilon) =
\frac{2d_\alpha}{\pi} \frac{d \delta_\alpha(\epsilon)}{d \epsilon},
\end{eqnarray}
where $d_\alpha$ is the dimension of the representation $\alpha$. A
resonance in
$\Delta N(\epsilon)$ will occur at an energy close to that at which the
phase
shift $\delta_\alpha(\epsilon)$ increases through $\pi/2$ with increasing
energy, i.e.,
where the real part of the determinant $D_\alpha$ vanishes. Close to this
energy, $\epsilon_\alpha$, the induced density of states takes a 
Lorentzian
line shape
\begin{eqnarray}
\Delta N(\epsilon) = \frac{d_\alpha\Delta_\alpha}{\pi} \frac{1}{
(\epsilon - \epsilon_\alpha)^{2} -
\left( {\Delta_\alpha}/2 \right)^{2}}.
\label{eq:AGN}
\end{eqnarray}
The width of the resonance, $\Delta_\alpha$, is
\begin{eqnarray}
\Delta_\alpha = \left[ \frac{2{\rm Im} D_\alpha(\epsilon)}{({d
{\rm Re} D}_\alpha (\epsilon)/d \epsilon)}
\right]_{\epsilon = \epsilon_\alpha}.
\end{eqnarray}

We note the close similarity between Eq. (\ref{eq:AGN}) and the
Anderson-Grimley-Newns model of chemisorption (Grimley, 1975;
Muscat and Newns, 1978, 1979).
On the lower energy side of the resonance, the phase of the reflected
wave
is
shifted such that electron density is accumulated in the region of the
adparticle-substrate bond, indicating that these states are bonding in
character. On
the
higher energy tail of the resonance, the electron density in the bond
region
is
reduced,
indicating that these states are antibonding in character (Lang and
Williams, 1978; Liebsch, 1978). However,
if
the interaction with the substrate is very strong,
bonding and  antibonding states will split apart:
A bound state (or resonance) is then formed
below the substrate band, which is bonding in character, and a broad
(antibonding) resonance will appear in the valence band (see
also the discussion of Fig. \ref{tb-DOS} in Section
\ref{adsorbate-substrate-interaction}).

\subsubsection{Energies}
\label{sec:energies}

We will assume that the dynamics of the electrons and the nuclei can
be decoupled and that whatever the dynamics of the nuclei are, the
electrons
are in the electronic ground state of the instantaneous geometry.
This is the Born-Oppenheimer approximation (Born and Oppenheimer, 1927;
Born and Huang, 1954),
which for adsorbates, and often also for chemical reactions,
is well justified; for some reactions, and in particular for
photo-chemistry,
important violations of the Born-Oppenheimer approximation occur, but
this
is not the subject of this chapter.
The {\em DFT total energy}, $E^{\rm total}(T, V, N^{\rm nuc.}_A,
N^{\rm nuc.}_B, \cdots,\{{\bf R}_I\})$,
at temperature $T$, volume $V$, and composition
$N^{\rm nuc.}_A, N^{\rm nuc.}_B, \cdots$,
when studied
as a function of the atomic coordinates, is often called the
potential-energy surface (PES) because it defines the potential-energy
landscape on which the nuclei $A, B, \cdots$ travel.
It is related to an experimentally accessible quantity only in a
restricted
way:  If, as is typically done, the self-consistent calculations are
performed at constant volume, the {\em DFT total energy} corresponds
to the
Helmholtz free energy at zero temperature and neglecting
zero-point vibrations. In general, the Helmholtz free energy is
\begin{eqnarray}
F(T, V, N^{\rm nuc.}_A, N^{\rm nuc.}_B, \cdots, \{ {\bf R}_I \} ) & =
& E^{\rm total}(T, V, N^{\rm nuc.}_A, N^{\rm nuc.}_B, \cdots,
\{{\bf R}_I \})
\nonumber \\
& & + E^{\rm vib.}(T, V, N^{\rm nuc.}_A, N^{\rm nuc.}_B, \cdots, 
\{ {\bf R}_I \})
\nonumber \\
& & - T\, S(T, V, N^{\rm nuc.}_A, N^{\rm nuc.}_B, \cdots, \{ 
{\bf R}_I \} )
\label{helmholtz}
\end{eqnarray}
with the vibrational contribution noted as $E^{\rm vib.}$ and
$S$ is the entropy. At a given volume $V$,  the atomic geometry of
stable
or metastable configurations is determined by
\begin{equation}
\left( \frac{\partial F(T, V, N^{\rm nuc.}_A, N^{\rm nuc.}_B, \cdots,
\{ {\bf R}_I \} ) }
{\partial {\bf R}_I } \right)_{T,\, V, N^{\rm nuc.}_A, N^{\rm nuc.}_B, 
\cdots} = 0,
\end{equation}
and for a given pressure $p$  it is determined by
\begin{equation}
\left( \frac{\partial G(T, p, N^{\rm nuc.}_A, N^{\rm nuc.}_B, \cdots,
\{ {\bf R}_I \} ) }
{ \partial {\bf R}_I } \right)_{T,\, p, N^{\rm nuc.}_A, N^{\rm nuc.}_B,
\cdots} = 0,
\end{equation}
where
\begin{eqnarray}
G(T, p, N^{\rm nuc.}_A, N^{\rm nuc.}_B, \cdots, \{ {\bf R}_I\}) & =
& F(T, V, N^{\rm nuc.}_A, N^{\rm nuc.}_B, \cdots, \{ {\bf R}_I\})
\nonumber \\
 & & + p\, V(T, p, N^{\rm nuc.}_A, N^{\rm nuc.}_B, \cdots, \{ {\bf R}_I\})
\label{gibbs}
\end{eqnarray}
is the Gibbs free energy. If the system is in contact with a particle
reservoir, for example, the sample is held in some gas phase, particles
can be exchanged between the system and the reservoir.
Then we have to add to Eq. (\ref{helmholtz}) a term
$- \sum_X \mu_X  N^{\rm nuc.}_X$, where
$\mu_X$ is the atom chemical potential of atom-type $X$, which
can be controlled by external reservoirs, i.e., by the
evironmental conditions (partial pressure and temperatur).

Later in this chapter we will study the  {\em adsorption energy
per adatom}. This is the difference of the total energy of the
adsorbate system and the total energy of the clean substrate
together with a corresponding number of free, neutral atoms.
For {\em on-surface} adsorption this reads
\begin{equation}
E_{\rm ad}^{\rm Na/Al(001)}  =  - \left( E^{\rm Na/Al(001)}
-  E^{\rm Al(001)} - N^{\rm nuc.}_{\rm Na} E^{\rm Na-atom} \right)/ N^{\rm nuc.}_{\rm Na},
\label{eq3_1}
\end{equation}
where $E^{\rm Na/Al(001)}$ is the total energy per adatom of the
adsorbate system, $E^{\rm Al(001)}$ is the total energy of the clean
Al(001) substrate, and $N^{\rm nuc.}_{\rm Na} E^{\rm Na-atom}$ is the
total
energy of  $N^{\rm nuc.}_{\rm Na}$ free Na atoms that take part in the
adsorption.
We have used here as an example indices which refer to the adsorption of 
Na
on Al(001), but translation to other systems is obvious.

Often adatoms are {\em not}
adsorbed on the surface with only slight modification of
the original surface structure, but instead  adsorption may occur
substitutionally. In this case the adatom kicks out an atom from the
surface and takes its site. In thermal equilibrium the kicked out atom
is then re-bound at a kink site at a step (the binding
energy at a kink site equals the bulk cohesive energy, which is shown in the next section).

For substitutional adsorption, the adsorption energy is defined
essentially the same way as above, only that the kicked off surface
atoms that are
re-bound at kink sites, have to be accounted for as well. Each rebound kicked out atom contributes an energy equal to that of a bulk atom (see Section \ref{sec:kink}).
Thus, for the substitutional adsorption the adsorption energy is
\begin{eqnarray}
E_{\rm ad}^{\rm Na/Al(001)-sub}
& = &  - ( E^{\rm Na/Al(001)-sub} + N^{\rm nuc.}_{\rm Na}
E^{\rm Al-bulk}
\nonumber \\
& &-  E^{\rm Al(001)} - N^{\rm nuc.}_{\rm Na} E^{\rm Na-atom} )/N^{\rm nuc.}_{\rm Na}.
\label{eq3_2}
\end{eqnarray}
$E^{\rm Al-bulk}$ is the energy of an atom in a bulk crystal of aluminum
and the quantity $E^{\rm Na/Al(001)-sub}$ in Eq. (\ref{eq3_2}) is the
total
energy of the slab with the adatoms adsorbed in substitutional sites.

We end this section with a warning that may be relevant for 
{\em Surface Science} studies more often than typically appreciated.
The energy quantities defined in this section are relevant
if thermal equilibrium conditions are attained, but often
this is not the case. Instead, a surface is studied which may be in a 
metastable state and this state hopefully corresponds to an 
equilibrium situation of the sample's history (at best).
This somewhat unclear situation will hopefully
improve in the future, when not just UHV studies are performed but 
also experiments under well controlled atmospheres.
We also note that structures in 
{\em local}  thermodynamic  
equilibrium, or metastable geometries can have a very long life time
-- just consider diamond (a metastable structure of carbon)
for which at room temperature the phase transition to its ground-state
crystal structure (graphite) is known to be rather slow.
In particular, for multi-component
materials, the stoichiometry at the surface requires exchange of atoms
or molecules with some reservoirs and this can be hampered by
significant energy barriers. Metal oxides represent an example
which comes to mind, as in UHV, oxygen can desorb into the chamber,
but with respect to the metal content at the surface, attaining the
thermal-equilibrium stoichiometry is hampered
(see, e.g., Wang et al., 1998).

\subsubsection{Binding energy at kink sites}
\label{sec:kink}

As mentioned above, kicked out surface atoms may be re-bound at kink
sites
at steps. We will see that this is in particular relevant when we
discuss below in this chapter the substitutional adsorption of alkali
metals atoms
and of cobalt. Therefore we now present a simple discussion of the
energetics of substrate atoms at kink sites. Although we use a
simplified model, we note that the conclusions are in fact valid
in general.

The total energy of a many-atom system can be written as the sum over
contributions assigned to the individual atoms,
\begin{equation}
E = \sum_{I=1}^{N^{\rm nuc.}} E_I.
\label{bond-cut}
\end{equation}
$E_I$ is the energy contribution due to atom $I$.
In an approximate way we can write $E_I = E_I(C_I)$,
where $C_I$ is the number of nearest-neighbor atoms of atom $I$.
This is a simplified presentation of an approach which has
many names (e.g., effective-medium theory, embedded-atom method,
Finnis-Sinclair potentials, glue model) and which we all call 
{\em bond-cutting
models}. The various implementations are similar but differ in the
way the function $E(C)$ is represented.
In the simplest, yet physically meaningful approach, the function
$E(C)$ is roughly proportional to $\sqrt{C}$ (see
Spanjaard and Desjonqu\`eres, 1990; Robertson et al., 1994;
Payne et al., 1996; Methfessel et al., 1992b; Christensen and Jacobsen,
1992).

\begin{figure}[b]
\unitlength1cm
   \begin{picture}(0,3.5)
 \centerline{
       \psfig{file=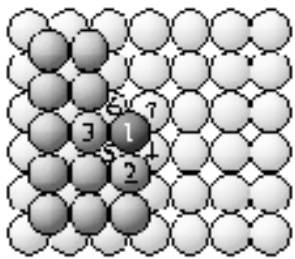,width=4.0cm}
}
\end{picture}
\vspace{-0.3cm}
   \caption{\small View of a kink site at a step on an fcc (001) surface.
The local coordination of an atom (labeled 1) adsorbed  at a kink
site $C_1$ changed from 0 (that of the free atom) to 6. For its
six neighbor atoms (labeled 2, 3, 4, 5, 6, 7) the coordination
numbers are increased by one, thus attaining the values:
$C_2 =$ 7,
$C_3 =$ 8,
$C_4 =$ 10,
$C_5 =$ 12,
$C_6 =$ 11,
$C_7 =$ 9.
}
\label{fig:kink}
\end{figure}

If we consider a single-element system and place
an atom at a kink site at a step of an fcc (001) surface, the changes of
the local coordination are as noted in Fig. \ref{fig:kink}.
Putting this information into Eq. (\ref{bond-cut}) and calculating the
{\em change} in the total energy by placing an atom ($I=1$) at a kink
site (i.e., $E^{\rm kink}_{\rm ad} = -E^{\rm after} + E^{\rm before}$)
gives
\begin{eqnarray}
E_{\rm ad}^{\rm kink} & =
& - \left( E_{1}(6) +E_{2}(7) +E_{3}(8)
     +E_{4}(10) +E_{5}(12) +E_{6}(11)
     +E_{7}(9) \right) \nonumber\\
& &  + E_{1}(0)+E_{2}(6)+E_{3}(7)+E_{4}(9) +E_{5}(11)
     +E_{6}(10) +E_{7}(8) \nonumber\\
& = & -E_{5}(12)+E_{1}(0).
\label{kink}
\end{eqnarray}
Thus, the adsorption energy at a kink site equals the cohesive energy.
This result is in fact plausible if one considers
that the adsorption at a kink site leaves the system essentially 
unchanged,
because the kink site is simply moved by one atom and thus, the
situation at the surface is physically not altered. Therefore
the difference between the original and final situations is
simply (and rigorously) the addition of one bulk atom, which has
the energy $E(12)$.

\subsubsection{The surface energy barrier}
\label{sec:barrier}
\begin{figure}[b]
\unitlength1cm
   \begin{picture}(0,8.0)
\centerline{
    \psfig{file=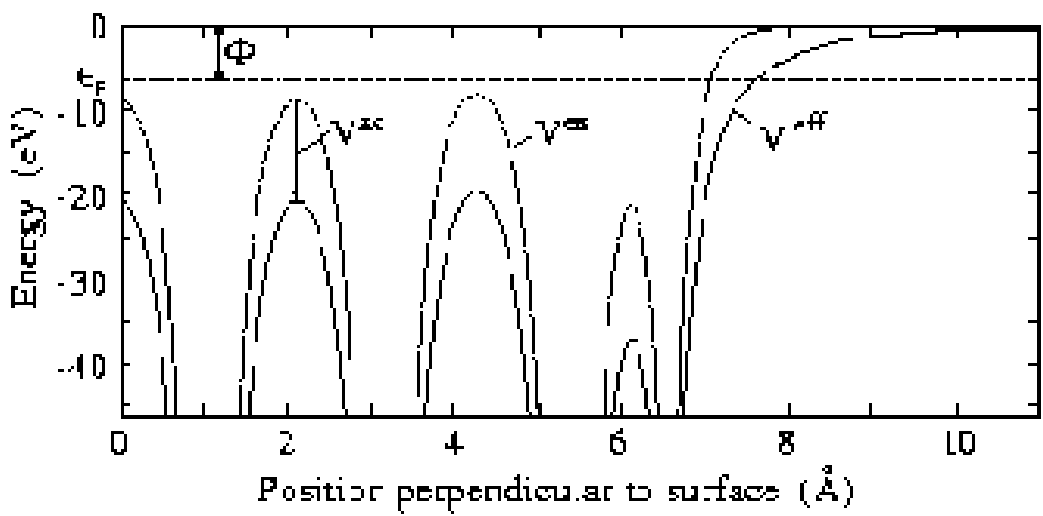,width=14.0cm}
}
    \end{picture}
\vspace{-0.3cm}
   \caption{\small The Kohn-Sham effective potential $V^{\rm eff}$ (solid line)
and the electrostatic potential $V^{\rm es}$ (broken line) at an
adsorbate-covered surface.  In the example, which shows results for
O/Ru(0001), the electrostatic potential in the bulk is below the
Fermi level, but for substrates with  low electron density it will be
above. }
\label{effective-potential}
\end{figure}

We add some remarks on the behavior of the effective Kohn-Sham
potential, $V^{\rm eff}$ (see Eq. (\ref{eq:T_s}) and Fig.
\ref{effective-potential}).
The increase in $V^{\rm eff}$
at the surface from its average bulk value up to the vacuum level is 
termed
the surface barrier. Inside the solid, the potential $V^{\rm eff}$ seen 
by
an electron becomes attractive,
due to the electrostatic potential of the ion cores, due to the
electrostatic field of the surface-dipole layer, and due to the lowering 
of
the electron energy by the formation of an exchange-correlation hole. For
an electron in the vacuum region, the potential is described by the
classical
image effect
\begin{eqnarray}
V^{\rm eff}(z) = - \frac{1}{4 \pi \varepsilon_0} 
\frac{(e^{-})^{2}}{4(z - z_{0})} \mbox{\quad \quad
for  $(z - z_0) 
\, \raisebox{.6ex}[-.6ex]{$>\!\!\!\!$}
\raisebox{-.6ex}[.6ex]{$\sim$}\, 2$ \AA},
\label{eq:image-potential}
\end{eqnarray}
where $z$ is the position of the electron,
$z_{0}$ is the position of the reference plane of
the image effect, and $\varepsilon_0$ is the vacuum 
dielectric function.
Figure \ref{effective-potential}
shows the electrostatic potential and
the effective one-particle potential at an adsorbate-covered surface. 
Here,
$\Phi$ is the work function, $V^{\rm es}$ is the electrostatic potential
due to the electron density $n({\bf r})$ and the nuclei,
and $V^{\rm xc}$ is the exchange-correlation potential. The latter is
attractive
and varies roughly as $n({\bf r})^{1/3}$. The only surface contribution
to the total height of the
surface barrier is due to the electrostatic potential; the
contribution of the exchange-correlation term to the height
is a property of the bulk and not of the surface.

Nevertheless, the behavior of the potential at the surface is largely
determined by
the distortion of the exchange-correlation hole which stays behind as an
electron passes through the surface region into the vacuum. In the 
local-density
approximation, the exchange-correlation hole is assumed to be spherically
symmetric and centered on the electron. Although both assumptions are, in
general, incorrect, this approximation affects the potential 
$V^{\rm eff}$
significantly only in the surface region. The local-density approximation
therefore yields an exponential decrease of the effective potential near 
the
surface rather than the l/$z$ behavior of the image effect. This
inaccuracy
of the local-density approximation appears to become noticeable well
outside
the surface region in the vacuum. Properties such as the ground-state
electron density, the work function, or the surface energy seem to be
relatively little affected (see, e.g., Methfessel et al., 1992a, b).
The detailed shape of the barrier thus appears to be of less importance
than its position and height.

The barrier affects individual electron wave functions. For example,
reflection of electrons from the inner side of the barrier can give
rise to surface resonances when the
electron is trapped between the barrier and the rest of the crystal. In
particular, the barrier position influences the energies of adsorbate
states.
For excited states, there are additional effects. Virtual surface 
resonances
will appear, which in LEED, for example, show up as narrow peaks in the
intensity versus voltage curves (McRae, 1971; Jennings, 1979).
Furthermore, the wavelength of an
electron is longer outside the crystal than inside. This yields the
well-known
refraction effect which broadens the angular range of emitted electrons
in the
vacuum region such that, at high polar angles, the electron current 
vanishes
(see, for example, Scheffler et al., 1978). Another effect is found
in the
interaction of light
with the surface: Photoabsorption, and hence photoemission, require a
gradient
in the potential. The surface barrier thus yields a special 
contribution to
this
excitation, called the surface photoeffect. For photoemission from
adsorbates on
transition metals, this contribution appears to be very small compared
to that
due to the ion core potentials. However, it can be important in systems
where
the valence electrons are nearly free electron-like, i.e., where their
interaction with the ion cores is small. For excited states, the inner
potential
will be modified (also becoming complex) due to inelastic 
electron-electron
interactions such as the
excitation of electron-hole pairs, plasmons or surface plasmons,
and dynamical corrections (e.g., a delay in the response of the 
electrons).

\subsection{The tight-binding picture of bonding}
\label{sec:tight}

\subsubsection{Adsorbate-substrate interaction}
\label{adsorbate-substrate-interaction}
\begin{figure}[b]
\unitlength1cm
   \begin{picture}(0,9.0)
\centerline{
       \psfig{file=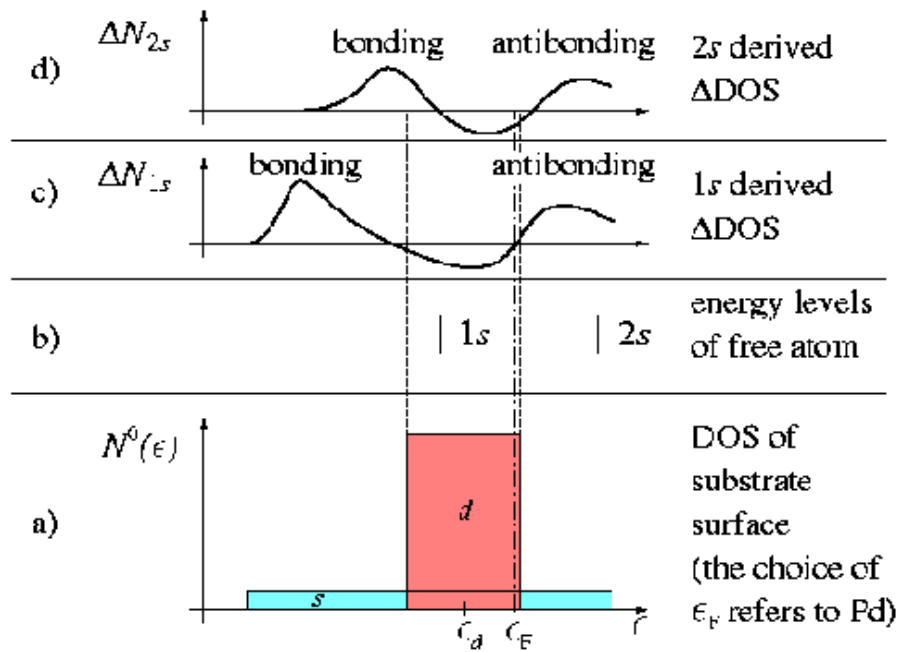,width=12.0cm}
}
\end{picture}
\vspace{-0.3cm}
   \caption{\small  Formation of adsorbate-induced peaks in the DOS.
The bottom panel (a) shows the density of states for a transition
metal before adsorption ($\epsilon_d$ is the center of the $d$-band).
Panel (b) shows the Kohn-Sham energy levels of a free atom (here as an
example, H). The interaction between the H $1s$-level and the substrate
$s$- and $d$-bands gives rise to a broadening and the formation of an
antibonding level (above the $d$-band) and a bonding level (below
the $d$-band) -- see panel (c).
Panel (d) shows that the interaction between the H $2s$-level with 
the substrate $s$- and $d$-bands gives rise to a broadening and the 
formation of a bonding level (at about  the lower edge of the $d$-band) 
and an antibonding level (at about the empty free atom $2s$-level).
}
\label{tb-DOS}

\end{figure}

When an atom and a surface start to interact, the respective
states mix and new states
are created which have energy levels usually broadened and shifted with
respect to the energy levels found in the uncoupled systems.
Typically, the new states can still be related to the original ones, and
Fig.  \ref{tb-DOS} shows a schematic tight-binding description of the
interaction of an atom with a transition metal surface.
The free atom's electronic structure is noted in panel (b)
of the figure, and the electronic structure of the  substrate
is sketched in panel (a).
In principle we could choose any adatom we like for this discussion,
but for ease  we take hydrogen. Therefore the two levels of relevance,
which result from solving the Kohn-Sham equation are the hydrogen
$1s$ and hydrogen $2s$ levels; for simplicity we neglect contributions
from the H $2p$ and other higher-lying states. We note that the highest
occupied DFT-LDA (and DFT-GGA) Kohn-Sham eigenvalue should not be 
confused
with ionization energy (see the discussion of Fig. \ref{Na3s} and
Eq. (\ref{eq-janak}) below). Instead, for partially occupied valence states it is roughly at the mid value of
the ionization and affinity energies.

Different transition-metal substrates mainly differ in the width of
the $d$-band, which increases from $3d$ to $4d$ to $5d$; and they differ
in the position of the Fermi level, which varies from the left to the
right of the periodic table as follows: For the $4d$ series it is at the
lower edge of the $d$-band for strontium, just below the top of the
$d$-band for palladium, and about 3 eV above the upper edge of the
$d$-band for silver. Thus, in the example of Fig. \ref{tb-DOS} we use
the Fermi level $\epsilon_{\rm F}$ corresponding to palladium.

At first we consider the role of the substrate $s$-band.
When the adsorbate and the substrate interact, the hybridization of
the adsorbate wave functions and the states of the substrate $s$-band
gives rise to a broadening of the adsorbate levels, and the atomic
levels will shift because the substrate Fermi level and the electron
chemical potential of the atom become aligned. The latter will result
in a fractional electron transfer (see the discussion of Fig. \ref{Na3s}
below). An analysis of the wave-function character in such a broadened
peak shows that the low-energy part of the peak belongs to states which
have an increased electron density between the adsorbate and the
substrate (such states are called ``bonding''),
and the high-energy part of the peak belongs to states which
have a node between the adsorbate and the substrate (such states are
called ``antibonding''). The substrate $s$-electrons spill out most
into the vacuum, and that is why the broadening (and shifting)
of electronic levels is the first change that happens when an atom is
brought toward a surface. Broadening implies a coupling of the formerly
localized electrons of the adatom to the substrate, thus a delocalization.
In  fact, still neglecting (for the moment) the interaction of the adatom
with the  $d$-band, there are three contributions which affect the adatom levels:
\begin{itemize}
\item[$(i)$] A shift which can go in either direction and which
is caused by charge transfer or charge redistribution (or
polarization) at the adatom. For partially occupied valence states it largely reflects an alignment of the adsorbate DOS with respect to the substrate Fermi level.

\item[$(ii)$] A shift toward
lower\footnote{Our wording is chosen as such that ``higher energy''
of a bound electronic state means closer to the vacuum level, i.e.,
with respect to Fig. \ref{tb-DOS} or \ref{Na-Si-Cl-DOS} the energy is 
more
to the right. Obviously, ``lower energy'' then refers to an energy more
to the left in these figures.\label{higher-lower}}
energies
because the potential at the surface is lower than that in vacuum (cf.
Fig. \ref{effective-potential} and/or Chapter 1 of this book).

\item[$(iii)$] A contribution which implies a shift to
lower$^{\ref{higher-lower}}$ energies because the self-interaction
(an artifact in DFT-LDA and DFT-GGA calculations) is  smaller for
the more extended states of the adsorbate than for the states in
the free atom.
\end{itemize}
These broadened and shifted energies are called ``renormalized atomic
levels''. The three contributions are indeed significant. As a
consequence, a self-consistent treatment is crucial for calculating
the adsorbate-substrate interaction.

We note that the self-interaction effect is indeed strong when
DFT-LDA energy levels are studied. But in total-energy {\em differences}
it nearly cancels out. This is demonstrated in Fig. \ref{Na3s}, which
shows the DFT-LDA
\begin{figure}[b]
\unitlength1cm
   \begin{picture}(0,6.5)
\centerline{
       \psfig{file=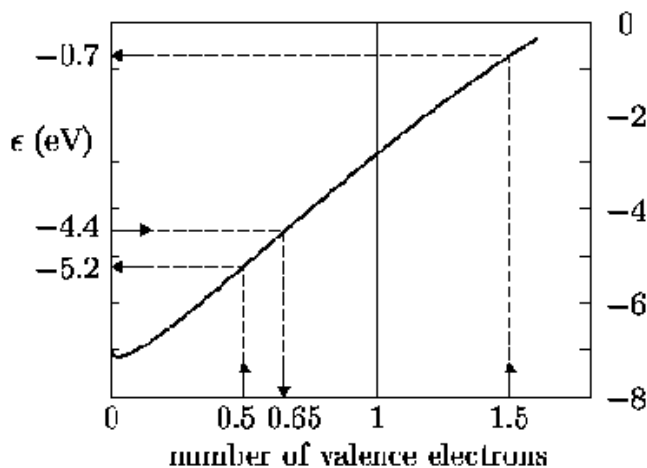,width=8.5cm}
}
    \end{picture}
\vspace{-0.3cm}
   \caption{\small  Kohn-Sham energy level (DFT-LDA) of the $3s$-state of Na as
function of the number of valence electron: $ f_{3s} = 0$ is the
Na$^+$ ion, $ f_{3s} = 1$ is the neutral atom, and $f_{3s} = 2$ is the
Na$^-$ ion. For $f_{3s} = 0.5$ the eigenvalue gives the ionization
energy (here 5.2 eV; the experimental result is 5.14 eV).
For $f_{3s} = 1.5$ the eigenvalue gives the electron affinity
(here 0.7 eV; the experimental result is 0.55 eV).
If the eigenvalue were fixed by an electron reservoir to
4.4 eV, which is the Fermi level of Al(001), an occupation of 0.65 of
the Na valence shell would result.
}
\label{Na3s}
\end{figure}
eigenvalue for the atomic Na $3s$-state as function of its occupation.
For the neutral Na atom the occupation is 1 and the energy level
is at $-$2.8 eV.
The extent to which a Kohn-Sham energy eigenvalue $\epsilon_k$
reflects the ionization energy, which is the minimum energy to remove
an electron from the $k$-th level, depends on how strongly the eigenvalue
depends on the occupation number $f(\epsilon_k, T)$.
If this dependence is negligible, then the negative of the energy 
eigenvalue
equals the excitation energy. This is the density-functional theory
analogue
of Koopmans' theorem in Hartree-Fock theory.

In general, the ionization energy ($I_{k}$) is defined by the
total-energy
difference of the neutral atom  and the positively charged
ion, and this can be read off from Fig. \ref{Na3s} as the energy at
occupation 0.5. This approach only employs the mean-value theorem
of integration
\begin{equation}
I_k = E^{N-1} - E^N =  \int_{N}^{N-1}
\frac{ d E^{N'}}{d N'} dN'
= - \int_{0}^{1} \epsilon_k(f_k) d f_k
\approx -\epsilon_k(f_k=0.5).
\label{eq-janak}
\end{equation}
It is called the Slater-Janak transition-state approach
of evaluating total-energy differences (Janak, 1978),
which works very well in LDA and GGA calculations
(for a discussion of the general proof see Perdew and Levy, 1997;
Kleinman, 1997; and references therein).
According to Fig. \ref{Na3s} the transition-state
gives $I_k \approx 5.2$ eV, which agrees well with the experimental
result for the ionization energy of 5.14 eV.
Analogously we note that the highest occupied Kohn-Sham eigenvalue
at occupation 1.0 agrees well with the mean value of the
ionization and affinity energies.

The result shown in Fig. \ref{Na3s} is typical for all atoms. It
demonstrates the importance of electron-electron correlation,
though it is also largely due to self-interaction, an LDA artifact.
Clearly, the measurable ionization energies (level at occupation 0.5)
and the theoretical Kohn-Sham eigenvalue (level at occupation 1.0)
are very different and should not be confused.
Figure \ref{Na3s} also shows that typically only  partial
electron transfer is to be expected upon adsorption.
If the substrate mainly
plays the role of providing an electron reservoir, thus,
the electron chemical potential is fixed by the substrate Fermi level,
which for Al(001) is at 4.4 eV below the vacuum level,
Fig. \ref{Na3s} then shows
that the occupancy of the Na $3s$-level will be adjusted to
0.65. Thus we obtain a partial ionization and a shift of
the Na $3s$-level, already from a study of  the free-atom eigenvalue.
This is fully in accord with what is suggested by the electronegativities
of the atoms: $\kappa_{\rm Na}=0.93$ and $\kappa_{\rm Al}=1.61$.

We  continue the discussion of Fig \ref{tb-DOS}.
At close distances of the adsorbate to the surface
the {\em renormalized atomic levels} [i.e., the levels which result
after the effects $(i), (ii)$, and $(iii)$] will interact with the more
localized $d$-states. Because the $d$-band
is rather narrow and its width comparable to the interaction strength,
the interaction will result in a splitting into the bonding states,
$ \psi_{\rm b} \approx (\varphi_{{\rm H}\,1s} + \varphi_{{\rm Pd}\,4d} )$
and the antibonding states, $ \psi_{\rm a} \approx 
(\varphi_{{\rm H}\,1s} -
\varphi_{{\rm Pd}\,4d} )$. The resulting adsorbate-induced  density of
states (DOS) for the H $1s$-state is shown in panel (c) of
Fig. \ref{tb-DOS}. Panel (d) shows the corresponding result induced by
the H $2s$ state. We note that the resulting peaks are close to the lower
and upper edge of the $d$-band and inside the $d$-band the density is
reduced, which simply reflects the fact that these states are shifted
from inside the $d$-band to higher and lower energy upon hybridization
with the adsorbate states.

\begin{figure}[b]
\unitlength1cm
   \begin{picture}(0,7.0)
\centerline{
       \psfig{file=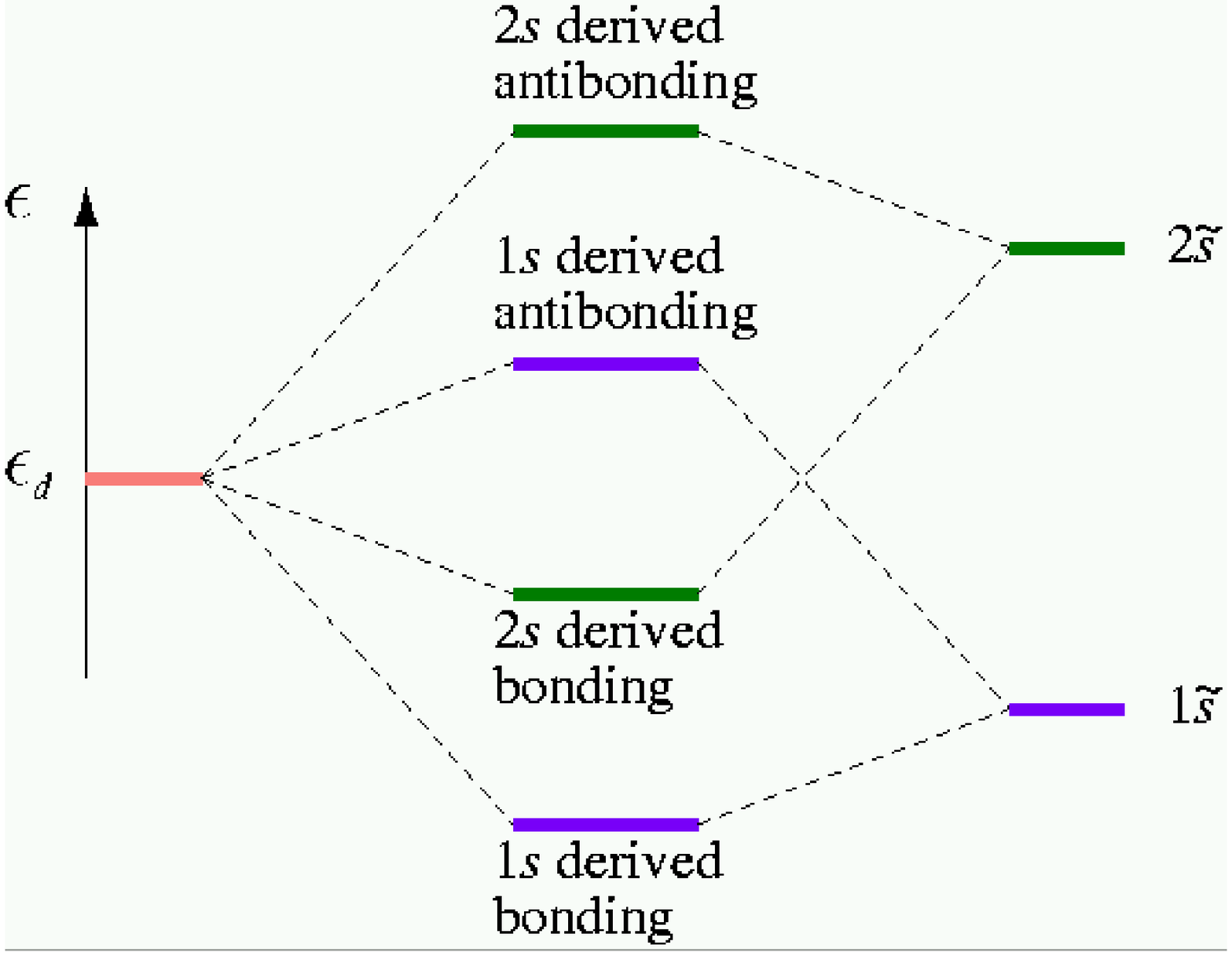,width=9.0cm}
}
\end{picture}
\vspace{-0.3cm}
   \caption{\small Schematic summary of the content of Fig. \ref{tb-DOS}.
Left:  the $d$-band-center  energy   of the clean substrate.
Right: the {\em renormalized energy levels} of the free atom
(i.e., after interaction with the $s$-band, but before interaction 
with the $d$-band).
Middle: the resulting bonding and antibonding H $1s$- and H
$2s$-derived levels on adsorption.}
\label{tb-levels}
\end{figure}

Obviously, the bonding is strongest when bonding states are occupied
and antibonding states remain empty. According to Fig. \ref{tb-DOS}
this will happen when the Fermi level is in the middle of the
$d$-band, as for example for Mo and Ru substrates.

Figure \ref{tb-levels} describes the same physics as Fig. \ref{tb-DOS},
but with an additional simplification:
Now the  broadening due to the substrate $s$-band is ignored and
the $d$-band is replaced by just a single level, the $d$-band center.
The energy levels of the ``free'' adatom are in fact the ``renormalized
levels'', and therefore we tagged them
in Fig. \ref{tb-levels} with a tilde:
$1\tilde{s}, 2\tilde{s}$.

\begin{figure}[b]

\unitlength1cm
   \begin{picture}(0,5.0)
\centerline{
       \psfig{file=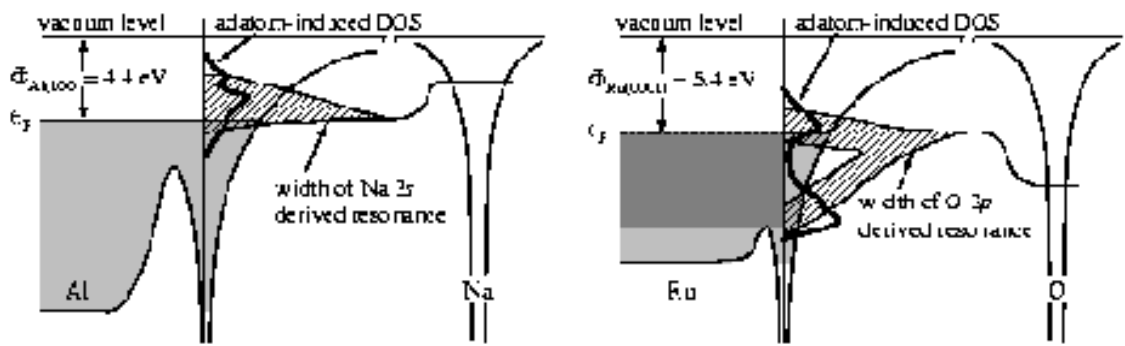,width=15.0cm}
}
\end{picture}
\vspace{-0.3cm}
   \caption{\small Side view of the surface,  and the adsorbate-induced
change in the density of states as a function of distance.
Two chemically distinct systems are displayed.
Left: Na on Al(001).
Right: O on Ru(0001).
The dark gray region in right figure marks the occupied part of
$d$ band of the Ru substrate.
The energy levels noted for the atoms are the mean values of the
ionization energy and electron affinity (Na: 2.8 eV, O: 7.5 eV),
see text for details.
}
\label{gurney}

\end{figure}

Deepening this discussion we show in Fig. \ref{gurney} results of actual
calculations where the distance dependence of the broadening and shift of
the energy level is displayed for two qualitatively different systems:
Na on Al(001) and O on Ru(0001).
For both adatoms, Na and O, the ionization energy of the free atoms
is below the Fermi level. In fact, the relevant ``energy level'' to
be considered is not the ionization energy and also not the electron
affinity, but the mean value of those two (or alternatively, the
Kohn-Sham eigenvalue of the highest occupied state; cf. the discussion
above).
This mean value is plotted in Fig. \ref{gurney}: For the Na atom
it is at (5.1+0.5)/2 eV =2.8 eV and for the O atom it is at
(13.6+1.4)/2 eV = 7.5 eV below vacuum.
When the adatoms approach the surface, the substrate Fermi level acts as
an electron reservoir. For Na this implies a partial charge transfer
from the adsorbate to the substrate (see the discussion of 
Fig. \ref{Na3s} above). But for oxygen, with its initially rather 
deep level, this implies
a partial charge transfer from the substrate to the adatom. Upon close
approach these partially ionized atoms interact with the substrate
orbitals. For Na/Al(001) the DFT-LDA calculations predict a double peak
structure, which has its main weight above the Fermi level, but with
a tail reaching into the occupied states
regime. For O/Ru(0001) the interaction is stronger because the adatom's
equilibrium position is very close to the surface and the substrate
$d$-states impart a strong covalent bond and a splitting into bonding and
antibonding states.
The left panel of Fig. \ref{gurney} nicely reflects the
Gurney description of alkali-metal adsorption, and the right
panel of  Fig. \ref{gurney}
displays the corresponding result for an ``opposite'' adsorbate,
namely one with high electronegativity and a strong covalent
adsorbate-substrate interaction.

The adsorbate-induced DOS (as shown in Fig. \ref{tb-DOS}c, d)
enables us to decide on the nature of the bond.
Assuming that the antibonding states are at least partially empty we have
chemical bonding of either covalent or ionic character.
When the states of the occupied peak are predominantly derived either
from adsorbate or substrate orbitals the bonding character is
called  ``ionic''.
When the wave-function character of the occupied states is derived by
about the same amount from the substrate and the adsorbate the
bond is called ``covalent''. The problem with this classification
is in the assumption that a substrate and an adsorbate region can be
defined and separated. However, for strong chemisorption situations
it is typically not obvious how to assign electron density to the
different atoms. Still the concepts are valuable
and we will  come back to them in Section \ref{sec:nature}.
We note in passing that an intriguing way to assign electron density
to individual atoms in a many-atom system was developed by Bader (1990, 1994, and references therein), but it is only now that these
concepts are being used in solid-state calculations.

The concept behind the qualitative discussion presented in this section
is essentially that of the Anderson-Grimley-Newns model for chemisorption
(see, for example, N{\o}rskov, 1990; Spanjaard and Desjonqu\`eres, 1990).
The main difference is that  free-atom levels are replaced
by {\em renormalized} levels, that we have not introduced a
restrictive assumption about electron correlation and the
localization of the adsorbate-substrate interactions,
and that we kept the
discussion at a qualitative level. We believe that the
approach sketched in Figs. \ref{tb-DOS}, \ref{tb-levels}
is useful but should not be overinterpreted. It
was used, for example, for a qualitative discussion of
H$_2$ dissociation at transition metal surfaces (see Hammer and 
Scheffler, 1995; and Section \ref{sec:reactions}). N{\o}rskov et al. 
(see Hammer and N{\o}rskov, 1997, and references
therein) refined the approach and used it
for a (semi-) quantitative discussion of adsorption at various
transition metal surfaces. The results confirm the simple
chemical picture that adsorption energies decrease with the $d$-band
filling  of the substrate (for $N_d \ge 5$). And when the $d$-band is
full, the $d$-electrons only play little role; thus the adsorption
energy is small, chemical activity is low, and the substrate is 
called noble. In fact, Hammer and N{\o}rskov  recently performed extensive
 DFT calculations of atomic and molecular adsorption at several
 metal surfaces and then used the above described approach to 
 explain the main trends of bonding to metal surfaces in terms of
 the $d$-band filling. The crucial term in their approach is the 
 adsorbate-substrate coupling matrix element.  The latter  depends
 on the adsorbate geometry (site and interatomic distances); 
 unfortunately, but also quite obviously, this cannot be obtained 
 from a simple Anderson-Gimley-Newns type of description.

As a warning remark we add that energy levels and the density of
states, $N(\epsilon)$, are important ingredients of the total energy, but
from the total-energy 
contribution
\begin{equation}
E^{\rm bands}= \int\limits^{+ \infty}_{- \infty} f(\epsilon)
N(\epsilon) \, \epsilon \, d\epsilon,
\label{bands}
\end{equation}
it is typically not possible to decide on the strength of 
chemical bonding.
The situation is different, when different geometries are
compared and a frozen-potential approximation applied
(the Andersen-Pettifor force theorem), as then (and only then)
the other terms in the DFT total-energy expression cancel and
{\em energy differences} based on Eq. (\ref{bands}) may give an
approximate description (see Skriver, 1985, and references therein).
In this context we note that many-atom systems tend to
assume a geometry with an electron density as hard as possible,
which means that either a band gap is opened or the density
of states at the Fermi level is reduced. This may be viewed as
a generalization of the Jahn-Teller theorem.
In this sense $N(\epsilon)$, in particular its behavior at the
Fermi level, is instructive.

\subsubsection{Adsorbate band structure}
\label{adsorbate-bands}
To discuss covalent interactions between adparticles on a surface, 
we consider an ordered adlayer which is in registry with a
crystalline substrate. Such a
system has two-dimensional translational symmetry and the 
eigenfunctions are two-dimensional Bloch states which are defined 
by their reduced ${\bf k}_{\parallel}$-vector (in the surface Brillouin 
zone) as well as by their energy.
\begin{figure}[b]
\unitlength1cm
   \begin{picture}(0,6.0)
\centerline{
      \psfig{file=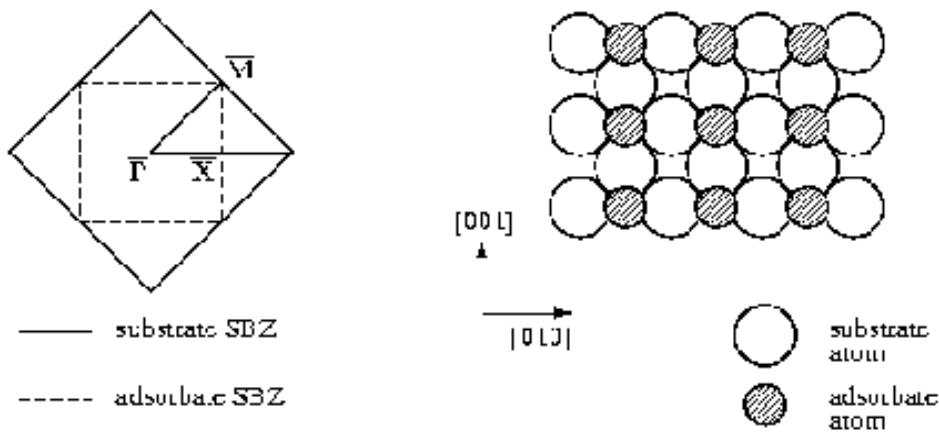,width=12.5cm}
}
    \end{picture}
\vspace{-0.3cm}
   \caption{\small The surface Brillouin zones (SBZs) and real-space lattice
of a c$(2 \times 2)$ adlayer on an fcc (001) surface. The lettering of
the symmetry points refers to the adsorbate SBZ.
}
\label{fig:SBZ-1}
\end{figure}
Figure \ref{fig:SBZ-1} shows the surface Brillouin zones for the clean
surface
and for a c$(2 \times 2)$ adsorbate overlayer on the (001) face of an fcc
metal. To obtain the bulk band structure projected onto the surface of a
clean semi-infinite system, it is necessary to integrate over $k_{z}$ 
inside
the bulk Brillouin zone and to shift regions outside the first
two-dimensional
Brillouin zone into its interior by a reciprocal lattice vector of the
(two-dimensional) surface structure. This folding back of the
${\bf k}_{\parallel}$-resolved density of states of a clean, 
unreconstructed
surface is indicated in Fig. \ref{fig:SBZ-2}a.
\begin{figure}[b]
\unitlength1cm
   \begin{picture}(0,5.5)
\centerline{
      \psfig{file=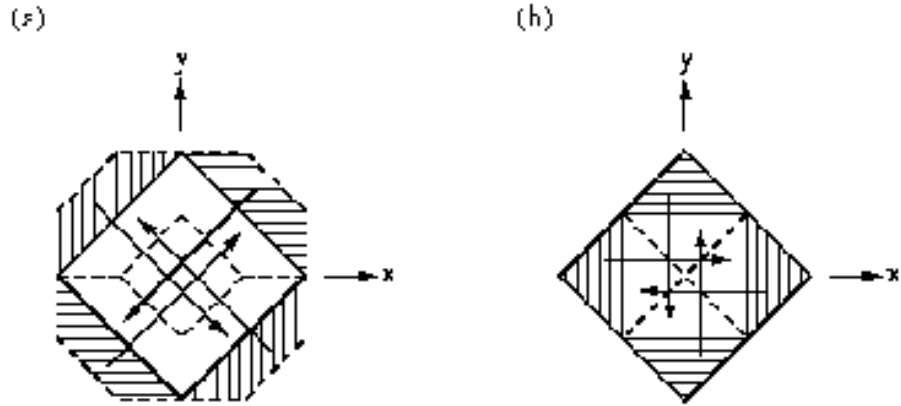,width=12.0cm}
}
    \end{picture}
\vspace{-0.3cm}
   \caption{\small (a) The relation between the surface Brillouin zone (full 
line)
and the projection of the bulk Brillouin zone of an fcc substrate 
onto the
(001) surface (broken line). Hatched regions have to be shifted by
a surface
reciprocal lattice vector into the first surface Brillouin zone, as
indicated  by arrows.
(b) Surface Brillouin zone for the c$(2 \times 2)$ adsorbate layer, 
showing
shift of hatched regions by (adsorbate) reciprocal lattice vectors.
} 
\label{fig:SBZ-2}
\end{figure}
The broken lines show the
projection of the bulk Brillouin zone. Even for the clean surface, it
is found
that the surface Brillouin zone is usually smaller than the projection
of the
three-dimensional bulk Brillouin zone on the surface. For a periodic
adlayer at
fractional coverage, the surface Brillouin zone is further reduced 
compared with
the clean surface, which again requires a shift of the 
${\bf k}_{\parallel}$-resolved density of states of regions outside
the new surface Brillouin zone
into its interior, as shown in  Fig. \ref{fig:SBZ-2}b. Because the
back-folded density of states adds to the original one, the change in the
${\bf k}_{\parallel}$-resolved density of states is obviously quite large.
We note, however, that this
folding back is a purely mathematical effect. It takes into account the
increase
of the periodicity parallel to the surface but it is independent of the
strength
and the type of the physical mechanism causing this change in periodicity.
The
physical importance of this effect, i.e., the degree of the mixing of the
wave functions involved in the surface region and in turn the question
how strongly
this effect will affect, for example, photoemission spectra, depends 
on the strength of the adsorbate-substrate interaction.
We return to this back-folding effect in the discussion of 
Figs. \ref{bands-1} and \ref{bands-2} of Section \ref{sec:LT-bands}
below.

The dispersion $\epsilon({\bf k}_{\parallel})$ of the 
chemisorption-induced bands can be described in a simple
tight-binding picture. It is appropriate to use a
LCMO (linear combination of molecular orbitals)
description where the MOs  contain the interaction with
the substrate as well as the intra-adparticle mixing of wave 
functions. Because this method is very simple and gives a reasonable
qualitative description, we consider it in more detail. Two-dimensional
Bloch states are used as a basis
\begin{eqnarray}
\chi_{\alpha}({\bf r}, {\bf k}_{\parallel}) = \frac{1}
{\sqrt{N^{\rm nuc.}}}
\sum^{N^{\rm nuc.}}_{{\bf R}_I}
e^{i {\bf k}_{\parallel} {\bf R}_{I}}
\phi_{\alpha}({\bf r} - {\bf R}_I),
\label{eq:basis}
\end{eqnarray}
where $\alpha$ labels the different (molecular) orbitals $\phi$
in the unit cell.
The ${\bf R}_{I}$ are two-dimensional lattice vectors. The wave 
functions  of the system are thus given by
\begin{eqnarray}
\varphi (\epsilon,{\bf k}_{\parallel},{\bf r}) = \sum_{\alpha}
C_{\alpha}(\epsilon, {\bf k}_{\parallel})
\chi_{\alpha}({\bf r},{\bf k}_{\parallel}).
\end{eqnarray}

For the Hamilton matrix we find
\begin{eqnarray}
H_{\alpha, \alpha^{'}}({\bf k}_{\parallel}) = \langle 
\chi_{\alpha}({\bf k}_{\parallel})|H|\chi_{\alpha^{'}}
({\bf k}_{\parallel}) \rangle =
\sum^{N^{\rm nuc.}}_{{\bf R}_I}  e^{i{\bf k}_{\parallel}{\bf R}_{I}} 
H_{\alpha,\alpha^{'}}({\bf R}_{I})
\end{eqnarray}
with
\begin{eqnarray}
H_{\alpha, \alpha^{'}}({\bf R}_{I}) = \langle \phi_{\alpha}
({\bf r})|H| \phi_{\alpha^{'}}({\bf r} - {\bf R}_{I}) \rangle .
\end{eqnarray}
Because the basis \ref{eq:basis} is usually not orthogonal, we get
an overlap matrix
\begin{eqnarray}
S_{\alpha, \alpha^{'}}({\bf k}_{\parallel}) = \langle 
\chi_{\alpha}({\bf k}_{\parallel}) | \chi_{\alpha^{'}}
({\bf k}_{\parallel}) \rangle =
\sum^{N^{\rm nuc.}}_{{\bf R}_I} e^{i{\bf k}_{\parallel}{\bf R}_{I}}
S_{\alpha, \alpha^{'}}({\bf R}_{I}) \quad
\end{eqnarray}
with
\begin{eqnarray}
S_{\alpha, \alpha^{'}}({\bf R}_{I}) = \langle \phi_{\alpha}
({\bf r})|\phi_{\alpha^{'}}({\bf r} - {\bf R}_{I}) \rangle .
\end{eqnarray}
The Schr{\"o}dinger equation is then
\begin{eqnarray}
\sum_{\alpha^{'}} \left\{ \sum^{N^{\rm nuc.}}_{{\bf R}_I}
e^{i{\bf k}_{\parallel}
{\bf R}_{I}} [H_{\alpha,\alpha^{'}}({\bf R}_{I}) -
\epsilon({\bf k}_{\parallel})S_{\alpha, \alpha^{'}}({\bf R}_{I})] 
\right\}
C_{\alpha^{'}}(\epsilon, {\bf k}_{\parallel}) =  0
\label{eq:matrix-SG}
\end{eqnarray}
and must be solved at each ${\bf k}_{\parallel}$. The zeros of the
determinant
of the matrix in the curly brackets in Eq. (\ref{eq:matrix-SG}) give the
dispersion $\epsilon({\bf k}_{\parallel})$.
The required matrix elements $H_{\alpha, \alpha^{'}}
({\bf R}_{I})$ and $S_{\alpha, \alpha^{'}}({\bf R}_{I})$ can be
calculated
numerically in some approximation (Bradshaw and Scheffler, 1979;
Horn et al., 1978; Scheffler et al., 1979; Jacobi et al., 1980).
%%ms  I have to check it:  maybe we should add a reference to
%%ms  the  chapter of Pollmann et al. in this book (** if ** he
%%ms  discusses the tb-approach. ??
Often, an empirical
tight-binding calculation might be sufficient, which introduces the
following three
assumptions: (1) Orbitals $\phi_{\alpha}$ at different centers are
orthogonal, the overlap matrix $S_{\alpha, \alpha^{'}}({\bf R}_{I})$  is
thus equal to one for $\alpha = \alpha^{'}$ and ${\bf R}_{I}$ = 
(0, 0) and zero
otherwise. (2) Only nearest neighbors (sometimes also second nearest
neighbors) are taken into account in the sum over ${\bf R}_{I}$ in
Eq. (\ref{eq:matrix-SG}). (3) The matrix elements
are fitted to experimental results; usually there are only very few
remaining.

\begin{figure}[b]
\unitlength1cm
   \begin{picture}(0,4.0)
\centerline{
       \psfig{file=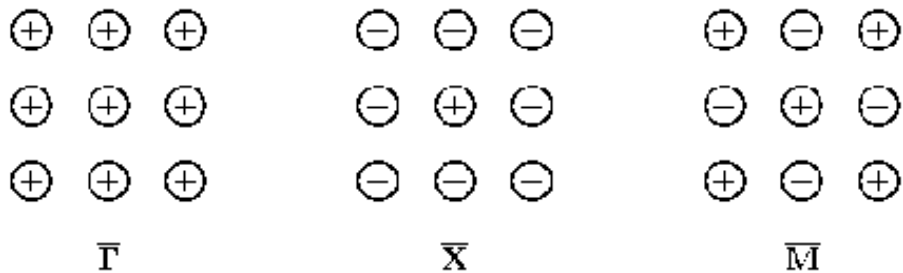,width=12.0cm}
}
    \end{picture}
\vspace{-0.3cm}
   \caption{\small Top view of a schematic representation of the two-dimensional
Bloch states formed from atomic or molecular orbitals belonging to the
$a_1$
representation of the $C_{4v}$ point group ($s$-like
orbitals) at the high-symmetry points of the surface Brillouin zone
(see Fig. \ref{fig:SBZ-1} and Eq. (\ref{eq:k-points})).
We display a region of 9 adatoms. The sign of the orbital in the center
${\bf R}_0$
is chosen +, and the signs of other orbitals are then
$e^{i{\bf k}_\parallel ({\bf R}_I -{\bf R}_0)}$ with
${\bf k}_\parallel$ definded by Eq. (\ref{eq:k-points}).
}
\label{fig:s-like-Bloch-states}
\end{figure}

We will illustrate this for the example of an adsorbate with $s$-like
states in a  c$(2 \times 2)$
overlayer on an fcc (001) surface ($p_z$-like states would behave the
same way).
The corresponding
adsorption-induced MOs $\phi_{\alpha}$ belong to the $a_1$
representations of the $C_{4v}$ point group
(we assume that the adsorbate occupies a fourfold symmetric
site). We then have two parameters in Eq. (\ref{eq:matrix-SG}):
$ H_{a_{1}, a_{1}}(0)$ and   $H_{a_{1}, a_{1}}({\bf R}_{I})$.
Figure \ref{fig:s-like-Bloch-states}
shows schematically the Bloch states according to Eq. (\ref{eq:basis})
for the three high symmetry points of the surface
Brillouin zone (see Fig. \ref{fig:SBZ-1})
\begin{eqnarray}
\overline{\Gamma} : {\bf k}_{\parallel} = (0,0), \quad 
\overline{\rm X} : {\bf
k}_{\parallel} =
\frac{g}{2}(1,0), \quad \overline{\rm M} :
{\bf
k}_{\parallel} = \frac{g}{2}(1,1),
\label{eq:k-points}
\end{eqnarray}
where $g$ is the length of the first two-dimensional reciprocal lattice
vector of the adlayer. Figure \ref{fig:s-like-Bloch-states} has been
constructed by placing orbitals
of $s$-like symmetry at all the atoms (or molecules)
of the adlayer (only 9 of them are shown
in the figure).
The phase factors of  orbitals at different sites is $e^{i{\bf
k}_{\parallel}{\bf R}_{I}}$ corresponding
to the appropriate ${\bf k}_{\parallel}$-point in the surface 
Brillouin zone. An analysis of this
figure already yields the qualitative band structure. At 
$\overline{\Gamma}$
the
$a_{1}$-derived two-dimensional Bloch state is completely bonding in the
adlayer, giving this state the lowest energy. At $\overline{\rm M}$ it is
completely antibonding (highest energy), and at $\overline{\rm X}$,
it is of mixed character. Thus, $s$-like adsorbate
states give rise to an energy band which has  the lowest energy
at $\overline{\Gamma}$ and the highest energy at $\overline{\rm M}$.
The analogous discussion for $p_x$-, $p_y$-like states
can be found in an earlier review article by Scheffler and Bradshaw 
(1983).

To discuss the interaction of an adsorbate with the $s$-band of the
substrate, we consider in
\begin{figure}[b]
\unitlength1cm
   \begin{picture}(0,8.0)
\centerline{
       \psfig{file=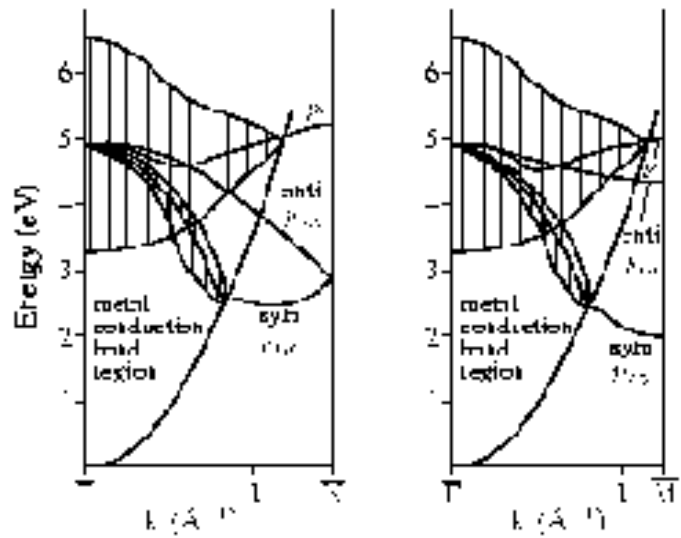,width=9.0cm}
}
    \end{picture}
\vspace{-0.3cm}
   \caption{\small Calculated band structure for a hexagonal $(1 \times 1)$
oxygen layer on a jellium substrate, simulating the Al(111) surface.
Reproduced after Hoffmann et al. (1979).}
\label{fig:O-on-jellium-bands}
\end{figure}
Fig. \ref{fig:O-on-jellium-bands} as an example the adsorption of oxygen 
on
jellium
corresponding to a ($1 \times 1$) overlayer on Al(111) (Hoffmann et al.,
1979), which has
been calculated using the two-dimensional KKR method (Kambe and
Scheffler, 1979). The
crystallographic point group is $C_{6v}$ because the jellium model 
neglects
the atomic structure of the substrate. Full lines show the center of the
peaks; hatched regions indicate the width of the peaks. The strong
dispersion of the levels is quite apparent. The width of the band 
structure (the dispersion of the full lines)
indicates the extent of the splitting between states which are bonding
with
respect to neighboring adparticles and those which are antibonding, 
just as
in the case of the tight-binding scheme discussed above. At
$\overline{\Gamma}$
the $p_{x}$- and the $p_{y}$-derived Bloch states are degenerate 
because of
the $C_{6v}$ symmetry and the splitting between the $p_{z}$- and $p_{x},
p_{y}$-derived states is very small. The
O $2p_{z}$-induced level is broad but that derived from $p_{x}, p_{y}$
is sharp. This different
behavior of the two levels can be explained in the following way. For an
ordered overlayer in registry with the substrate, two-dimensional Bloch
states of the overlayer can only hybridize with those of the substrate
which
have the same ${\bf k}_{\parallel}$ vector. The resulting wave functions
inside the jellium can thus be given as a sum over plane waves
propagating
toward the surface and the reflected waves.
\begin{eqnarray}
\varphi(\epsilon, {\bf k}_{\parallel},{\bf r}) = \sum_{\bf g}
U^{+}_{\bf g} e^{i{\bf k}^{+}_{\bf g} {\bf r}} +
U^{-}_{\bf g} e^{i{\bf k}^{-}_{\bf g} {\bf r}}.
\end{eqnarray}
Because the potential is constant inside the jellium, the ${\bf
k}^{\pm}_{\bf
g}$-vectors of these plane waves are given by
\begin{eqnarray}
{\bf k}^{\pm}_{\bf g} = \left( {\bf k}_{\parallel} + {\bf g} ,
\pm \sqrt{\frac{2m(\epsilon - V_{0})}{\hbar^2}} - ({\bf k}_{\parallel} +
{\bf g})^2 \right),
\end{eqnarray}
where ${\bf g}$ are reciprocal lattice vectors of the (two-dimensional)
surface
Brillouin zone. Obviously, for ${\bf k}_{\parallel}$ = (0, 0) and a
wave
function which
is antisymmetric with respect to a mirror plane of the system, the
coefficients $U^{+}_{(0, 0)}$ and $U^{-}_{(0, 0)}$ vanish. At energies
close
to the bottom of the conduction band of the substrate, all the remaining
terms have an imaginary value for $k_z$, which implies that these states
decay exponentially normal to the surface. These are true surface states 
and
have a sharp energy. In other words, we could say that at the bottom 
of the
band the substrate states are $s$-like and thus do not have the correct
symmetry in order to hybridize with $p_{x}, p_{y}$-like adsorbate levels 
at
${\bf k}_{\parallel}$ = (0, 0). Only at higher energies could these states also couple to bulk
bands
and broaden. Figure \ref{fig:O-on-jellium-bands} also shows that the
levels
change their width with
${\bf k}_{\parallel}$ and become discrete when they lie outside the
metal conduction band. We note the behavior of the symmetric $p_{x},
p_{y}$-derived band at the point at which it begins to hybridize with the
substrate wave functions: just outside the metal conduction band region
it
bends slightly to lower$^4$
energies. Here it is purely bonding in character. At the same
${\bf k}_{\parallel}$ value, we note that
the corresponding antibonding (broad)
level occurs in the conduction band. When the band enters the 
conduction band, the
bonding and the antibonding levels form one broad peak as mentioned
above.

\subsection{Adsorption of isolated adatoms}
\label{sec:nature}
Calculations of isolated adatoms afford an analysis of the nature of the
adsorbate-substrate bond without it being obscured by the influence of
other adsorbates. In this section we will therefore summarize 
characteristic
results of the adsorption of isolated atoms.

The calculations presented in the remaining part of this section were
performed with the
Green-function method,
which provides the most accurate and efficient approach for calculating
properties of truly isolated adatoms on extended substrates.
For technical details of the method we refer the interested reader to the
original papers (see in particular Bormet et al., 1994a; Lang and
Williams, 1978; Bormet et al., 1994b; Wenzien et al., 1995; Scheffler
et al., 1991; and references therein).

We consider
group I, group IV, and group VII adsorbates, namely Na, Si, and Cl. 
By such a trend study of atoms from the left to the right side of 
the periodic table,
a classification of the nature of the bond becomes rather clear, 
but we will also emphasize (again) the limitation and/or danger of
``nature-of-bond'' concepts. Three different substrates
will be considered, namely,
\begin{itemize}
   \item  jellium  with an electron density corresponding to aluminum,
   \item  Al(111),
   \item  Cu(111),
\end{itemize}
which enables us to identify the role of substrate $s$-, $p$-,
and $d$-states.
For the Al(111) and Cu(111) systems the adsorbates were placed in the
fcc-hollow site. In this section we discuss on-surface adsorption;
substitutional adsorption, where the adatom replaces a surface atom,
will be treated in Section \ref{sec:substitutional}, and subsurface
adsorption will be briefly touched upon in Section \ref{sec:catalytic}.

\subsubsection{Geometry}
Table \ref{tab:radii} gives the adsorbate heights $Z$ and
effective
\begin{table}[b]
\begin{center}
\begin{tabular}{|l|ccc|ccc|ccc|}
\hline
Substrate                 &   \multicolumn{3}{c|}{Na}
                          &   \multicolumn{3}{c|}{Si}
                          &   \multicolumn{3}{c|}{Cl}\\
                          &   $Z$ (\AA) & $R$ (\AA) & $\Delta R/R_{\rm J}
$
                          &   $Z$ (\AA) & $R$ (\AA) & $\Delta R/R_{\rm J}
$
                          &   $Z$ (\AA) & $R$ (\AA) & $\Delta R/R_{\rm J}
$\\
\hline
Jellium   & 2.79  & 1.82 &   0   & 2.37 & 1.46 &   0   & 2.52 & 1.59 &   
0
\\
Al(111)  & 2.69  & 1.73 & --5\% & 1.95 & 1.13 &--22\% & 2.09 & 1.24 &--22
\%
\\
Cu(111)  & 2.43  & 1.56 &--14\% & 1.68 & 0.96 &--34\% & 1.78 & 1.03 &--35
\%
\\
\hline
\end{tabular}
\end{center}
\caption{\small
Calculated geometrical parameters for adsorbed atoms from the left to
the right side of the periodic table, and for different substrates.
The height $Z$ is defined with respect to the center of the top substrate
layer.
The effective radii of the adatoms $R$ are
evaluated by subtracting from the calculated bond lengths the
radius of a substrate atom (as given by the inter-atomic distances in the
bulk). Thus, we use 
$R_{\rm J} = R_{\rm Al} = 1.41$ \AA \,
and   $R_{\rm Cu} = 1.27$ \AA.
For the jellium substrate we assume the geometry of Al(111).
Also noted as a percentage is the difference of the adatom radii with
respect
to that of the jellium calculation:
$( R_{\rm J}-R ) /  R_{\rm J}$. The jellium results are from
Lang and Williams (1978), the Al(111) results are from Bormet
et al. (1994a), and the Cu(111) results are from
Yang et al. (1994).\vspace{0.2cm} }

\label{tab:radii}
\end{table}
%%%%%%%%%%%%%%%%%%%%%%%%%%%%%%%%%%%%%%%%%%%%%%%%%%%%%%%%%%%%%%%%%%%%%%%%%
%
radii $R$ of the adsorbates, as obtained from the distances between the
adatom and its nearest-neighbor substrate atoms, using the substrate
atomic radii from the bulk. Both the calculated heights and the
effective radii (the bond length is given by the sum $R + R_{\rm Al}$
or   $R + R_{\rm Cu}$)
are found to be noticeably smaller for  Al(111)
than for the jellium substrate.
This is due to the fact that jellium is devoid of any information
concerning the atomic structure. Thus, the shell-like property
of electrons,  i.e., their $s$-, $p$-, $d$-like character
is missing which hinders the formation of directional orbitals in
the substrate; and
aluminum, despite its reputation of being a jellium-like
system, is in fact a material
with noticeable ability for covalent bond formation.
On comparison of the results for adsorption on the Cu and Al substrates,
it is found that the trend from jellium to aluminum continues:
The effective radius of the studied adsorbates
on Cu(111) is smaller than on Al(111).
The smaller values reflect that bonding
of the adsorbates is stronger  on  Cu(111) than on Al(111),
which was identified as being due to the Cu $d$-electrons
(Yang et al., 1994);
although the top of the Cu $d$-band is about 2 eV below the Fermi level,
the $d$-states play a noticeable role.

The results demonstrate that stronger bonds go together with shorter bondlengths. As noted above in Table \ref{tab:radii} we considered a 
threefold
coordinated
adsorption site. For other sites with lower coordination the strength
per individual bond will typically increase, because the
same number of adsorbate electrons have to be distributed (by 
tunneling or
hopping) over fewer bonds. As a consequence, the bond length 
will typically
decrease.
This does not mean that the binding energy will increase as well,
because this is determined by {\em all}
bonds. This correlation between local coordination and
bond strength, and the correlation between bond strength and bond length
is well known (see,  e.g., Pauling, 1960; Methfessel et al., 1992a).
But we also emphasize that when significant changes in hybridization
occur for different geometries, and/or when the system cannot attain the
geometry of optimum bond angles, this simple picture breaks down.

\subsubsection{Density of states $\Delta N(\epsilon)$}
\label{sec:Delta-N}
As emphasized in Sections
\ref{sec:intro-1},
\ref{sec:DOS}, and
\ref{sec:tight},
inspection of the adsorbate-induced change in the  density
of states $\Delta N(\epsilon)$ is particularly  informative.
\begin{figure}[b]
\unitlength1cm
   \begin{picture}(0,11.0)
\centerline{
       \psfig{file=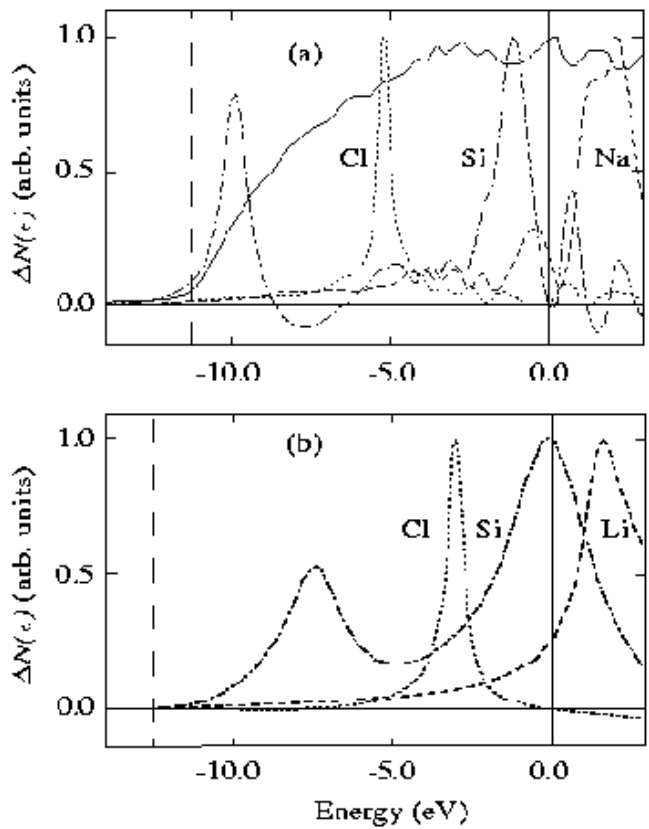,height=11.0cm}
}
    \end{picture}
\vspace{-0.3cm}
   \caption{\small   Adsorbate-induced change of the density of states
(cf. Eq. (\ref{eq:Delta-DOS})) for
three different adatoms (group I, IV, and VII of the periodic
table) on an Al(111) substrate (top: a) and on jellium with an electron
density corresponding to Al (bottom: b). For the Al substrate
also the bulk density of states is displayed (the full line in
figure a). The results are from
Bormet et al., 1994a (top), and from Lang and Williams, 1978 (bottom).
The Fermi level $\epsilon_{\rm F}$ is at the energy equal to zero.
The vertical dashed line indicates the bottom of the band of
the substrate.}
\label{Na-Si-Cl-DOS}

\end{figure}
Together with the know\-ledge of the
position of the Fermi energy,  it tells us whether the electronic states
which are formed upon adsorption are occupied, unoccupied, or
partially occupied, and this enables us to discern
the nature of the chemical bond.
Figures \ref{Na-Si-Cl-DOS}a and b show such results
for adsorbates on Al(111) and on jellium.
The adsorbates investigated in the work of Bormet  et al. (1994a)
(Fig. \ref{Na-Si-Cl-DOS}a)
were Na, Si, and Cl, and in that of  Lang and Williams (1978)
(Fig. \ref{Na-Si-Cl-DOS}b)
they were Li, Si, and Cl. The results of both of these calculations
agree qualitatively and show the following: For both alkalis (Na
and Li) the adsorbate resonance lies well above the Fermi level and is
thus largely unoccupied. This indicates that the valence electron of the
alkali metal atom (or part thereof) has been transferred to the substrate
and the adatom is partially positively charged.
In an opposite manner, on adsorption of chlorine, the resonance in the
curve corresponding to the Cl $3p$ resonance lies  5\,eV below the
Fermi level, and the Cl 3$s$ peak is positioned even below the
substrate valence band at about $-18$\,eV, i.e., outside the energy range
displayed in Fig. \ref{Na-Si-Cl-DOS}.  This result implies that
a transfer of electron density from the substrate to the Cl
adatom has taken place; the adsorbed Cl atom is partially negatively
charged. Thus, Li and Na constitute examples of positive ionic
chemisorption and Cl is an example of negative ionic chemisorption.

For the adsorption of an isolated Si atom it can be seen from the jellium
calculations that the Si $3p$ resonance lies just at the Fermi level,
which implies that it is about half occupied.
As noted in Sections \ref{sec:DOS} and
\ref{adsorbate-substrate-interaction},
the states of the energetically
lower half of the resonance are bonding between the adatom
and the substrate and the energetically higher states
are antibonding. Because the Fermi level cuts the $p$-resonance
approximately
at its maximum, the bonding nature of Si is covalent,  i.e., the
bonding states
are filled and the antibonding states are empty.
The results for atomistic (Al and Cu) substrates also show  that the bond
is covalent.
In this case, however, more structure occurs in the DOS than in the
jellium calculations. This arises because  the atomic structure
of the substrate leads to band-structure effects,
clearly reflected by the structure of the bulk DOS at 
$\epsilon_{\rm F}$ in Fig. \ref{Na-Si-Cl-DOS}a.
As a consequence, adsorption of a covalent atom, such as Si, results
in a splitting of the bonding and antibonding states,  and the adatom
density of states exhibits a minimum at the Fermi level 
(Bormet et al., 1994a).
Similarly to the jellium substrate, also for Si/Al(111) the Fermi level
cuts the Si $3p$-induced DOS roughly in the middle.
We  note in passing that for Si on Al(111) (cf. Fig.
\ref{Na-Si-Cl-DOS}a) the structure of the $p$-like adsorbate density
of states is largely determined by the clean-substrate density of
states: With the Green function ${\cal G}^0$ of the clean substrate,
we find at maxima of ${\rm Im}\left\{ {\rm Tr}\left( {\cal
G}^0(\epsilon)\right)\right\}$  minima of $\Delta N(\epsilon)$,
and at minima of ${\rm Im}\left\{ {\rm Tr}\left( 
{\cal G}^0(\epsilon)\right)\right\}$ we
find maxima (see Bormet et al., 1994a, for more details).

Despite the differences, which clearly exist between jellium and Al(111),
we find for both systems that the  adsorbate-induced change in the
DOS confirms the expected picture for a chemisorbed adsorbate on
a metal surface:
\begin{itemize}
   \item  Atomic levels of the adsorbate are broadened due to the
hybridization with the extended substrate states (in particular the
substrate $s$-states).
   \item Compared to the free atom DFT-LDA level\footnote{For 
open-shell systems, which are discussed here,
the DFT-LDA Kohn-Sham eigenvalue of the highest occupied level
is a good estimate of the mean value of the
ionization energy and the electron affinity (cf.
Figs. \ref{Na3s}, \ref{gurney} and their discussion).}
the adsorbate-induced peak is found at
higher$^{\ref{higher-lower}}$
energy for adsorbed Na,  at about
the same energy for adsorbed Si, and at lower energy for adsorbed Cl.
The calculated shifts are
$\Delta \epsilon({\rm Na}\, 3s) \approx +0.6$ eV,
$\Delta \epsilon({\rm Si}\, 3p) \approx +0$ eV,
$\Delta \epsilon({\rm Cl}\, 3p) \approx -0.5$ eV.
\end{itemize}
This trend also conforms to the finding that
the ionic character of the adatoms changes from plus to minus when
going from Na to Si to  Cl. It reflects that partially occupied levels have to align with respect to the substrate Fermi level (contribution $(i)$ discussed in Section \ref{sec:definitions}). But fully occupied levels (such as the $3p$ states of the Cl$^-$ ion) are more affected by the surface potential (contribution $(ii)$ discussed in Section \ref{sec:definitions}).

%%A discussion such as above should be used with caution, because
%%the energies of adsorbate-like states are typically
%%affected by more than one contribution.
%%Apparently, the contributions $(ii)$  [due to the surface
%%potential]
%%and $(iii)$ [due to the self-interaction], which were noted in
%%Section \ref{adsorbate-substrate-interaction},
%%are small for the {\em valence} states in the
%%discussed examples.

For a core (or semi-core) state the shift of an adsorbate level
upon adsorption  mainly reflects
the effects $(i)$ (due to the changed electrostatic potential caused by transfer or redistribution of valence electron density) and $(ii)$ (due to the substrate surface potential).
With respect to the latter, we note that core levels are affected
more by the electrostatic part of the potential than by the full
effective potential, because the change in the
exchange-correlation potential for core electrons due to the substrate electrons
is negligible: At the high electron density in the core region, the
exchange-correlation potential is only weakly affected by the relatively
slight increase in electron density due to the substrate valence
electrons.
The shift of core levels due to effects $(i)$ and $(ii)$ of Section
\ref{adsorbate-substrate-interaction}
is the
initial-state contribution of X-ray spectroscopy of the adsorbate
core-level shift.
The results of Fig. \ref{Na-Si-Cl-DOS} reveal that the  Si $3s$
semi-core level is shifted
by 3.4 eV and the Cl $3s$-level  by only 1.67 eV toward
lower$^{\ref{higher-lower}}$
energy. For Si the shift largely reflects how the adatom core states
feel the substrate potential.
The shift of Cl is smaller because the Cl is positioned further
away from the surface than Si, and because the electron transfer 
toward  the Cl adatom implies a repulsive potential for the core
and semi-core states
implying a contribution shifting their energy levels to
higher$^{\ref{higher-lower}}$ energies.

\subsubsection{Electron density: $n({\bf r})$,
$\Delta n({\bf r})$, and $n^\Delta ({\bf r})$ }

The trend seen in the results of Fig. \ref{Na-Si-Cl-DOS} is
in accordance with what is expected from electronegativity
considerations: Na is electropositive with respect to the neighboring
Al atoms,  i.e., it gives up an electron more readily than Al.
Cl, on the other hand, is strongly electronegative on Al and
electron transfer from Al to Cl should occur. Silicon has nearly
the same (or slightly higher) electronegativity as Al.

\begin{figure}[b]
\unitlength1cm
   \begin{picture}(0,5.0)
\centerline{
       \psfig{file=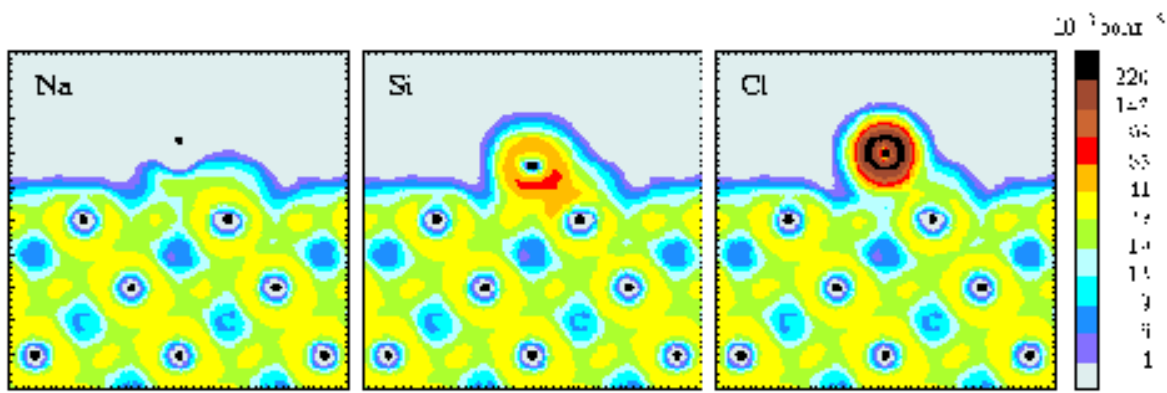,width=15.5cm}
}
    \end{picture}
\vspace{-0.2cm}
   \caption{\small  Electron density (valence only), cf. Eq. (\ref{eq:n(r)}),
for three different adatoms
(groups I, IV, and VII of the periodic table) on an Al(111) substrate.
We display a cut along the $(1 \overline{1} 0)$ plane, perpendicular
to the surface. The results are from Bormet et al. (1994a).}
\label{Na-Si-Cl-e-density}
\end{figure}

The formation of bonding and antibonding levels, together with the
position
of the Fermi level, will be reflected in the electron density
$n(\bf r)$. With this hierarchy in mind it is useful to inspect
{\em in addition} to the DOS (Fig. \ref{Na-Si-Cl-DOS})
the electron density and, what is more
instructive, the electron-density {\em change},  i.e.,
comparing the adsorbate system and the uncoupled systems.
Figure \ref{Na-Si-Cl-e-density} shows the electron density of the
valence states.

\begin{figure}[b]
\unitlength1cm
   \begin{picture}(0,9.0)
\centerline{
       \psfig{file=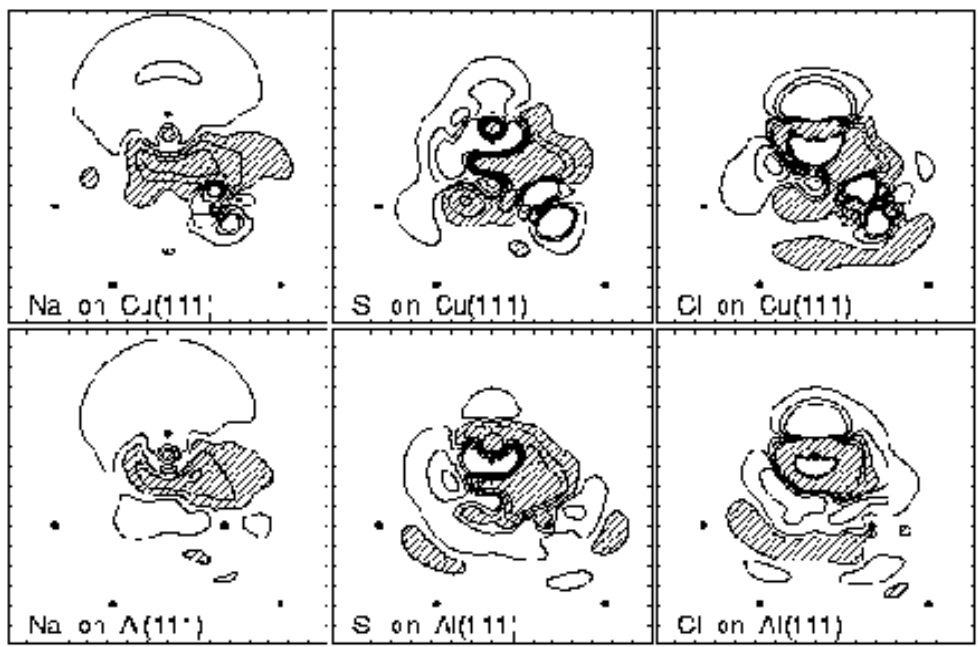,width=13.0cm}
}
    \end{picture}
\vspace{-0.3cm}
  \caption{\small Electron {\em difference density}
$n^\Delta({\bf r})$  (see Eq. (\ref{n-Delta})), considering for
the free adatom the neutral charge state, for
three different adatoms (group I, IV, and VII of the periodic
table) on a Cu(111) substrate (top) and on an Al(111) substrate
(bottom). We display a cut along the $(1 \overline{1} 0)$ plane,
perpendicular to the surface.
Densely hatched areas indicate accumulation of electron density
[positive $n^{\Delta}({\rm {\bf r}})$]. Non-hatched areas correspond
to electron depletion [negative $n^{\Delta}({\rm {\bf r}})$].
The positions of nuclei are marked by dots.
Results are from Yang et al., 1994, (top), and
Bormet et al., 1994a, (bottom).}

\label{Na-Si-Cl-delta-e-density}

\end{figure}

In the case of sodium, the charge transfer from the adsorbate toward the
substrate is clearly visible. From the vacuum side the sodium looks
practically naked. Figure  \ref{Na-Si-Cl-e-density} may overemphasize 
this
impression because it shows a wide range of electron density in order
to be able to compare atoms from the left to the right of the periodic
table. The first
displayed contour has a very low value which supports (again)
the description of Na as being a (partially) ionized adatom and that
particles which approach the adsorbed Na from the vacuum side
will experience the ``naked'' side of the adsorbate.
Figure \ref{Na-Si-Cl-e-density} also shows that the electron density
between the Na adatom and the Al substrate is increased. Thus, charge
has been displaced from the vacuum side of the Na atom toward the 
substrate side.
The details of this charge transfer are more clearly visible in the
electron {\em difference density} $n^\Delta ({\bf r})$, which is
the difference of the density of the adsorbate system,
displayed in Fig.  \ref{Na-Si-Cl-e-density}, of the density
of the clean surface, and of the free atoms (cf.
Eq. (\ref{n-Delta})). This {\em difference density}
is shown in Fig.   \ref{Na-Si-Cl-delta-e-density}.
The maximum of $n^\Delta({\bf r})$ is located between the adsorbate
and the substrate. In a more detailed discussion given in
Section \ref{sec:LT-screening} we show that the shape of this induced
charge density is the quantum-mechanical realization of the classical
image effect which is actuated by the partially ionized adsorbate.

For Si, as expected from the adsorbate-induced DOS,
a directional covalent bond is present between the Si adatom
and the nearest-neighbor substrate atom (see Figs.
\ref{Na-Si-Cl-e-density}
and \ref{Na-Si-Cl-delta-e-density}). Furthermore,
it can be seen that
the maximum of the charge density of the chemisorption bond is
closer to the more electronegative Si atom.
We also see a typical increase of electron density on the opposite
side of the bonding hybrid.

In the case of Cl on Al the charge density distribution around
the adsorbate is almost spherical, again supporting the
picture we had derived from the DOS in Fig. \ref{Na-Si-Cl-DOS}
of a (partially) negatively charged adatom.

The results for the Al and Cu substrates
show a number of similar features, and some interesting differences.
Firstly, it can be noted that in each case the perturbation to the system
caused by the adsorbates does not reach very far into the metal substrate.
The interior is essentially identical to that of the clean surface for
layers deeper than the second. We emphasize that this localization
holds for the electron density {\em perturbation} but not for individual
wave functions. For Na adsorption the results of Fig.
\ref{Na-Si-Cl-delta-e-density} imply the building up of
a surface dipole which locally decreases the work
function.  The opposite situation is found for Cl,  which is negatively
charged and sits on an adsorption site that is positively charged.
In this case charge has moved from the substrate toward the Cl atom,
and the local work function therefore will be increased relative to the
value of the clean surface. For both substrates, Si appears covalently
bound
with a slight electron transfer toward the adatom. Comparing in more
detail the results for Al and Cu, we see that there is more charge
between
the Na adatom and the top  layer of the Cu substrate. Also, there is a
greater depletion of charge at the vacuum side. These effects are
consistent with  the fact that the electronegativity difference between
Na
and Cu  $(0.93 - 1.9 = -0.97)$ is larger than that between Na and
Al $(0.93 - 1.61 = -0.68)$. The Si atom, which clearly forms a covalent
bond with both substrates, is slightly more electronegative than Al
(by 0.27). In this respect, it can be seen that more charge resides on 
the
Si atom when adsorbed on Al than when on Cu for which the 
electronegativity
difference
is zero.
%%ms the following, taken out:
%%ms The result for Cl on Cu displays some features similar to
%%ms Si  on Cu, namely, the pile up of charge density between 
%%ms the Cl adatom and the nearest Cu atom. This structure is 
%%ms practically absent for Cl on Al.
In all cases for the adsorbates studied, the adsorption on Cu
exhibits some structure in the valence-electron density change
near the nucleus of the Cu atom closest to the adsorbate, which
reflects the  participation of the Cu $d$-electrons in the bonding.

\subsubsection{Surface dipole moments}
A further interesting quantity obtainable in adsorption 
calculations is the change in the adsorbate induced dipole 
moment as a function
\begin{table}[b]
\begin{center}
\begin{tabular}{lccc}
\hline
\vspace{0.1cm}
Substrate      &   \multicolumn{3}{c}{Adsorbate}\\
               &    Na        &    Si     &      Cl \\
\hline
Jellium        &   +0.4       &   +0.0    &  $  -0.5$\\
Al(111)       &   +0.4       &  $-0.1$   & $  -0.5$\\[0.1cm]
\hline
\end{tabular}
\end {center}
\caption{\small Dynamic charge $d \mu/ d Z_{\rm ad}$,
as obtained for isolated adatoms on jellium,
and from supercell calculations for adsorbates on Al(111) at a
low coverage of $\Theta=1/16$. The units are electron charges. The
results are from Bormet et al. (1994a).}
\label{tab:dyn_mu}
\end{table}
of adsorbate height. In a naive picture one would expect that for
ionic bonding the dipole moment $\mu$ changes linearly with the adatom
height $Z_{\rm ad}$. For a covalent bond the dipole moment should be
approximately
constant (for small variations of the adsorbate height). Indeed we find
that this picture applies.  For Na we obtain a nearly linear increase of 
the
dipole moment
with increasing adsorbate height, and for Cl we obtain a decrease.
The dynamical charge, which is the slope of $\mu(Z_{\rm ad})$, is 
given in Table \ref{tab:dyn_mu} and is in good agreement with the 
jellium calculations of
Lang and Williams. These results support again the usefulness of the
ionic pictures for Na and for Cl, and of the mainly covalent
description of Si adatoms on Al(111).

A homogeneously distributed layer of adatoms induces an electric field
due to the adatom induced dipoles.  The dipole strength is
related to the adsorbate induced change in the work function,
$\Delta\Phi_{\rm ad}$,
by the Helmholtz equation:
\begin{eqnarray}
\Delta\Phi_{\rm ad} =  \Theta \, \mu(\Theta)/ \varepsilon_0.
\label{eq:type-3}
\end{eqnarray}

We have seen (cf. Figs. \ref{Na-Si-Cl-e-density} and
\ref{Na-Si-Cl-delta-e-density} )
that adsorption on a metal surface only
significantly affects the electron density of the bare metal substrate
in the outermost layers. The spatial distribution of the electronic 
charge with respect to the adsorbate nuclei gives rise to a dipole 
moment which changes the work function.
Furthermore, charge transfer toward (or away from) the adparticle, as
well as a permanent
dipole moment give rise to a long range repulsive electrostatic 
interaction
between different adparticles. If this is the dominant lateral
interaction,
which is true in some cases at low coverage, it will further result in a
depolarization of the adsorption-induced dipole moment with increasing
coverage (Antoniewicz, 1978; Topping, 1927; Kohn and Lau 1976).

It is convenient to illustrate this depolarization effect by considering the
physisorption of rare gases. For these systems there is practically no charge transfer, and we only have to consider the static adsorbate dipole.
 For not too small adsorbate-substrate separation, we can
calculate, using classical electrostatics, the dipole moment $\mu$ which
enters Eq. (\ref{eq:type-3}). The dipole moment on each atom as a function
of coverage is written as
\begin{eqnarray}
\mu(\Theta) = \mu_{\rm st} + \alpha {\cal E}_z (\Theta),
\label{eq:p-theta}
\end{eqnarray}
where $\mu_{\rm st}$ is the static dipole moment of the adparticle 
alone, i.e., without considering the
effect of screening by the metal conduction electrons and for
$\Theta \rightarrow 0$. ${\cal E}_z(\Theta)$
is the component normal to the surface of the microscopic electric 
field at the
adparticle under consideration and $\alpha$ is its polarisability. To
calculate the field ${\cal E}_z(\Theta)$, it is necessary to take the
image dipoles into account. Thus, as well as the direct 
dipole-dipole interaction, the
dipole-image and image-image interactions must be considered.
The actual dipole moment at coverage $\Theta$ then is
\begin{figure}[tb]
\unitlength1cm
   \begin{picture}(0,8.0)
\centerline{
      \psfig{file=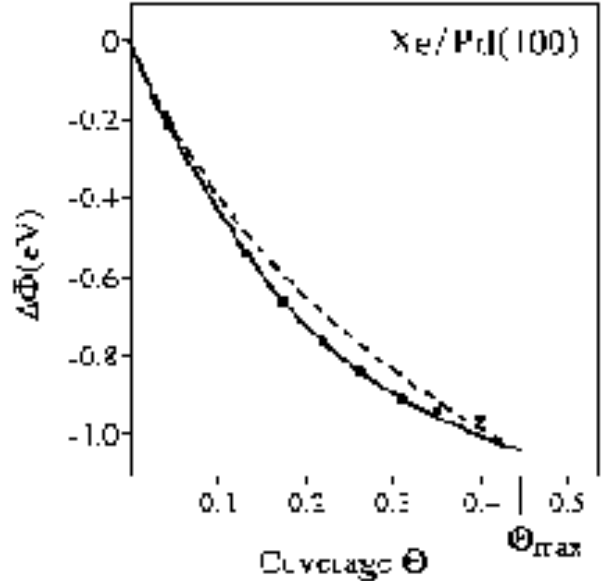,width=8.0cm}
}
    \end{picture}
\vspace{-0.5cm}
\caption{\small Calculated work function change for the adsorption system
Xe/Pd(001)
as a function of coverage.$^2$ Solid line: lattice sums proportional to
$\Theta^{3/2}$. Broken line: lattice sums proportional to $\Theta$. The
points represent the experimental results of Palmberg (1971).
After Bradshaw and Scheffler (1979).
}
\label{fig:Xe-Pd}
\end{figure}
\begin{eqnarray}
\mu(\Theta) = \frac{\mu_{\rm st}}{1 + \alpha [T + V - 1/4 D^3]},
\label{eq:p-theta-2}
\end{eqnarray}
where $D$ is the distance of the adparticle dipole to the effective image
plane, and $T$ and $V$ are lattice sums of the direct and the indirect in
teractions (see Scheffler and Bradshaw, 1983, for details).
The dipole moment of an adparticle at zero coverage is thus given
by
\begin{eqnarray}
\mu_0 = \frac{\mu_{\rm st}}{(1 - \alpha/4 D^3)}.
\label{eq:p-0}
\end{eqnarray}
Due to the screening by the substrate, the adatom-induced dipole
moment ist thus enhanced from the value $\mu_{\rm st}$ 
to $\mu_0$.
Equation (\ref{eq:p-theta-2}) shows that with increasing coverage, i.e.,
increasing value of $T + V$, the
particle is depolarized. Expressions similar to Eq. (\ref{eq:p-theta-2}),
but without the image
terms $(V - 1/4 D^3)$ were proposed by Topping (1927) and Miller (1946).
Equation (\ref{eq:p-theta-2}) has
been applied to the physisorption
system Xe/Pd(001) (Bradshaw and Scheffler, 1979). Figure \ref{fig:Xe-Pd}
shows measured and calculated changes in
work function. $D$ has been taken to be 1.5 \AA,
and for the polarizability the gas
phase value, $\alpha$ =  4 \AA$^3$, was used. For the monolayer
obtained at $\Theta$ = 0.44, the lattice sums are given by 
$T + V$ = 0.17 \AA$^{-3}$.
The only remaining parameter is the dipole moment of the single
adsorbed Xe atom, $\mu_{0}$. For the best fit to the experimental 
data, this is given by 0.93 Debye
with the positive end away from the surface. At coverages between
zero and
$\Theta_{\rm max}$, the lattice sums are either proportional to
$\Theta^{3/2}$ (Topping, 1927), if an ordered array is present at
every coverage, or
proportional to $\Theta$ (Miller, 1946), if the adlayer is disordered.
Both cases
yield similar behavior and are shown in Fig. \ref{fig:Xe-Pd}. We thus
see that the work function does not change linearly with coverage 
but that depolarization causes a less rapid change at higher coverages.

\subsection{Alkali-metal adsorption: the traditional picture of
 {\em \bf on-surface} adsorption}
\label{sec:LT-alkali}

In this section we discuss the traditional picture of alkali-metal
adsorption. This implies that we consider situations where
alkali-metal atoms adsorb {\em on the surface} without disrupting the
substrate very much, which is in contrast to {\em substitutional} 
adsorption
(see Section \ref{sec:substitutional}) where the adatom
kicks out an atom from the substrate and takes its place.
We recall that we discuss only the closed-packed
substrate surfaces, namely fcc (111) and fcc (001). For more open 
surfaces,
adsorbates, in particular alkali-metal atoms, can induce significant
surface reconstructions.
That substitutional adsorption at close-packed surfaces (and in more
general terms, surface alloy formation) can happen even with
adsorbates that do not mix in the substrate bulk, was first found by
Schmalz
et al. (1991) in a combined study of
surface extended X-ray absorption fine structure (SEXAFS)
experiments and DFT total-energy calculations. This subject will be
discussed
in Section \ref{sec:substitutional} below. In this section we discuss
{\em on-surface} adsorption and note only in passing that in the past,
several experiments have been interpreted inappropriately by incorrectly
assuming an on-surface adsorbate geometry. We will take care that only
those experiments are discussed below for which the adatom indeed sits
on the surface. In fact, attaining the substitutional adsorbate geometry
is hampered by an energy barrier, and therefore  for
low-temperature adsorption, adatoms are likely to stay on the surface.
For the example discussed below, in which  Na is at Al(001), the transition
temperature is $T = 160$~K.

Substitutional adsorption has been observed for Na, K, and Rb on
Al(111), where
the tendency toward the substitutional adsorption seems to decrease
with increasing adatom radius. Thus, for Cs substitutional adsorption
has not been found.

\subsubsection{The Langmuir-Gurney picture}
\label{sec:gurney}

In the seminal work of Taylor and Langmuir in (1933)
for the Cs/W system, it was shown that the work function of the clean
surface is reduced by  several electronvolts  on Cs adsorption and that
for thermal desorption at low coverage, almost all the Cs adatoms leave
the
surface as positive ions.
In the light of such results, the alkali metal bond to the substrate was
viewed as a spectacular example of ionic bonding akin to that in alkali
halides (see also Naumovets, 1994). As  discussed above (see Fig.
\ref{nacl}), even in NaCl
the nature of bonding is debatable, and it is therefore not a surprise
that the question whether or not alkali-metal adsorption should be
described in terms of ionic bonding has sometimes been raised.
With all the warnings mentioned in Section \ref{sec:intro} in mind,
we argue (and already did so in Section \ref{sec:nature})
that the ionic picture is  appropriate.
For a more detailed
analysis of the controversy about the ionic nature of
the bond we refer to the publications of Scheffler et al. (1991) and
Bormet et al. (1994a).

A  simple picture of the interaction between alkali-metal atoms and
a metal
surface,  and of the resulting chemisorption bond was proposed by
Langmuir (1932).
He assumed that the alkali-metal atom transfers its valence $s$-electron
completely to the substrate.
In a more rigorous theoretical description of alkali-metal adsorption
Gurney, in 1935, proposed a  quantum-mechanical picture applicable at low
coverages, where the discrete $s$-level of the free alkali metal atom
broadens and shifts, becoming partially emptied as a result of the
interaction with the substrate states as it approaches the surface. 
Figure
\ref{gurney} (left) summarizes this view.
The partially positively charged adatom then induces a negative screening
charge density in the substrate surface giving rise to an
adsorbate-induced dipole moment which naturally explains the reduction
in the work function of clean surfaces.

The building up of such adsorbate-induced dipole moments also leads 
to the
understanding that the dominant interaction between the
adsorbates is repulsive and that the adatoms should form a structure with
the largest possible interatomic distances compatible with the coverage. 
In this description it is expected that with increasing coverage the
adsorbate-adsorbate distance gradually  decreases, and the electrostatic
repulsion between the adatoms increases.
To weaken this repulsion, i.e., to lower the total energy, some fraction
of the valence electrons
flows back from the Fermi level of the metal to the adsorbate.
Thus, a reduction of the adsorbate-induced dipole moment,
i.e., depolarization takes place.
This picture of Langmuir and Gurney is fully supported by systematic
studies
of Lang and  Williams, Bormet  et al., and others, some of which were
discussed in Section \ref{sec:nature} above.
If the coverage-dependent depolarization is strong and/or the
nature of the adsorbate-substrate binding changes with the local
coverage,
it is also possible (and observed) that with increasing coverage
a phase separation occurs into close-packed islands (also called
``condensation'') (Neugebauer and Scheffler, 1993; Over et al., 1995).

\subsubsection{Coverage dependence of the work function}

\label{sec:LT-work-function}
\begin{figure}[tb]
\unitlength1cm
   \begin{picture}(0,6.5)
\centerline{
      \psfig{file=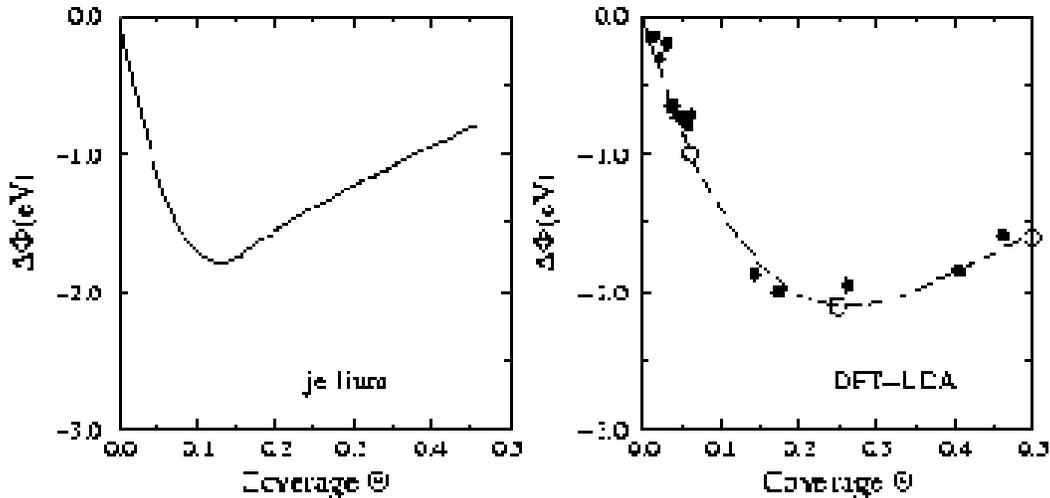,width=14.0cm}
}
    \end{picture}
\vspace{-0.5cm}
   \caption{\small Change in work function with increasing coverage
(cf. footnote \ref{coverage}). Left: Results of Lang (1971)
using a jellium on jellium model with parameters corresponding to Na
on Al(001). Right: DFT-LDA calculations (open circles) for periodic
Na adlayers on Al (001). Closed circles are experimental
results from Porteus (1974) and Paul (1987), obtained by adsorption
at low temperature ($T=100$ K).
}
\label{fig:workfunction}
\end{figure}

In his earlier work, Lang (1971 and 1973) studied the
alkali metal-induced work-function change using the jellium model for
the  substrate
and for the adsorbate layer. Despite the extremely approximate nature of
this approach, the results demonstrated that many
ground-state properties of metal surfaces and metal-adatom systems can
be described in a physically transparent manner. As an example, we show
in Fig.~\ref{fig:workfunction}
the change in work function
with increasing coverage (cf. footnote \ref{coverage}) for parameters
corresponding to Na on aluminum
(in the jellium approach; left side). The right side of the figure 
shows a calculation
for Na on Al(001), which takes the atomic structure into account.
We note that for coverages $ \Theta_{\rm Na}\,\,
\raisebox{.6ex}[-.6ex]{$>\!\!\!\!$}\raisebox{-.6ex}[.6ex]{$\sim$}\,\,
0.15$
the on-surface
geometry is a metastable structure which only exists at temperatures
below 160~K (Andersen et al. 1992; Aminpirooz et al., 1992).

The coverage dependence of the work function possesses a form similar to
that often observed experimentally. It is explained as a consequence
of the above-mentioned depolarization of the alkali metal-induced
surface dipole-moment induced by continuous reduction of the
adsorbate-adsorbate distance and corresponds to a rapidly decreasing work
function at low coverage, reaching a minimum at about $\Theta=0.1$
(Fig.~\ref{fig:workfunction}, left)
or $\Theta=0.25$
(Fig.~\ref{fig:workfunction}, right),
and subsequently rising toward the value of the pure alkali metal.

\subsubsection{Ionization of the adsorbate and screening by the substrate
electrons}
\label{sec:LT-screening}

We will now, for the moment, assume that we can neglect any effect due
to the chemical interaction (i.e., hybridization of orbitals) upon
adsorption
and that the substrate mainly
plays the role of providing an electron reservoir,  i.e., fixing the
electron chemical potential to its Fermi level. Using the Fermi-level
position of Al(001), Fig.~\ref{Na3s}
shows that then the occupancy of the Na $3s$-level will be
65\%. Thus 35\% of the Na $3s$-electron is now in the substrate.

A partially ionized atom in front of a metal surface
will not leave the metal electrons unaffected.
To discuss this, we will at first
sketch the quantum mechanical concept of
screening at a metal surface which asymptotically, for large
distances, includes the classical image  effect.
When a charged particle approaches a metal surface a screening
charge will be
created at the surface.  The quantum-mechanical realization of
the classical image effect is an induced electron density
which has its center of gravity at the ``effective metal
surface'',  i.e., the image plane (cf. Lang and Williams, 1978).
Thus, whereas in the classical description the electrostatic screening of a
{\em point charge} at a distance $D$ in front of a metal surface is modeled by an image point charge
inside the metal at a distance $-D$ away from the surface,
the truth is that there is no image point charge inside the substrate
but a charge density right at the surface (see Fig. \ref{screening}b),
which creates the same
electrostatic field in the vacuum as the (fictitious) point charge
of classical electrostatics.

\begin{figure}[b]
\unitlength1cm
   \begin{picture}(0,6.0)
\centerline{
      \psfig{file=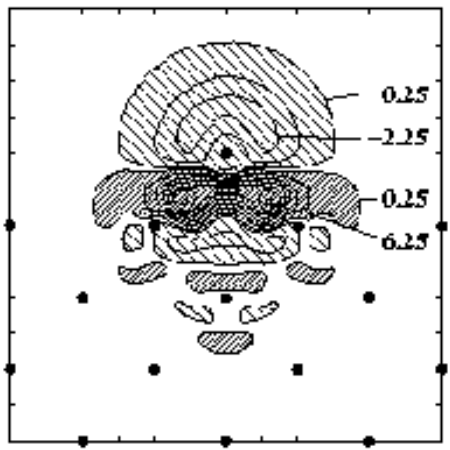,width=6.0cm}
}
    \end{picture}
\vspace{-0.1cm}
\caption{\small Electron {\em difference density} $n^{\Delta}({\rm {\bf
r}})$ for Na adsorbed on Al(001). For the definition of
$n^{\Delta}$ see Eq.~(\ref{n-Delta}); $n^{\rm Na}$ is taken from a free,
neutral atom. Units are 10$^{\rm -3}$ bohr$^{\rm -3}$. The
contour line spacing is
10$^{\rm -3}$ bohr$^{\rm -3}$. Densely hatched areas
indicate accumulation of electron density [positive
$n^{\Delta}({\rm {\bf r}})$]. Sparsely hatched areas correspond
to electron depletion [negative $n^{\Delta}({\rm {\bf r}})$].
The positions of nuclei are marked by dots (from Scheffler et al., 1991).
}
\label{Na-Al(001)}
\end{figure}

Considering close-packed surfaces, the image-plane position
(see Eq. (\ref{eq:image-potential}))
is on the vacuum side of the center of the outermost atomic layer,
because electrons spill out into the vacuum to lower their
kinetic energy.
At small distances (typically $ d < 2$\,{\AA}) the justification of
the image-plane concept gradually breaks down. However, it is still approximately
valid, however, because metallic screening is still important.  
The concept
can be approximately justified by assuming that for a charged system
close to the surface, the image plane ($z_0$ in Eq.
(\ref{eq:image-potential}))  is displaced toward the substrate.
This is what we have in mind when we use the term ``image effect'' in the
following discussion.
For a deeper discussion of this problem we refer to Scheffler and
Bradshaw (1983), Finnis et al. (1995), and references therein.

Figure \ref{Na-Al(001)} shows that the description
of partial ionization of the Na adatom and building up a screening charge
in the substrate remains {\em qualitatively} valid when the
alkali-metal atom is adsorbed.
Figure \ref{Na-Al(001)} displays the {\em difference density}
$n^\Delta({\bf r})$ (cf. Eq. (\ref{n-Delta})) using
a neutral atom ($f_{3s} = 1$).
It can be seen that the upper half of the adsorbed Na atom has a negative
$n^{\Delta}$ thus less electron density than the free, neutral alkali
atom, and on the substrate side of the adatom there is an increase
of electron density.

Figure~\ref{screening}~a shows a plot of
$n^{\Delta}(f_{3s}=0.1)$,
where $n^{\rm Na}$ (cf. Eq.~(\ref{n-Delta})) is now taken from a
self-consistent calculation for the partially  ionized free atom.
\begin{figure}[b]

\unitlength1cm
   \begin{picture}(0,6.0)
\centerline{
       \psfig{file=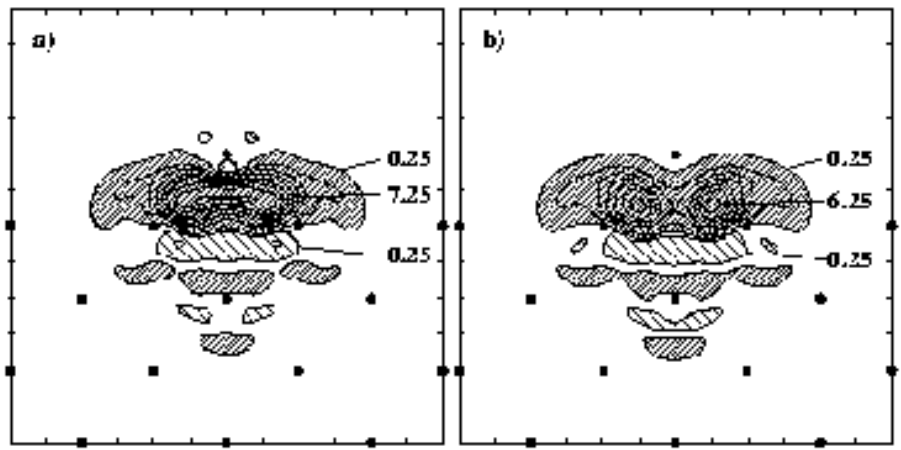,width=12.0cm}
}
    \end{picture}
\vspace{-0.1cm}
   \caption{\small Panel (a): Electron {\em difference density}
$n^{\Delta}({\rm {\bf
r}})$ for Na adsorbed on Al(001). For the definition of
$n^{\Delta}$ see Eq.~(\ref{n-Delta}); $n^{\rm Na}$ is taken from a
partially ionized, free Na atom ($ f_{3 s} = 0.1$).
Units are 10$^{-3}$ bohr$^{-3}$. The contour
line spacing is  10$^{-3}$ bohr$^{-3}$.
Panel (b) shows the screening charge density of an external, positive point
charge of $-0.09e^-$. Here the units are 10$^{-4}$ bohr$^{-3}$.
Densely hatched areas indicate accumulation of electron density [positive
$n^{\Delta}({\rm {\bf r}})$]. Sparsely hatched areas correspond
to electron depletion [negative $n^{\Delta}({\rm {\bf r}})$].
The positions of nuclei are marked by dots (from Scheffler et al., 1991).
}

\label{screening}

\end{figure}
This picture looks very
similar to the pure screening charge density which we obtain for an
external point charge (see Fig.~\ref{screening}b). The shape of
this screening charge density is very similar for a negative point
charge and for a positive
point charge, as long as the latter is sufficiently weak so
that it cannot bind an electron. The magnitude of the contour
lines nearly scale linearly with the magnitude of the external
point charge. The similarity between Figs. \ref{screening}~a and
\ref{screening}~b is quite
surprising, because a Na-ion is -- of course -- different
from a point charge. Nevertheless, the response of the metal is
similar for both cases.
The slight differences between Figs. \ref{screening}~a and
b  may be
interpreted as an indication that some covalency and some $s$-$p_{z}$
mixing are present in the
Na-Al(001) interaction, but these contributions are not
very large. From the similarity of Figs. \ref{screening}~a and b
we conclude that
the ``charge transfer picture'' is indeed useful and
appropriate to
describe the physics of alkali adsorption at low coverage.
However, charge transfer alone is not sufficient to understand
the adsorbate-substrate interaction because the charge transfer
actuates a significant change in the substrate surface electron density.
We also note that the
metal screening charge is largely located in front of the metal,
i.e., between the adsorbate and the substrate
(see Fig.~\ref{screening}b). As a consequence,
the adsorbate-induced electron-density change
cannot be divided {\em directly} into adsorbate and metal
contributions.

\subsubsection{Surface band structure}
\label{sec:LT-bands}
\begin{figure}[b]

\unitlength1cm
   \begin{picture}(0,13.0)
\centerline{
       \psfig{file=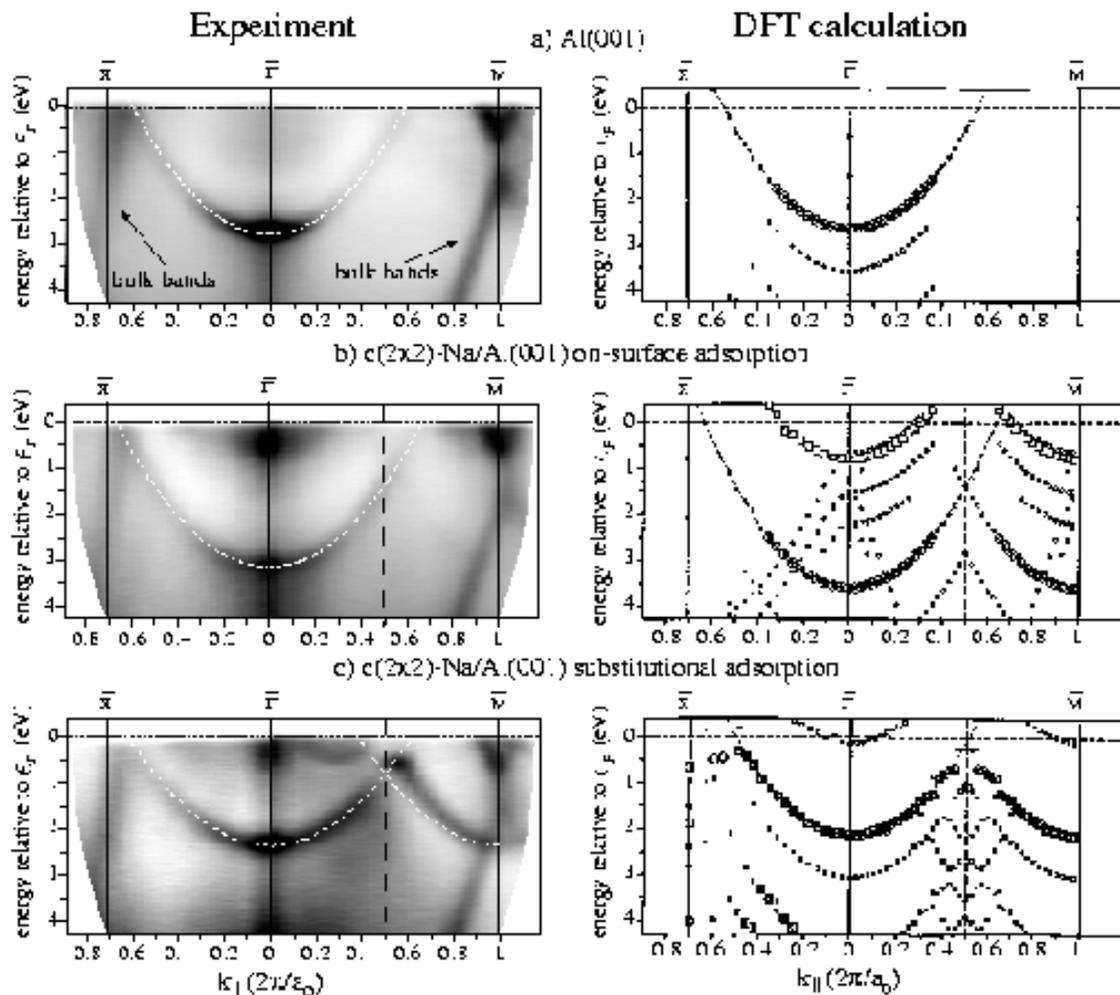,width=15.0cm}
}
\end{picture}
\vspace{-.5cm}
\caption{\small Comparison of experimental (left panels) and calculated
(right panels) surface band structures. The symbols used are such that
squares and circles represent Na- and Al-derived bands, respectively
(for a precise definition see the paper by Stampfl et al., 1998).
Panels (a) show results for the clean  Al(001) surface,
panels (b) the on-surface hollow site,
and panels (c) display results for the substitutional adsorption.
The results are taken from Stampfl et al. (1998). }
\label{bands-1}
\end{figure}

At low coverages
the interaction between adsorbed alkali-metal atoms will originate
from their
strong dipole moments. As the coverage increases, wave functions
will overlap and a surface electronic band structure will form.
This band is of $s$-character and its dispersion was discussed in 
Section \ref{adsorbate-bands} above.
In Fig.~\ref{bands-1} we show experimental (left panels) and
calculated (right panels) surface states/resonances for (a) the clean
Al(001) surface, (b) the on-surface c$(2 \times 2)$-Na/Al(001) structure,
and (c) the substitutional c$(2 \times 2)$-Na/Al(001) adsorbate
structure (discussed
in more detail in Section~\ref{sec:Na-Al(001)} below).
Firstly, we note that there is
good overall agreement between theory and
experiment.
The main, lower-lying band in each case is Al-derived,
as indicated by the circles in the theoretical plot.
The bands represented by squares are Na-derived.
Compared to the position of the surface-state band of the
clean surface, it can be seen that the Al-derived band of the on-surface
adsorbate phase
lies around half an eV lower in energy.
The mechanism giving rise to
this downward shift is discussed below.
Interestingly, the experimental results show clearly that this
state
does not have the c$(2 \times 2)$ periodicity of the adsorbates,
but rather has kept the
$(1 \times 1)$ periodicity of the clean surface
(cf. Fig.~\ref{bands-1}a).
Indeed, inspection of the electron density reveals that
the on-surface adsorbate represents only a modest
perturbation of the surface electronic structure, which cannot
be resolved
in the photoemission experiments (see the discussion of 
Fig. \ref{fig:SBZ-2}
above).

\begin{figure}[tb]
\centerline{
      \psfig{file=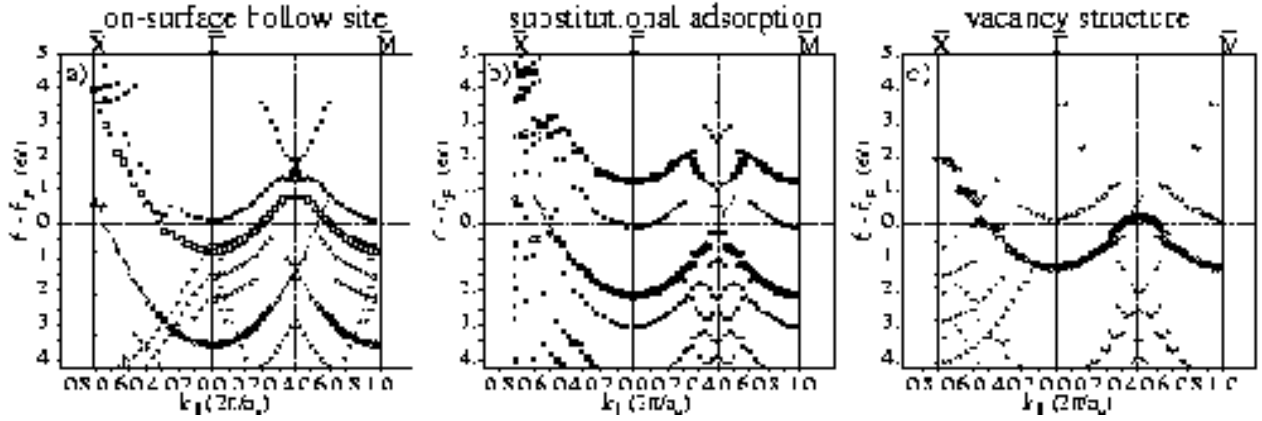,height=5.5cm}
}
\vspace{-.3cm}
\caption{\small Surface band structures of c$(2 \times 2)$ Na on Al(001) for
(a) on-surface adsorption, (b) substitutional adsorption,
and (c) the vacancy structure, including the energy range above
the Fermi level.  The symbols used are such that
squares and circles represent Na- and Al-derived bands, respectively
(for a precise definition see the paper by Stampfl et al., 1998).
}
\label{bands-2}
\end{figure}

From the calculated bands of the on-surface adsorbate phase 
(right middle
panel of Fig.~\ref{bands-1}), it can be seen that
there is also a surface-state/resonance with an energy of
about 0.7 eV below the Fermi level at $\overline{\Gamma}$. As
indicated by
the open squares, it is a Na-derived band (of Na 3$s$ character).
It crosses the Fermi level and is
partly occupied. From Fig.~\ref{bands-2}, which
shows the calculated band structure also
in the region {\em above} the Fermi level, it can be seen
that the part of the Na-derived band above $\epsilon_{\rm F}$
exhibits a band structure with c$(2 \times 2)$ periodicity of
the adsorbed
Na layer.
Thus the higher coverage of Na present in this phase, as compared
to the situation of an isolated adatom,  has given rise
to the development of significant adsorbate-adsorbate bonding and
band formation.

From construction of electron ``difference density'' plots (cf. Eq. (\ref{n-Delta})), that is, subtracting
off from the total valence electron
density of the system, the superposition of the electron densities of the
clean Al substrate and the Na atoms (arranged in the same
c$(2 \times 2)$ periodicity), it is found
that due to Na adsorption there is an increase of electron density
at the position of the surface-state band (the band of parabolic shape
which has a lower energy at 2.8 eV in Fig. \ref{bands-1}a, left and
3.2 eV in Fig. \ref{bands-1}b, left),
and from the region between the Na atoms electron density is depleted.
These results, together with comparison of the band structure of a
{\em free} c$(2 \times 2)$-Na monolayer which has an electron occupancy
{\em larger} than the Na-derived band seen in Fig.~\ref{bands-1},
suggests electron transfer occurs
from the Na atoms directly into the pre-existing surface states
of Al(001). Thus, the Na-induced structures seen in 
Fig. \ref{bands-1} b, left
(see the dark regions at $\overline{\Gamma}$)
originate from a coupling between the
3{\em s}-derived band of the free c$(2 \times 2)$ monolayer of Na
and the surface-state band of the clean Al(001) surface.
This may be understood
as being due to the formation of bonding and antibonding
states, leading to the downward shift of the Al-derived band
with an increase in population and the
upward shift of the Na-derived band with a
decrease in population.
Nevertheless,
the electron density of the Al-derived
band lies well below the smeared-out density of the Na-derived
band, and also below the position of the Na atoms.
Therefore the traditional picture of a thin metallic film covering
a metallic substrate, as in the jellium model,
remains qualitatively valid.
We will see  in the next  section that this is {\em not} the case for the
substitutional-adsorbate phase.

\subsection{Substitutional adsorption and formation of surface alloys}
\label{sec:substitutional}

Recent experimental and theoretical studies of the adsorption and
co-adsorption of alkali metals on Al surfaces have shown that the
traditional view of alkali-metal adsorption outlined in Section
\ref{sec:LT-alkali} above, is in fact, only  part of the whole picture
and that phenomena such as the following may occur:
\begin{enumerate}
\item The alkali-metal atoms may not necessarily assume  highly
      coordinated sites on the surface.
\item The alkali-metal adatom may kick out  surface substrate atoms and
      adsorb substitutionally. Substitutional adsorption has been shown
      in three cases to occur as the result of an irreversible
      phase transition from an ``on-surface'' site by warming to room
      temperature, without change in the periodicity of the surface unit
      cell or of the coverage, i.e., order-preserving phase
      transformations
      between metastable and stable structures occur.
\item The alkali-metal atom may switch site on variation of coverage.
\item Island formation may occur.
\item There may be a strong intermixing of the alkali-metal atom with
      the substrate surface, as for example the formation of a four-layer
      ordered binary surface alloy which was identified for Na on
      Al(111) at a coverage $\Theta_{\rm Na} = 0.5$ (Stampfl and 
      Scheffler,
      1994c). Also ternary surface alloys form for the co-adsorption
      of Na
      ($\Theta_{\rm Na}$=0.25) with either of K, Rb, or Cs on Al(111)
      (Christensen et al., 1996).
\end{enumerate}
Only some of these aspects will be discussed below. A comprehensive
description of the updated, modern view of alkali-metal adsorption can
be found in a special issue of Surface Reviews and Letters (SRL, 1995)
which collected review papers of the key groups which were working 
on this
subject. There, and also in Adams (1996), it is shown in greater detail
than possible here that the adsorption and co-adsorption of alkali metals
on Al surfaces exhibit a wealth of previously unexpected phenomena.

\begin{figure}[tb]
\unitlength1cm
   \begin{picture}(0,4.0)
\centerline{
      \psfig{file=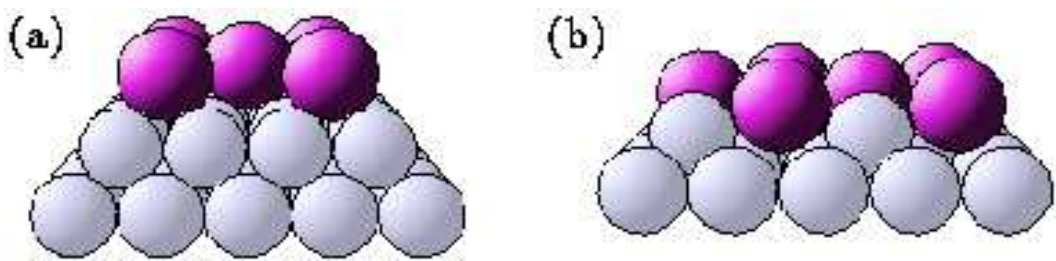,width=14.0cm}
}
    \end{picture}
\vspace{-.3cm}
\caption{\small Perspective view of the atomic structure of
         c$(2 \protect\times 2)$-Na on Al(001). Dark and light gray
         circles represent, respectively, Na atoms and Al atoms.
         (a) On-surface hollow site geometry and
         (b) substitutional geometry, where every second
         Al atom in the top unreconstructed Al layer has been kicked out.
 }
\label{Na-Al(001)geometry}
\end{figure}
\begin{figure}[tb]
\unitlength1cm
   \begin{picture}(0.0,4.0)
\centerline{
       \psfig{file=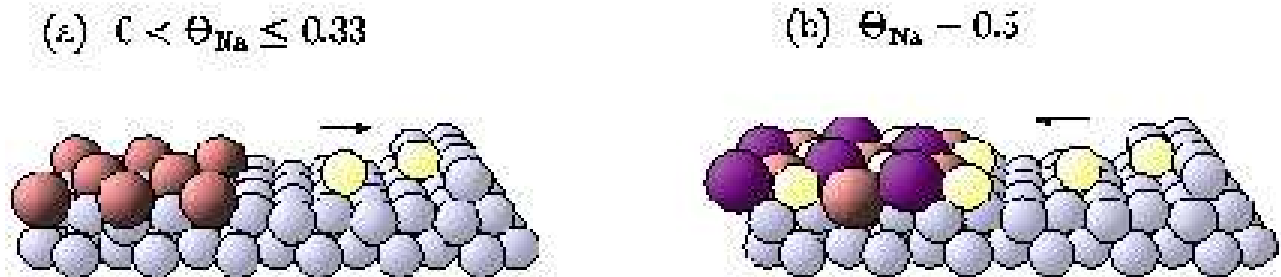,width=17.0cm}
}
    \end{picture}
\vspace{-0.5cm}
   \caption{\small Possible mass-transport scenario for substitutional
adsorption
and the formation of a surface alloy for the system Na/Al(111).
(a) Island formation occurs with a
$(\protect\sqrt{3} \protect\times \protect\sqrt{3})R30^{\protect\circ}$
structure with Na atoms in substitutional sites, for coverages
$0 < \Theta_{\rm Na} \protect\leq  1/3$: Every third Al atom is
``missing''; Al atoms diffuse to steps. (b) Surface alloy formation
with a $(2 \times 2)$ geometry implies diffusion of Al atoms back from
steps.
}

\label{diffusion-and-steps}

\end{figure}

Although it is well known that many, though not all, materials mix and
form alloys, up until recently intermixing was not considered to be very
relevant for adsorption on close-packed surfaces.
Thus, adsorbates on such surfaces were assumed to occupy on-surface sites.
Figure \ref{Na-Al(001)geometry}a
gives
an example for Na on Al(001). It was also assumed that this view is
justified,
in particular, for systems where the adatom is insoluble in the bulk
substrate.
This standard picture was questioned in 1991 by a combined study of
surface-extended X-ray adsorption fine-structure (SEXAFS)
experiments and DFT total-energy calculations for the system
Na on Al(111) (Schmalz et al., 1991);
by now  it is well established that often adatoms do not adsorb on the
surface, but instead, it can be energetically favorable that they
kick out
atoms from the substrate and take their sites. The kicked out atoms then
diffuse to a step (see Fig. \ref{diffusion-and-steps}a). The 
process may be
kinetically hindered, but this hindrance may be overcome even at
rather low temperatures,
e.g., for Na on Al(001) for $T \ge 160$
K (Andersen et al. 1992; Aminpirooz et al., 1992)  and/or by the heat
of adsorption. Whether the
solubility in the substrate bulk is low or even zero is of no relevance
at all for substitutional adsorption. Atoms, which are too big
to fit into a bulk vacancy, can still prefer to take a substitutional
site, because at the surface bigger atoms can simply sit somewhat
above the center of the created surface vacancy. Since 1991 many examples of
substitutional
adsorption have been reported,  as for example
K on Al(111)   (Neugebauer and Scheffler, 1992, 1993;
                 Stampfl et al., 1992, 1994a),
Na on Al(001) (Stampfl et al. 1994b),
Au on Ni(110)    (Pleth Nielsen, 1993),
Sb on Ag(111)    (Oppo et al., 1993),
Co on Cu(111)    (Pedersen et al., 1997),
Mn on Cu(001)    (Rader et al., 1997),
Co on Cu(001)    (Nouvertn\'e et al., 1999),
to name a few.
Thus, the phenomenon is not at all exotic, but rather general.

In this section we will discuss in particular three systems,
Na on Al(001),
Na on Al(111), and
Co on Cu(001), as these exhibit  qualitatively
different behavior.

Alkali metal-induced surface reconstructions are well known on the
more open surfaces (Somorjai and Van Hove, 1989;
Barnes, 1994; Behm, 1989)  as these clean surfaces are
close  to a structural instability (Heine and Marks, 1986).
But that significant
reconstruction can occur on close-packed fcc (111) and (001) 
surfaces had not been expected
previously.\footnote{The late $5d$ transition metals (Ir, Pt, Au) are
exceptions to this rule. In these systems relativistic effects give 
rise to a substantial  lowering of the $s$-band and a rather high 
surface stress. For the (001) surfaces, the desire to achieve a 
higher coordination
in the top layer is indeed stronger than the cost of breaking
(or stretching) some bonds between the first and second layer
(see Fiorentini et al., 1993).}
As a warning for future work we mention
that for alkali-metal adsorbates, several studies,
had found ``good agreement''  between LEED experiments and
theory as well as band-structure calculations and photoemission
experiments (this also was the case for the example discussed in
this section). However, it is now known that the reported agreement 
was purely coincidental, and the concluded physical and chemical 
properties were grossly incorrect because the alkali-metal atoms
were not sitting on the surface (cf. Fig. \ref{Na-Al(001)geometry}a),
but in substitutional sites (cf. Fig.  \ref{Na-Al(001)geometry}b).
Consequently, the nature of the adsorbate chemical bond, the surface
electronic structure, and the origin of the coverage dependence of the
work function are in fact qualitatively different to what was assumed
before 1992 (Andersen et al., 1992;
Aminpirooz et al., 1992;
Stampfl et al., 1994b;
Berndt et al., 1995).
Thus, good agreement between theory and
experiment and/or a convincing physical picture are no guarantee that
the description and trusted understanding is indeed correct.

The associated electronic properties of
these surface atomic arrangements,
perhaps not surprisingly, also exhibit  behavior
deviating from expectations based on early ideas.
For example, experimental measurements of the change in work function
as a function of alkali-metal coverage can be quite different
to the ``expected'' form of Fig.~\ref{fig:workfunction}
and the density of states
induced by alkali-metal adsorbates may not correspond to that
expected from the model of Gurney.

\subsubsection{Na on Al(001)}
\label{sec:Na-Al(001)}

\begin{figure}[b]
\unitlength1cm
   \begin{picture}(0,6.0)
\centerline{
      \psfig{file=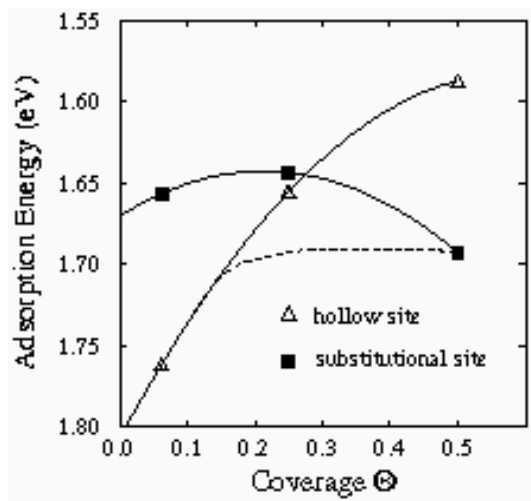,width=7.0cm}
}
    \end{picture}
\vspace{-.3cm}
\caption{\small
 Adsorption energy versus coverage
    for Na on Al(001) in the on-surface hollow site
    and in the surface substitutional site
     (from Stampfl and Scheffler, 1995).
}
\label{Na-Al(001)-energy}
\end{figure}
For Na on Al(001), the adsorption energy for the on-surface hollow site
and the substitutional site as a function of adsorbate coverage is shown
in Fig.~\ref{Na-Al(001)-energy}. It can be seen that the on-surface
hollow site is clearly preferred over the substitutional site  at low
coverages. Its adsorption energy rapidly decreases with increasing
coverage indicating a strong repulsive interaction between the Na atoms.
The adsorption energy for  Na in the surface substitutional site
depends much more weakly on coverage than
for the on-surface site, and, in fact,
the adsorption energy for the $\Theta=0.5$ substitutional structure
is more favorable than even
that of the lower coverage substitutional structures.

These results are interpreted as follows:
At low coverages,
the on-surface adsorption in a homogeneous adlayer is the 
stable structure
for
low as well as high temperature, but for higher
coverages, on-surface adsorption becomes metastable.
For high temperature  adsorption, or warming the substrate,
the adatoms then switch to substitutional sites,
forming islands with a c$(2 \times 2)$ structure.
The phase transition from the on-surface hollow site
to substitutional adsorption is indicated in Fig. 
\ref{Na-Al(001)-energy}
by the dashed line.
This predicted behavior is in good accord with experimental studies,
see e.g., Fasel et al. (1996).

The widely differing adsorbate geometries and the strong
dependence of them on coverage
and temperature, as described above,
means that the type of bonding and chemical properties of the adsorption
system will vary significantly depending
on these factors.
In this section we examine the electronic structure and bonding
nature of the c$(2 \times 2)$ substitutional structure of
Na on Al(001); that of the c$(2 \times 2)$ on-surface hollow structure
was briefly touched upon already in Section~\ref{sec:LT-bands}.

Firstly, it can be noted from Fig.~\ref{bands-1}c for substitutional
adsorption that the main band in the experimental
results clearly exhibits a
c$(2 \times 2)$ periodicity, in contrast to the on-surface
hollow structure.
This is due to the significantly reconstructed
Al(001) surface.
It can also be noticed that this band is higher in energy than
the surface state band of the clean Al(001) surface.
Furthermore, the Na-derived band, as clearly observed for the
on-surface phase in Fig.~\ref{bands-1}b in a greater region
around $\overline{\Gamma}$ below  $\epsilon_{\rm F}$,
is nearly absent in the substitutional phase.
However, from Fig.~\ref{bands-2}b which shows the same calculated 
band structure as in Fig.~\ref{bands-1}c, but where the energy region
extends higher into the positive range, it can be seen that a 
significant
Na-derived band (tagged by square symbols)
lies well above the Fermi level,  i.e., it is
unoccupied. Figure~\ref{bands-2}c also shows the calculated surface state
bands of the artificial c$(2 \times 2)$ vacancy structure. Here a strong
Al-derived band can clearly be seen, its  position being higher in
energy than that of the clean surface. Similarly to the clean Al(001)
surface, the maxima of charge density of the surface states for the
vacancy structure lie on top of the uppermost surface Al
atoms, having this time the c$(2 \times 2)$ periodicity.
\begin{figure}[tb]
\centerline{
      \psfig{figure=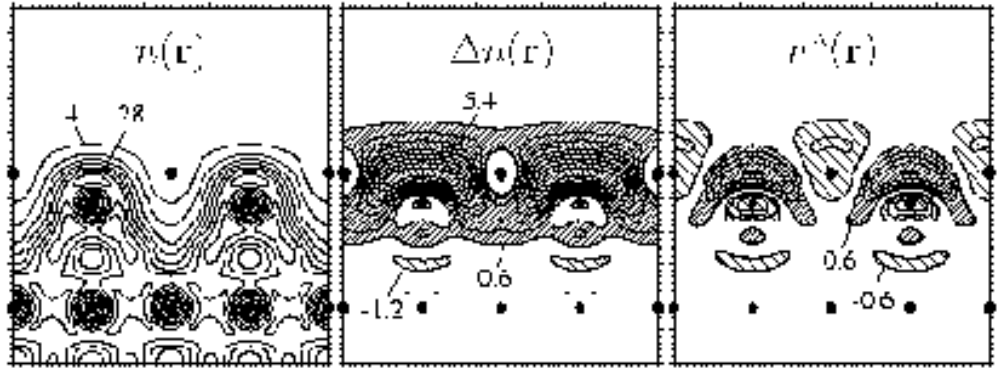,height=5.0cm}
}
\vspace{-.3cm}
\caption{\small Total valence electron density $n({\bf r})$
(left panel), density difference $\Delta n({\bf r})$ (middle panel),
and the {\em difference density} $n^\Delta ({\bf r})$ for
$f_{3s}=1$ (cf. Eq. (\ref{n-Delta})) (right panel),
of the substitutional geometry of the Na/Al(001) adsorbate system
at $\Theta=0.5$.
The units are $10^{-3}$  bohr$^{-3}$.
The results are from Stampfl et al. (1998).
 }
\label{den-diff-ht}
\end{figure}
On adsorption of Na, as for the on-surface phase, this band shifts
down in energy. The character of this Al-derived band is only slightly
changed compared to that of the vacancy structure,
which explains the well developed c$(2 \times 2)$ character of the
band found in Fig.~\ref{bands-1}c.

In this case electron transfer occurs from the Na
atoms into the surface state/resonance of the {\em vacancy structure}.
Figure~\ref{den-diff-ht} (right panel) shows that
electron charge has been transferred mainly from the region on top
of the Na atoms into the region above the Al atoms, corresponding to
the position of the surface states. Compared to the Na-Al bond length
of Na in the hollow site, that in the substitutional 
geometry is
approximately 4\% (from DFT-LDA) and 11\% (from LEED) shorter, 
indicating a larger ionicity of the bonding.
The results discussed above, together with Fig.~\ref{den-diff-ht} 
(middle panel), which shows the regions where electron density
has been increased due to Na adsorption, demonstrates that for the 
{\em substitutional} adsorption phase
the Na adlayer  {\em cannot}
be regarded as a simple metal film on a metallic substrate, and the 
jellium model is not valid.

Another interesting property which yields insight into the
bonding nature are the work function change and surface
dipole moment.
As discussed in Sections~\ref{sec:gurney} and 
\ref{sec:LT-work-function},
for {\em on-surface} adsorption the characteristic change in
the work function $\Delta \Phi (\Theta)$ with coverage is typically explained in terms of the Gurney picture and
the ``usual form'' can be described reasonably
well by assuming a jellium model.
\begin{figure}[tb]
\unitlength1cm
   \begin{picture}(0,6.0)
\centerline{
       \psfig{file=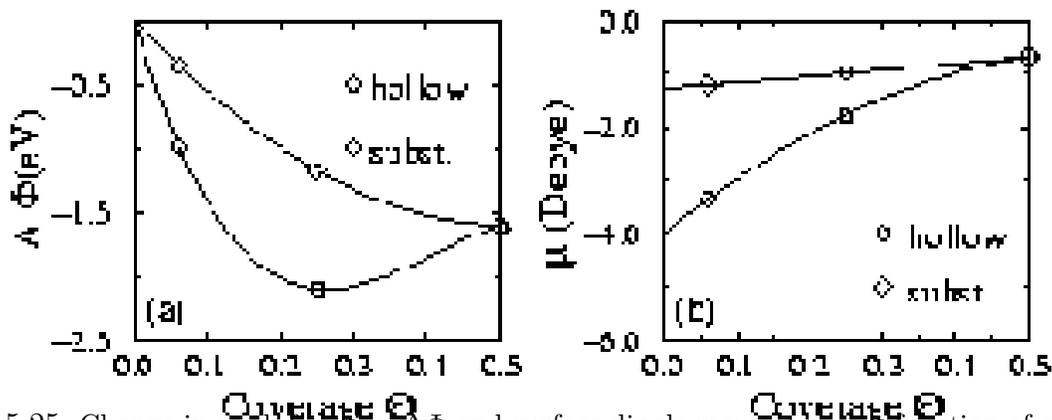,width=14.0cm}
}
    \end{picture}
\vspace{-.8cm}
\caption{\small Change in work function $\Delta \Phi$ and
surface dipole moment $\mu$ as a function
of coverage for Na in the substitutional (diamonds) and on-surface
hollow sites (circles) for Na on Al(001) (from Stampfl and 
Scheffler, 1995).
}
\label{na-al-sub-wf}
\end{figure}
The work function change $\Delta \Phi$  and surface dipole
moment $\mu $
are
shown in Figs.~\ref{na-al-sub-wf}a  and \ref{na-al-sub-wf}b,
respectively, as a function of coverage.
For comparison, results for Na in the on-surface hollow structures
are also displayed.
As is consistent with the traditional picture of alkali-metal adsorption,
there is a significant decrease of
the surface dipole moment with increasing
coverage for the on-surface site and the typical form of
$\Delta \Phi (\Theta)$ is observed.
The substitutional adsorption, on the other hand, exhibits a much
weaker dependence.
This reflects the fact that in the substitutional site the Na atoms sit
lower in the surface and the  repulsive
adsorbate-adsorbate interaction is screened better.

As discussed above (cf. Fig. \ref{Na-Al(001)-energy}), under not too
low temperature conditions a phase transition to
c$(2 \times 2)$ islands
occurs at coverage $\Theta \approx 0.15$; thus the local dipole
moment will be fixed at the value of the c$(2 \times 2)$ phase and
the work function change will vary linearly as $\Delta \Phi = \Theta
\mu/\epsilon_{0}$.
At $\Theta$=0.5 it is found that the values of $\mu$
(and $\Delta \Phi$) are the same for the on-surface and
substitutional adsorption.
This is in agreement with experiment,
i.e., in all cases (theory and experiment) the value of
$\Delta \Phi$ is 1.6~eV~ (Porteus, 1974; Paul, 1987).
%%Regarding the work function change as a product of electron charge and
%%distance,
%%the similarity can be understood in that for the substitutional
%%adsorption the ``amount'' of charge transfer is greater and the distance
%%smaller, and for the hollow site, the opposite (by about the same degree)
%%is true (Stampfl et al., 1998).

\subsubsection{Na on Al(111)}
\label{sec:Na-Al(111)}

The adsorption of Na on Al(111) was the first alkali metal on
close-packed metal system that was discovered to assume substitutional
adsorption. Further interesting and unanticipated phenomena were
found to occur for higher Na coverages on this surface, in particular,
the formation of a ``four layer''
surface alloy; the complex atomic geometry of which had
foiled initial experimental attempts at its determination.
This difficulty was related to its relative complexity, but was also
due to conceptual barriers since such a structure was not expected
to occur.

\begin{figure}[b]
\unitlength1cm
   \begin{picture}(0,7.0)
\centerline{
      \psfig{file=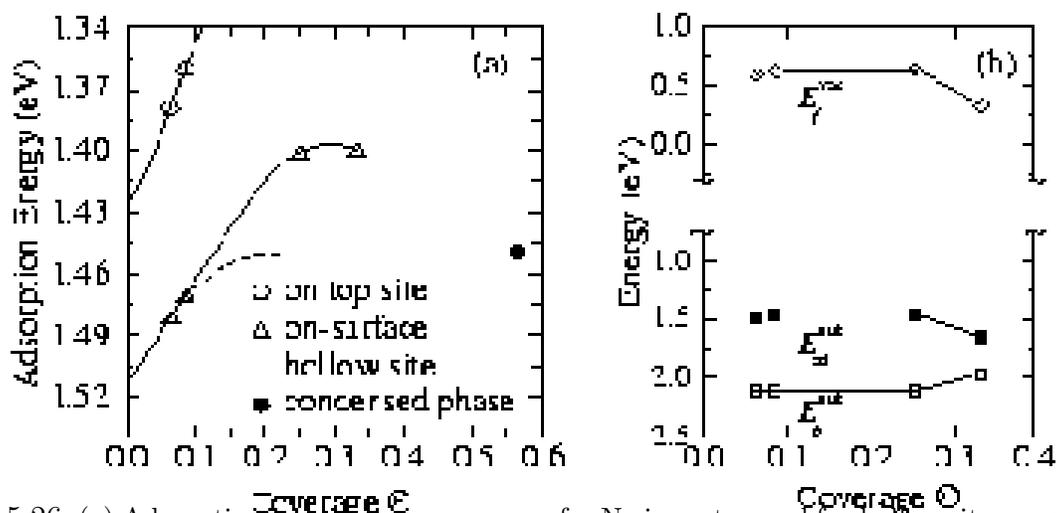,width=14.0cm}
}
    \end{picture}
\vspace{-.8cm}
\caption{\small (a) Adsorption energy versus coverage for Na in on top and
fcc-hollow
sites on Al(111). The dashed line marks the phase transition from the
homogeneous adlayer into adatom islands with a condensed structure.
  (b) Adsorption energy \protect$E_{\rm ad}^{\rm sub}$
and binding energy $E_{b}^{\rm sub}$
 for substitutional adsorption (cf. Eq. (\ref{eq3_2})), and 
vacancy formation energy $E_{f}^{\rm vac}$
(from Neugebauer and Scheffler, 1993).}
\label{na-al111-ad-en}
\end{figure}

In Fig.~\ref{na-al111-ad-en} the adsorption energy of Na on Al(111)
is displayed for various structures and coverages of Na in on-surface
geometries (Fig. \ref{na-al111-ad-en}a)
and in the substitutional site (Fig. \ref{na-al111-ad-en}b).
For the latter, the adsorption energy is split into its constituents,
namely, the binding energy and the surface vacancy formation energy.
For on-surface adsorption, the theory indicates that for low
coverages the
hollow site  is  energetically most favorable and
strong repulsive adsorbate-adsorbate interactions exist.
A condensed $(4 \times 4)$ structure is seen to be energetically more
favorable than homogeneous adlayers of Na
for coverages larger
than approximately $\Theta_{\rm Na}=0.1$.
In the condensed phase, the coverage is $\Theta_{\rm Na}=9/16$
and the structure represents a densely packed hexagonal adlayer
with nine Na adatoms per surface unit cell, all occupying 
different (mostly
low symmetry)  on-surface sites.
Thus, the picture is that for very low coverages,
the adsorbates occupy on-surface hollow sites and are uniformly
distributed
over the surface (homogeneous adlayers)
but for coverages greater than about $\Theta_{\rm Na}=0.1$, 
island formation
with the condensed structure and a $(4 \times 4)$ periodicity
occurs as indicated by the dashed line. In this case it is energetically
favorable to build up a metallic-like bonding between the
adatoms and to reduce the ionic character of the adatom-substrate
bonding.

\begin{figure}[tb]
\unitlength1cm
   \begin{picture}(0,4.5)
\centerline{
      \psfig{file=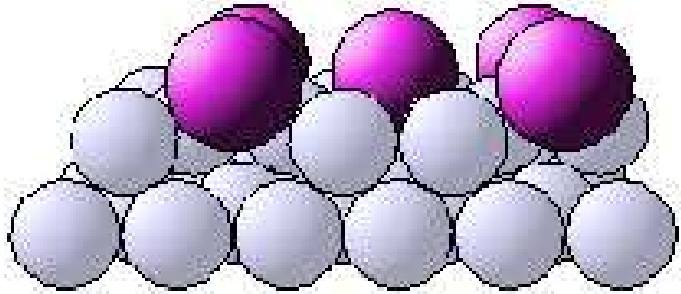,width=9.0cm}
}
    \end{picture}
\vspace{-.3cm}
\caption{\small Atomic geometry of the $(\sqrt{3} \times \sqrt{3})R30^{\circ}$
substitutional structure of Na on Al(111). Dark and light gray circles
represent Na and Al atoms, respectively.}
\label{na-al111-geom}
\end{figure}

For the case of substitutional adsorption it can be seen from
Fig.~\ref{na-al111-ad-en}b that
the adsorption energy of the substitutional geometry is
the most favorable for
{\em all} structures investigated (note the different energy scales of
Figs.~\ref{na-al111-ad-en}a and \ref{na-al111-ad-en}b). In particular,
substitutional adsorption with a
$(\sqrt{3} \times \sqrt{3})R30^{\circ}$
periodicity
has {\em the} most favorable adsorption energy.
It is therefore expected
that a condensation into $(\sqrt{3} \times \sqrt{3})R30^{\circ}$
islands with the  Na atoms in substitutional sites  occurs, beginning
at very low coverages.
The atomic structure is displayed in Fig.~\ref{na-al111-geom}.
The repulsive character of the binding energy for the
$(\sqrt{3} \times \sqrt{3})R30^{\circ}$ phase
can be seen (cf. Fig. \ref{na-al111-ad-en}b)  to be over-compensated for
by the attractive interaction of the surface vacancies.

\begin{figure}[tb]
\unitlength1cm
   \begin{picture}(0,5.5)
\centerline{
       \psfig{file=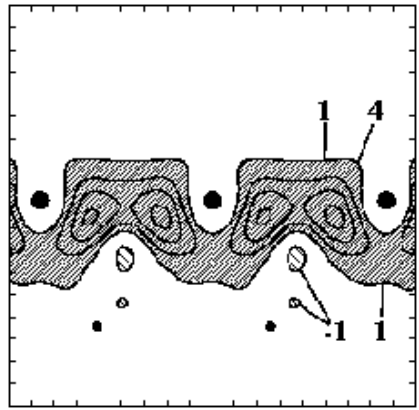,width=5.5cm}
}
   \end{picture}
\vspace{-.3cm}
   \caption{\small
Change of the electron density for
$(\protect\sqrt{3} \protect\times
\protect\sqrt{3})R30^{\protect\circ}$-Na on Al(111) with Na in the
substitutional site.
The reference system is the
$(\protect\sqrt{3} \protect\times \protect\sqrt{3})R30^{\protect\circ}$
surface vacancy structure plus a free standing Na layer. The contours are
displayed in a  $(1 \overline{2} 1)$ plane. Substrate
atoms are represented by small dots and Na atoms by large dots.
The units are $10^{- 3}$ bohr$^{-3}$
(from Neugebauer and Scheffler, 1992).}
\label{na-sub-ddiff}
\end{figure}

In Fig.~\ref{na-sub-ddiff} the difference between the electron
density of the $(\sqrt{3} \times \sqrt{3})R30^{\circ}$-Na/Al(111) phase
and that of the underlying vacancy structure plus a free standing Na
layer with also $(\sqrt{3} \times \sqrt{3})R30^{\circ}$ periodicity
is shown.  It can be seen
that due to Na adsorption, electron density has been displaced
from the Na atoms toward the Al atoms of the substrate.
As noted above, the reason for the favorable adsorption energy
of Na in this structure
is due to the particularly low vacancy formation energy;
the reason for this has been
attributed to the formation of covalent-like, in-plane bonding
between the remaining top-layer Al atoms, the honey-comb arrangement of
which is similar to that of graphite.
The electronic structure of the reconstructed surface
is found to be largely responsible for that of the
$(\sqrt{3} \times \sqrt{3})R30^{\circ}$-Na/Al(111) phase.
In particular, new states close to the
bottom of the Al valence band are found as well as broad unoccupied
bands. In this case, the role of Na is apparently mainly to create the
vacancy but not to modify very much the electronic structure of the
vacancies (Wenzien et al., 1993).

On further deposition of 1/6 of a monolayer of Na onto the substitutional
$(\sqrt{3} \times \sqrt{3})R30^{\circ}$-Na surface, a $(2 \times 2)$
structure forms with two Na atoms per unit cell. As discussed
above, the atomic
geometry of this phase proved initially difficult to determine.
Its correct structure
was first proposed on the basis of
DFT-LDA calculations (Stampfl and Scheffler, 1994c)
and was subsequently confirmed
by a LEED intensity analysis (Burchhardt et al., 1995). From
consideration of
the
atomic structure of the
$(2 \times 2)$ phase, it would seem that no mass
transport is necessary in its formation,  i.e.,
there is an Al atom missing due to the substitutional adsorption
of Na, but there is an additional Al atom embedded between the Na atoms.
However, the lower coverage
$(\sqrt{3} \times \sqrt{3})R30^{\circ}$
substitutional structure involves displacement of 1/3 of a monolayer
of Al
atoms, which are assumed to diffuse across the surface
to be re-bound at steps.
The results indicate, therefore, that formation of
the $(2 \times 2)$ structure from the
$(\sqrt{3} \times \sqrt{3})R30^{\circ}$
structure involves the reverse process, that is, diffusion of 1/3 of a
monolayer
of Al atoms back from the steps which are used in the
formation of
the $(2 \times 2)$ structure.
This process is depicted in Fig.~\ref{diffusion-and-steps}b.

Interestingly, in
a manner similar to that discussed
above for the two substitutional structures of Na on Al(001) and
Al(111),
the {\em occupied} part of the
surface band structure of the ($2 \times 2$) Na-Al surface alloy
can be explained largely in terms of the
underlying Al structure. The latter
corresponds to the reconstructed Al $(2 \times 2)$
vacancy layer {\em plus} the Al atom in the hcp-hollow
site on this structure.
In particular, a significant peak is observed
at approximately 2~eV below the Fermi level.
Analyzing the wave functions of this state
at $\overline{\Gamma}$ shows that it is localized on top of the
uppermost hcp-hollow Al atoms. For the surface alloy, in addition,
unoccupied features are identified which are associated with
the Na atoms, and at $\overline{\Gamma}$,
they are centered above the uppermost Na atoms (Stampfl and 
Scheffler, 1994c).

\subsubsection{Co on Cu(001)}
\label{sec:Co-Cu(001)}
In the above sections it was shown that the substitutional adsorption
of alkali-metal atoms is driven by the strong dipole moment of adatoms,
a rather low formation energy of surface vacancies, and the fact that the
adsorbate-adsorbate repulsion is reduced in the substitutional
geometry. Thus, substitutional adsorption does not occur for single
adatoms, but only when the coverage has reached a critical value where
the adatom-adatom interaction becomes significant (cf. Fig.
\ref{Na-Al(001)-energy}).

For cobalt adsorbed on Cu(001) the situation is different.
Here the adatoms can assume a substitutional geometry at the lowest
coverages and with increasing coverage a structural phase transition
\begin{figure}[tb]
\unitlength1cm
   \begin{picture}(0,6.5)
\centerline{
       \psfig{file=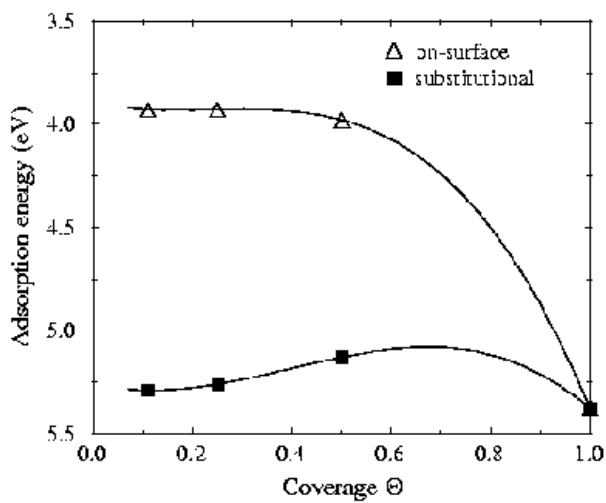,width=8.0cm}
}
    \end{picture}
\vspace{-.3cm}
\caption{\small
    Adsorption energy versus coverage
    for Co on Cu(001) in the on-surface hollow site
    and in the surface substitutional site
    (from Pentcheva and Scheffler, 2000).
}
\label{Co-Cu-energy}
\end{figure}
occurs  toward the formation of close-packed
islands.  The difference
compared to alkali-metal adsorption is due to the fact that
cobalt, as a transition metal from the middle of the $3d$ series, likes
to involve its $d$-electrons in the chemical binding. This implies
that Co likes to assume a highly coordinated site.
A single Co adatom on a Cu(001) surface in
the on-surface hollow site has four Cu neighbors. However, in the
substitutional site it is embedded in the electron density 
provided by eight
Cu neighbors. Thus adsorption in a substitutional site (i.e., in a
surface vacancy) is clearly more favorable than adsorption in an
on-surface site. In fact, the energy difference between on-surface
adsorption and into-surface-vacancy adsorption is larger than the
energy to create a surface vacancy. Thus, single Co adatoms tend to
kick out Cu atoms from the surface and assume their sites.
We note that the  Cu atoms taken out of the surface are re-bound at
kink sites at steps where they attain the Cu-bulk cohesive energy
(cf. Section \ref{sec:kink}).
In this context it is also relevant that Cu adatoms on
Cu(001) have a higher mobility than Co adatoms, which implies
that thermal equilibrium with kink sites can be attained easily for Cu
adatoms,
but is hindered for Co adatoms.

Figure \ref{Co-Cu-energy} shows the adsorption energies for
on-surface and
substitutional adsorption of Co. For low coverages substitutional
adsorption is energetically favorable. For higher coverage
it is still more energetically favorable for an open adlayer
to adsorb substitutionally than on-surface, but the energetically
lowest configuration is that of a close-packed Co islands.
In other words, Fig. \ref{Co-Cu-energy} tells us that substitutionally
adsorbed Co atoms form strong bonds with their eight Cu neighbors.
However, the strongest bonds are achieved when Co adatoms form
close-packed Co islands. Thus, for higher coverages, and/or
higher temperature, when Co adatoms  perceive the existence of
other Co atoms on the surface, isolated, substitutionally adsorbed Co
atoms are predicted to leave their site and close-packed
Co islands will be formed. In fact, recent DFT calculations predict
that these islands preferentially will have a thickness of 2-3 Co
layers and are capped by a Cu layer (Pentcheva and Scheffler, 2000).

\subsection{Adsorption of CO at transition metal surfaces \\
-- a model system for a simple molecular adsorbate}
\label{sec:CO}

The adsorption of a diatomic molecule on a surface represents
the next degree of complexity  with respect to the adsorption of a single
atom and
serves as a link to understanding the behavior of more complex molecular
adsorbates, as well as to the important area of carbonyl chemistry.
\begin{figure}[tb]
\unitlength1cm
   \begin{picture}(0,5.5)
\centerline{
      \psfig{file=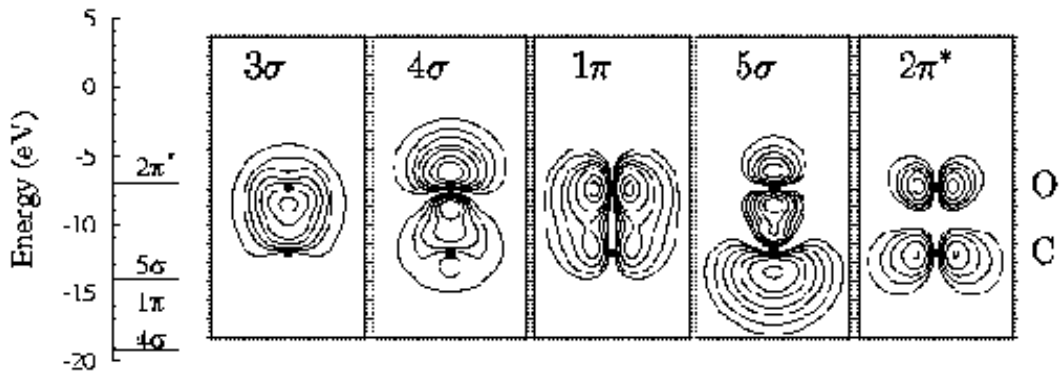,width=14.0cm}
}
   \end{picture}
\vspace{-.3cm}
   \caption{\small Electron density of the valence
molecular orbitals of a free CO molecule and  their DFT-GGA Kohn-Sham
eigenvalues (far left) with respect to the vacuum level.
The lower and upper small black dots
represent the positions of the C and O atoms, respectively.
The first contour lines are at 8 $\times$ $10^{-3}$ bohr$^{-3}$,
except for the $2\pi^{*}$ orbital
where it is 15 $\times$ $10^{-3}$ bohr$^{-3}$, and the
highest-valued contour lines are
at 0.5, 0.3, 0.2, 0.15, and 0.15 bohr$^{-3}$ for the
3$\sigma$, 4$\sigma$, 1$\pi$, 5$\sigma$, and 2$\pi^{*}$ orbitals,
respectively.}
\label{co-free-states}
\end{figure}
As such, CO adsorption has become a paradigm for the
study of a simple molecular adsorbate on a surface and has been
extensively studied both experimentally and theoretically,
see, e.g., Hermann et al. (1987), Hoffmann (1988), Campuzano 
(1990),
and references therein.
This interest  in CO adsorption also originates  from the
technological importance of oxidation catalysis (e.g., the car 
exhaust catalytic converter).

\begin{figure}[b]
\unitlength1cm
   \begin{picture}(0,5.0)
\centerline{
     \psfig{file=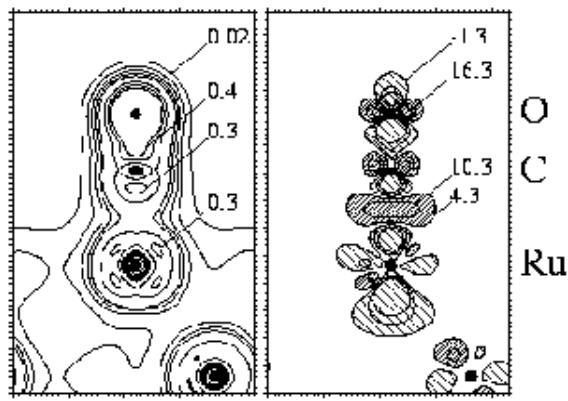,width=7.5cm}
}
   \end{picture}
\vspace{-.3cm}
  \caption{\small Valence electron density (left) and
{\em difference density} $n^\Delta({\bf r})$ (cf. Eq.
(\ref{n-Delta})) for the adsorption of CO in the on top site
on Ru(0001). Units are bohr$^{-3}$ in the left panel
and $10^{-3}$ bohr$^{-3}$ in the right panel.}
\label{co-on-ru}
\end{figure}

\begin{figure}[b]
\unitlength1cm
   \begin{picture}(0,5.5)
\centerline{
      \psfig{file=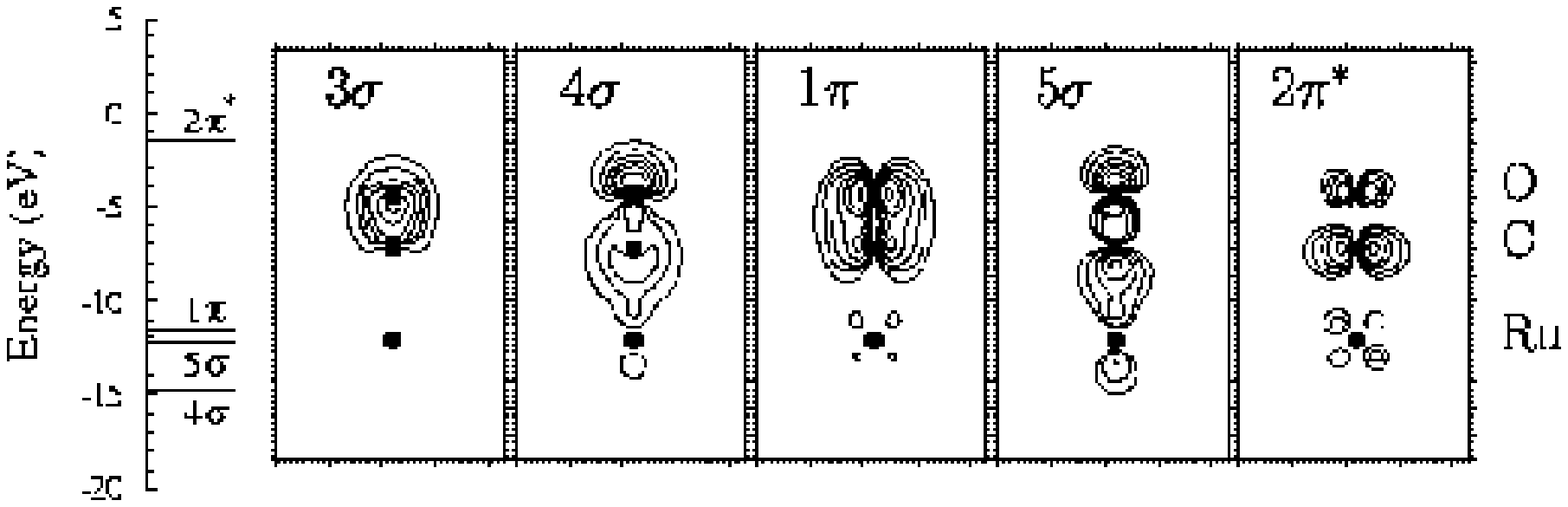,width=15.0cm}
}
   \end{picture}
\vspace{-0.3cm}
   \caption{\small Electron density distribution of the CO-derived  states for
CO adsorbed in the on-top site of Ru(0001) and their DFT-GGA
Kohn-Sham eigenvalues (far left) with respect to the vacuum level.
The black dots represent the
positions of the O, C, and Ru atoms. The first contour lines are at
8 $\times$ $10^{-3}$ bohr$^{-3}$, except for the 1$\pi$ orbital
where it is
4 $\times$ $10^{-3}$ bohr$^{-3}$. The highest-valued contour lines 
are at 0.5 for the 3$\sigma$ and 4$\sigma$ orbitals and at 0.2, 0.3,
and 0.09 bohr$^{-3}$  for the
1$\pi$, 5$\sigma$, and 2$\pi^{*}$ orbitals, respectively.
}
\label{co-states}
\end{figure}

The three outer valence orbitals of a free CO molecule are sketched
in Fig. \ref{co-free-states}. With
decreasing ionization energies, these are the 5$\sigma$ orbital 
(largely C $2s$,
C $2p_{z}$), the doubly degenerate 1$\pi$ orbital (largely 
C $2p_{x}$, $p_{y}$, O $2p_{x}$, $p_{y}$) and the 4$\sigma$ orbital 
(largely O $2s$, O $2p_{z}$). The first  unoccupied state, 
shown at the far right
in the figure, is the antibonding C $2p_{x}, p_{y}$,  O $2p_{x}, p_{y}
(2\pi^{*})$ orbital.  The two
most important orbitals are the $5\sigma$ and the $2\pi^*$ orbitals which
correspond to the HOMO and LUMO, respectively (see
Fig.~\ref{co-free-states}).
The notation here is that ``$\sigma$'' indicates orbitals that
are rotationally invariant with respect to
the inter-nuclear axis and ``$\pi$'' represents orbitals that
are lacking this symmetry.

When CO is brought toward a metal surface the CO 5$\sigma$ is significantly perturbed by the hybridization with the
substrate $d$-electrons. The energy of the 5$\sigma$ orbital changes most
strongly
because the bonding to the substrate is governed by the interaction
of this
orbital. This gives rise to charge transfer from the CO
5$\sigma$ orbital to the metal, but the metal gives charge back 
into the antibonding  2$\pi^{*}$-CO orbital. This is called the
donor-acceptor model
for CO bonding (Blyholder, 1964, 1975), which is known from the metal
carbonyls and is
similar to the results for H$_{2}$ adsorption (see
the discussion of Fig. \ref{fig:H2-tight-binding} in Section
\ref{sec:H2} below). The back donation from the substrate into the
2$\pi^*$-CO orbital weakens the bonding within the CO molecule and
strengthens the bond to the substrate.
At close distances the ordering
of the 1$\pi$- and the 5$\sigma$-derived levels is reversed compared with
the gas phase (see also the discussion of CO on Ni by Hermann and Bagus
(1977), and of CO on Cu by Hermann et al.
(1987)). In the Blyholder
model the lower lying 4$\sigma$ and
1$\pi$ MOs (as well as the 3$\sigma$ and of course the core states)
are assumed not to play an important role in the CO-metal bond
formation.

In the following we will use Ru(0001) as the substrate for the
discussion of CO adsorption but note, that the basic results are valid
valid for other transition-metal substrates as well.
In the left panel of Fig.~\ref{co-on-ru} the valence electron density
of CO in the on-top site on the Ru(0001) surface is shown, and in
the right panel of Fig.~\ref{co-on-ru} is the difference between
the electron density of the CO/Ru(0001) system and
the superposition of Ru(0001) and free molecular CO. From the latter, the
electron
redistribution can be seen to be in good general agreement with the
Blyholder donor-acceptor model: Depletion is clearly seen from the 
$\sigma$
orbitals
of CO and an increase in electron density into the 2$\pi^*$ orbitals.
Depletion can also be clearly noted from Ru states with $d_{z^{2}}$-like
character, as well as a significant increase in electron density in the
region of the adsorbate-substrate bond,  i.e., between the C and Ru atoms.
A similar behavior was found for CO on other substrates
(see, e.g., Wimmer et al., 1985 and  Bagus et al., 1986).
It is pointed out that in addition, there is  participation to the
CO-metal bonding by the Ru atoms in the {\em second} layer.

The Kohn-Sham eigenvalues of the free
CO molecule shift noticeably upon adsorption on the Ru(0001) surface;
in particular, a significant downward shift of the $5\sigma$ 
orbital energy due to hybridization with Ru  states, and also a
small downward shift  of the
$4\sigma$ level is observed. The $1\pi$-level  is changed only little,
and the $2\pi^{*}$-level moves up in energy reflecting the
increased occupation. 
Also, correspondingly, the development of 
antibonding states occurs. Thus, the behavior follows that of
Section 
\ref{adsorbate-substrate-interaction}; and we realize that
the effects $(ii)$ and $(iii)$ 
are apparently small for the on-top adsorbed CO.
In Fig.~\ref{co-states} the spatial distribution of some of 
these CO-derived
states are displayed. It can be seen that the $3\sigma$ orbital
remains unperturbed (compare with Fig. \ref{co-free-states})
by CO adsorption on
the substrate since it
lies significantly lower in energy and away from the surface.
The $4\sigma$ orbital interacts with
Ru $d_{z^{2}}$-like states, as does the strongly interacting
$5\sigma$ orbital.
The $1\pi$ orbital on the other hand interacts only weakly 
with the substrate.
The unoccupied $2\pi^*$-orbital hybridizes with
Ru states of $d_{xz}$ and $d_{yz}$-character. Certain of
these adsorbate-substrate bonding states also have an 
antibonding partner  (not
shown), the weight of which resides predominately in the substrate.

We see therefore that the first-principles calculations support
in general the
Blyholder model, but that the details of the bonding are
somewhat more complicated;
similar observations have been pointed out
and discussed in more detail by Hu et al. (1995) for first-principles
studies of CO on Pd(110) and from experiments by Nilsson et al. (1997).

\subsection{Co-adsorption [the example CO plus O on Ru(0001)]}
\label{sec:co-adsorption}

\begin{figure}[b]
\unitlength1cm
   \begin{picture}(0,10.5)
\centerline{
       \psfig{file=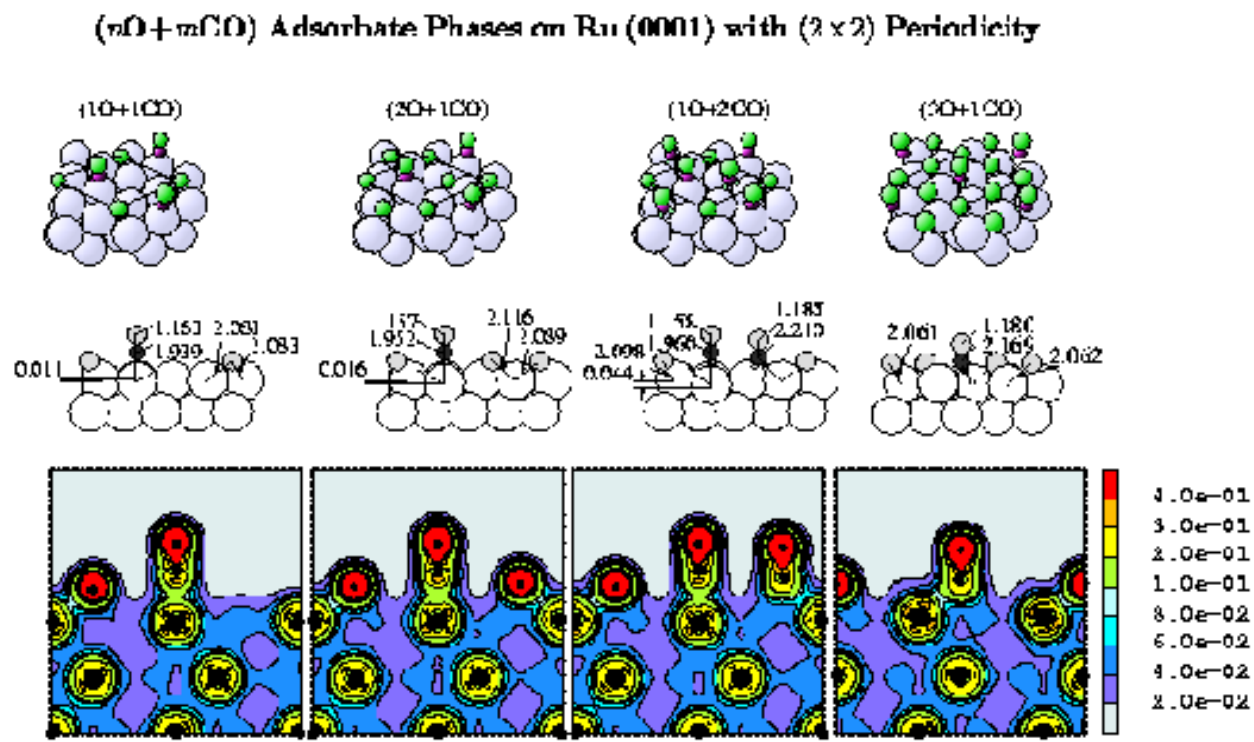,width=16.5cm}
}
   \end{picture}
\vspace{-0.3cm}
   \caption{\small
Perspective and  side views of the
various  phases of O and CO on Ru(0001).
Large and small (green and red) circles represent Ru, O, and C atoms,
respectively.
The lower panel shows the electron density
of the valence states.
The contour lines are in bohr$^{-3}$ and
distances are in \AA\, (from Stampfl and Scheffler, 1998).}

\label{co+o}
\end{figure}
A prerequisite to understanding heterogeneous catalytic reactions
is knowledge of the behavior of the various reactants
(e.g., the adsorption sites and binding energies), as well as their
mutual interactions.
The co-adsorption system ($m$CO+$n$O)/Ru(0001) represents
a well-studied model system, not least  due to
the drive aimed at obtaining an understanding of the catalytic oxidation
of CO by O$_{2}$, but also as a ``simple'' model system for oxidation catalysis in general.
Despite the considerable interest, it is only recently that
the detailed atomic structure of some of the phases of ($m$CO+$n$O) on Ru(0001)
have been determined, and new ones discovered.
Depending on the experimental conditions, co-adsorption of
CO and O on Ru(0001) can form the following phases:
$(2 \times 2)$-(1O + 1CO) [Kostov et al., 1992; Narloch et al., 1995],
$(2 \times 2)$-(2O + 1CO) [Narloch et al., 1994],
and $(2 \times 2)$-(1O + 2CO) [Schiffer et al., 1997].
These mixed ($m$CO+$n$O)/Ru(0001) surface
structures are depicted in the top panel of Fig.~\ref{co+o}.

For the first two structures,
low-energy electron diffraction
(LEED) intensity analyses have been performed: In the first phase,
the O atoms occupy hcp sites and the CO molecule adsorbs in the
on-top site.
In the second phase, a restructuring induced by CO adsorption
of the O atoms of the $(2 \times 1)$ (Lindroos et al., 1989)
phase occurs: Half of the O atoms,
initially occupying the hcp sites, switch to fcc sites and CO
adsorbs again in the favored on-top site.
For the (1O + 2CO) structure, there has been no
LEED intensity analysis, but infrared absorption spectroscopy
(IRAS) and X-ray photoelectron spectroscopy (XPS)
experiments (Schiffer et al., 1997) indicate
that the O atoms occupy hcp sites and the CO molecules occupy
on-top and fcc sites.

To obtain insight into the behavior of these co-adsorption systems,
DFT calculations have been carried out.
The calculated atomic geometries are
displayed in the middle section of Fig.~\ref{co+o}.
Good agreement with the LEED determined geometry was found
for the first two of these
phases for which comparison is possible (Stampfl and Scheffler, 1998).
The calculations show that for the (2O + 1CO) structure, it is indeed
energetically
more favorable (by 0.59~eV) for half of the O atoms
to occupy the less favorable fcc sites and CO to adsorb in the on-top
site rather than maintaining the $(2 \times 1)$-O arrangement and CO
adsorbing in a hollow site.
In addition to those phases that have been experimentally identified,
the theory predicts the stability of another phase
(Stampfl and Scheffler, 1998),
namely $(2 \times 2)$-(3O + 1CO)/Ru(0001), seen in the
far right-hand-side of Fig.~\ref{co+o}. The adsorption energy of
CO in this structure is notably weaker than for CO in the on-top site;
it is,
however, still appreciably exothermic with a value of 0.85~eV.
An important consideration concerning whether a structure
can in fact form, is the kinetics.
The possibility of kinetic hindering due to
energy barriers induced by the adsorbed O atoms was investigated
by calculating the total energy of CO at various distances above
the hcp-hollow adsorption site,  i.e., above the
vacant  O site of the $(2 \times 2)$-3O/Ru(0001)
structure.
Incidentally, this structure has recently been shown to
represent a new stable phase of O on
Ru(0001) (Kostov et al., 1997; Kim et al., 1998; Gsell et al., 1998).
The calculations show that there is an energy
barrier to adsorption of $\approx$0.35~eV.
This implies that rather high CO pressures would be
required in order to realize this structure experimentally.
Similar calculations were carried out for the (1O + 2CO) phase for
the CO molecule above the fcc site. It was found in this case there
is also an energy barrier, but slightly smaller of about 0.2~eV, thus
(at least
partially) explaining the low sticking coefficient and the
necessary high exposures found experimentally in order to
create this phase (Schiffer et al., 1997).

The valence electron density of the various phases are
also shown in Fig.~\ref{co+o}.
The oxygen atoms appear as the red (i.e., high electron density),
almost spherical features.
Both CO and O can be seen to
induce a significant redistribution of the electron density of the
top-layer Ru atoms.
For CO in the hollow sites, it is apparent that the bond
strength (per bond) is
weaker than that for CO in the on-top site. In the hollow sites, however,
CO forms {\em three} bonds with the metal surface so it is expected
that they be longer and weaker.
The calculations show nevertheless that the adsorption energy is
significantly
weaker in the hollow sites than in the on-top site for these
structures; this is not the case for the clean surface where the
energy difference is only about 0.04~eV. Thus, this
significant energy difference is a consequence
of the co-adsorbates.
These ($m$CO+$n$O)/Ru structures depicted in Fig.~\ref{co+o}, each
posessing the same periodicity but with
varying numbers of species and adsorption sites,
represent an ideal model co-adsorption series for study by
first-principles calculations. From analysis of the results
of such calculations much can be learnt about the various
interaction mechanisms at play.

\subsection{Chemical reactions at metal surfaces}
\label{sec:reactions}
%%ms  THINK !!
%%ms  here I mention words, which are probably well defined
%%ms  in the chemical literature and just often used badly.
%%ms  Thus, I should also give the "chemical" references??
%%CS  perhaps we could cite Somorjai's book on
%%CS  calalysis ? I am not sure of the reference: it might be
%%CS  G.A. Somorjai; Introduction to Surface Chemistry and Catalysis,
%%CS  New York: John Wiley & Sons, Inc., 1994.
%%CS  (I think this may not be exactly what you had in mind, but
%%CS   still it could be good)
%%ms  ...   yes, it's not exactly what I had in mind ...but
%%ms   I will think about it
This section summarizes some basic aspects of the present understanding
of the reactivity
of surfaces. Here the term ``reactivity'' usually refers to the surfaces'
ability to break bonds of an approaching molecule and to adsorb the
fragments, which is  often the rate limiting step in catalytic reactions.
For example, in the ammonia synthesis it is the dissociation
of N$_2$, and for various examples of oxidation catalysis (e.g., the
catalytic oxidation of CO) it is the dissociation of O$_2$ (see
Section \ref{sec:catalytic} below). Just having referred
to ``catalysis'' a
word of warning is appropriate because industrial catalysis involves
many more aspects than just dissociation of a certain molecule. Other
aspects are ``selectivity'', which means that only the desired reaction
should take place, and competing reactions yielding unwanted products
are suppressed. Also the buffering of intermediate chemical products
is important, as is the self-maintenance of the catalyst, the possible
role of the catalyst's support and of promotors.
And (often) it is important that no poisonous by-products are released.

\subsubsection{The problem with ``the'' transition state}
\label{sec:transition-state}

We will discuss the surface reactivity in terms of  molecules approaching
the surface considering all relevant atomic coordinates. The total energy
as function of the atomic coordinates is called the ``potential-energy
surface'' (PES), cf. Section \ref{sec:energies}. It represents
the energy surface on which the atoms will move.
Whereas the electrons are assumed to be in their ground state of the
instantaneous geometry, the  wave functions of the nuclei describe 
the details
of the atoms' dynamics, i.e., the vibrations, rotations, center-of-mass
translation, the scattering at the surface, the dissociation,
and the surface diffusion of the fragments.

In order to keep the discussion simple we will discuss the situation
of a molecular beam which is sent toward the surface, and in which the
molecules have a certain center-of-mass kinetic energy, and are in a
well defined vibrationally and rotationally excited state or the
ground state.
The probability of dissociation then is considered to be a
measure for the surface reactivity. It contains
\begin{itemize}
\item information concerning  the surface electronic structure,  i.e.,
      on the relevance of so-called frontier
orbitals (Wilke et al., 1996,
      and references   therein),
\item information about the high-dimensional PES on which
      the approaching molecule travels toward the surface,
\item the statistical average over many trajectories which
      finally determines with what probability the active
      sites at the surface are found and the molecule will
      dissociate, or if it will be reflected into the gas phase.
\end{itemize}

To discuss the dissociation probability  of incoming molecules,
it is often assumed that the reaction proceeds along
\begin{figure}[b]
\unitlength1cm
   \begin{picture}(0,6.0)
\centerline{
       \psfig{file=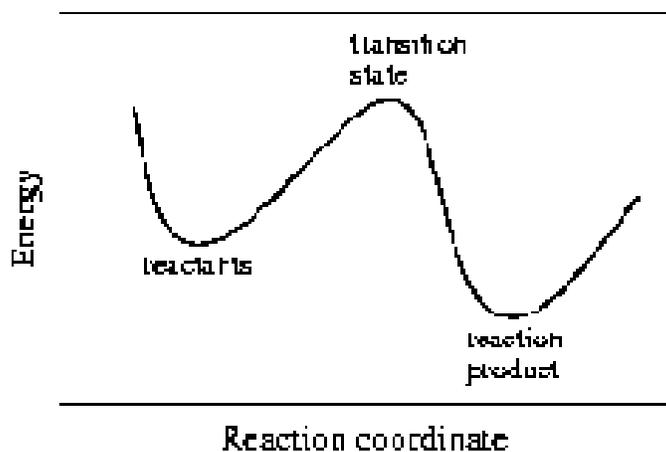,width=9.0cm}
}
   \end{picture}
\vspace{-0.5cm}
   \caption{\small
Energetics of a chemical reaction in which reactants
(their energy is that of the left minimum)
reach the reaction product (energy of the right minimum)
via a well defined transition state.}
\label{transition-state}
\end{figure}
a one-dimensional (or low-dimensional) reaction coordinate
and that it will
cross a well
defined ``transition state'' (cf. Fig. \ref{transition-state}). Then
the reaction probability is given by an Arrhenius behavior with
the energy barrier given by this transition state. While the concept
behind Fig. \ref{transition-state} has
proven useful (at least often) in gas-phase chemistry, it may
be misleading for the description of  surface chemical
reactions. In particular, we note that the phase space for a
molecule-surface
reaction has  very high dimensionality. For example, even in the
simplest surface reaction, i.e., when H$_2$ molecules are sent
toward the surface and if the substrate atoms do not move during
the scattering event,
the translations, vibrations and rotations of the two H atoms 
take place  in
a 6-dimensional configuration space, i.e., a 12-dimensional 
phase space.
As a consequence, the assumption of ``the'' transition state, 
as depicted in Fig. \ref{transition-state} can be grossly misleading. 
Instead, many
transition states exist, and which of them is taken with what probability
depends on the details of the incident H$_2$ dynamics, i.e.,
the H$_2$ translational kinetic energy, and the vibrations and rotations.
Consequently,
a good treatment of the statistics of the many possible trajectories
is mandatory.

As noted in Section \ref{sec:energies} we will restrict the discussion
in this chapter to situations where the Born-Oppenheimer approximation is
justified. Still, a severe problem remains, namely that knowledge of the
PES  barely exists. Up until recently only rough,
and as we now know, often incorrect semi-empirical models were used,
and these earlier studies of the dissociation dynamics
were restricted in
their dimensionality. For example, in the past the dissociation of H$_2$
was described in terms of  only 2 or 3 coordinates (out of the six
important
coordinates),
and the dependence of the PES
on the other coordinates was simply neglected.
The only reason for this simplification was that evaluating a
PES using good-quality electronic structure theory
is elaborate. Since about 1994 the situation has changed,  i.e.,
several groups started to take into account the higher
dimensionality (see Gross and Scheffler, 1998,
and references therein).
Though involved, even treating the six-dimensions of the two 
hydrogen atoms
is not yet complete because in general it could happen that
electronic excitations play a role in the scattering event
which is outside the  Born-Oppenheimer approximation, and furthermore, it
is well possible that the dynamics of the substrate atoms will
play a role.
These concerns may not be very important for the systems studied so far,
but for other systems they may well be relevant.
With respect to the validity of the Born-Oppenheimer approximation
we are not aware
of a serious breakdown, although sometimes it has been speculated.
For adsorbates at metal substrates, levels are typically broadened, which
implies that excited states have a short life time and thus
relaxation into the ground state configuration will be (nearly)
instantaneous on the time scale of the nuclear motion.
Obviously, for insulators and semiconductors the situation is different
(see, e.g., the discussion in  Gross et al., 1997).
Also, the situation would be different if laser excitations are involved
(i.e., photo-chemistry),
but this is not the topic of this chapter.

\subsubsection{Dissociative adsorption and associative desorption
of H$_2$ at transition metals}
\label{sec:H2}
Dissociative adsorption (or the time-reversed process, which is
associative desorption) is a dynamical process, and because of the high
dimensionality of the PES, a proper treatment of the dynamics is indeed
crucial.
Typically the dynamics of atoms is treated classically, i.e., by Newton's
equation of motions, but the underlying PES and the forces acting on the
atoms are (or could be  and should be) calculated by DFT.
This is what is called ``{\em ab initio} molecular dynamics''. It started
with
the seminal work of Car and Parrinello (1985). The elegance of their
approach
appears to imply that typically it is not very efficient,
and since their original paper several alternative (and numerically more
efficient) formulations have been developed (see, e.g., Payne et al., 
1992;
Kresse and Furthm{\"u}ller, 1996; Bockstedte et al., 1997; and references
therein).

When the moving nuclei are hydrogen atoms, it is often necessary to
treat also the nuclei as quantum particles. For such problems
a rather involved, high-dimensional
``{\em ab initio} quantum dynamics'' method
has been implemented (Gross et al., 1995, 1998; Kroes et al.,
1997; and references therein).
This is probably the most advanced approach and for heavy particles
(obviously) it becomes identical to {\em ab initio} molecular dynamics.

In the following section we describe some general features of the
PES and then we show examples which
demonstrate the importance of a quantum dynamical treatment of 
scattering and
dissociation of molecules at surfaces.
The dissociative adsorption of H$_2$ appears to be a very simple reaction.
However, due to the quantum nature of the hydrogen nuclei, the actual processes are
rather complex. Again, as always in this chapter, we will keep the
discussion simple, and for more details we refer to papers by Gross and
Scheffler (1998),
Gross (1998) and Kroes (1999).\\

\begin{figure}[tb]
\unitlength1cm
   \begin{picture}(0,10.0)
\centerline{
     \psfig{file=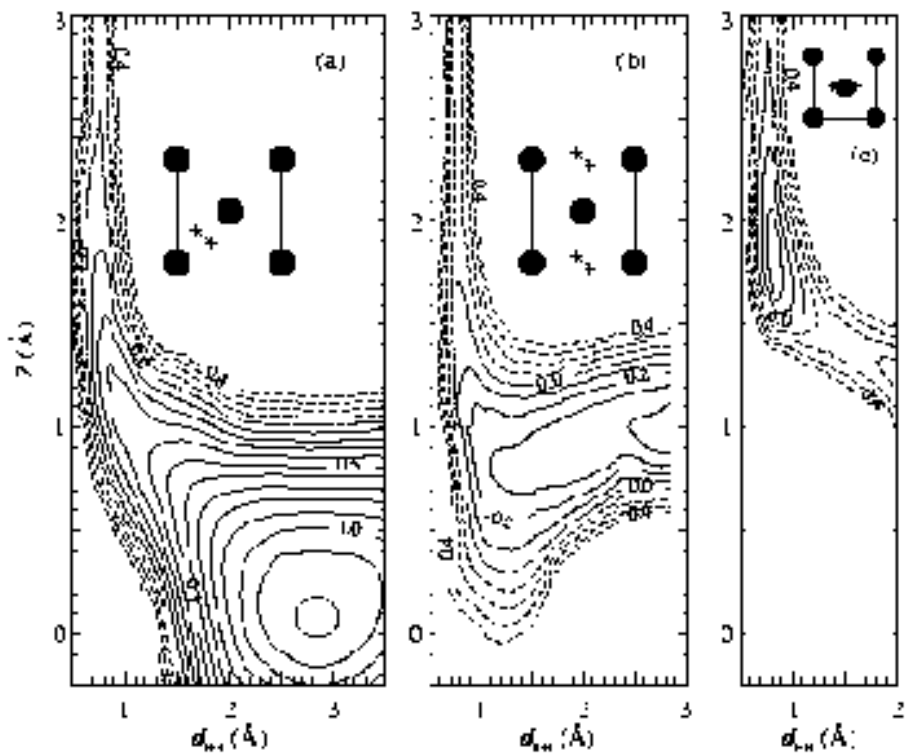,width=12.0cm}
}
   \end{picture}
\vspace{-0.3cm}
  \caption{\small
Cut through the six-dimensional potential energy surface (PES) of an
H$_{2}$ molecule above of Pd(001). We display ``elbow plots'' where 
$Z$ is
the height of the H$_{2}$ center of mass over the surface, and
$d_{\rm{H-H}}$ is the distance between the two hydrogen atoms.
Each cut is defined by the lateral H$_2$ center-of-mass coordinates in
the surface unit cell and the orientation of the molecular axis,
i.e., the coordinates $X, Y, \theta_{{\rm H}_2}$,
and $\phi_{{\rm H}_2}$:
The molecule is kept
parallel to the surface ($\theta_{{\rm H}_2}= 90^{\rm o})$
at an azimuthal orientation, $\phi_{{\rm H}_2}$, and $X, Y$ coordinates
shown in the insets.  The units of the potential energy are eV and
the interval between adjacent contour lines is 0.1 eV.
The results are from Wilke and Scheffler (1996).
}
\label{fig:H2-Pd-elbows}
\end{figure}

%%\subsubsection
\noindent
{\bf 5.9.2.1~~The potential-energy surface of H$_2$ at transition-metal surfaces}\\
A good knowledge of the high-dimensional PES of the molecule-surface
system is mandatory for a detailed understanding,
as the PES rules the scattering and the dissociation.
The high-dimensionality implies that the dynamics of the problem will be
complex and therefore typically it will be impossible to analyze the 
PES by simply
looking at it. In fact, looking at a PES with dimensionality  equal
to or higher than six is only possible in terms of
cuts along planes in configuration space. Figure 
\ref{fig:H2-Pd-elbows} shows
three examples of what is usually called an ``elbow plot'' because it
often looks like an elbow. The planes in configuration space are 
identified
by the insets in the figure. Along these planes the height of the
molecule
$Z$ and the H--H distance $d_{\rm H-H}$ is varied, and lines of
constant
potential energy are displayed. Obviously, the cut through
the PES shown in
Fig. \ref{fig:H2-Pd-elbows}a
looks very different to those in
Figs. \ref{fig:H2-Pd-elbows}b, c.

If the H$_2$-surface distance (i.e., $Z$) is large the contour lines in
Fig. \ref{fig:H2-Pd-elbows} reflect
the energetics of a free H$_2$ molecule, which has the equilibrium
separation of 0.75 \AA. We can also read off the nearly harmonic
potential
underlying the H-H vibration, which implies an   H$_2$ zero-point energy 
of
0.26 eV,  and a vibrational excitation energy of 0.52 eV.

Although the knowledge of one elbow plot is already much better than just
knowing the transition state, as the latter is just one point in the 
whole
configurations space, restricting the world to only one elbow
ignores the fact that two H-atoms have six degrees of freedom, not just
two.
It is obvious from Fig. \ref{fig:H2-Pd-elbows} that neglecting
the high dimensionality, i.e., assuming that the elbows for different
choices of $(X, Y, \theta_{{\rm H}_2}, \phi_{{\rm H}_2})$ are similar,
is by no means justified.

We will now discuss one aspect of the PES which explains the
general trend of the reactivity of different
transition-metal surfaces. However we already like to add the warning
that an appropriate (and reliable) analysis of chemical reactivity
has to include a good {\em statistical treatment} of the {\em dynamics}
of the atoms, which will be discussed in the next section.

Figure \ref{fig:H2-Pd-elbows}a shows that on Pd(001) there is at least
one pathway toward dissociative adsorption
which is not hindered by an energy barrier. In fact, if we inspect the
PES for Rh (the left neighbor of Pd in the periodic table) [see Eichler 
et al., (1999a)] one sees that
here the PES offers even
more pathways without barriers: For molecules with their
axis parallel to the surface almost all dissociation paths
are non-activated.
On the other hand, for Ag (the right neighbor of Pd) we
find that all pathways toward dissociative adsorption are
hindered by a significant energy barrier (Eichler et al., 1999a).
This result simply reflects the higher chemical activity
of the true transition metals compared to the noble metal silver.
\begin{figure}[tb]
\unitlength1cm
   \begin{picture}(0,8.5)
\centerline{
       \psfig{file=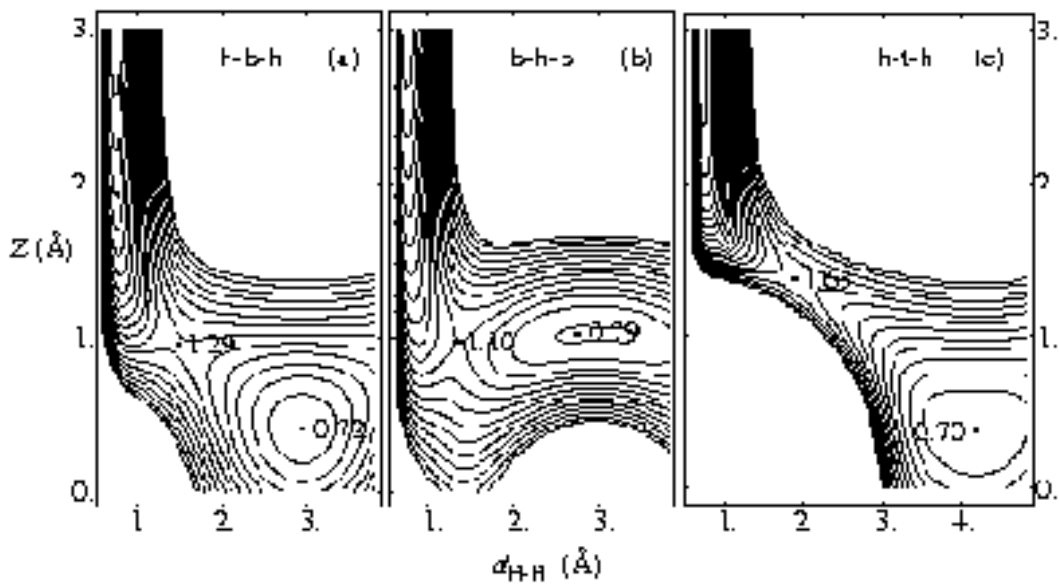,width=14.0cm}
}
   \end{picture}
\vspace{-0.3cm}
   \caption{\small Same as Fig. \ref{fig:H2-Pd-elbows} but for
H$_2$ at Ag(001). The results are after Eichler et al. (1999a)
and private communication.
}
\label{fig:H2-Ag-elbows}
\end{figure}
Figure \ref{fig:H2-Ag-elbows} shows the PES of H$_2$ at Ag(001).
The important aspect in comparing the (a) panels of
Figs. \ref{fig:H2-Pd-elbows} and \ref{fig:H2-Ag-elbows}
is not just that there is an energy barrier for the silver substrate,
but {\em where this barrier is located} in configuration space: The
lowest
barrier (see Fig. \ref{fig:H2-Ag-elbows})
is found very close to the surface and at a H--H distance which is
significantly stretched (by about 100\%) compared to the free molecule.
We also note in passing that the adsorption state of H on Ag(001) is only
metastable, i.e., at a higher energy than the free H$_2$ molecule.

Keeping the
problems with ``the'' transition state in mind (we will come back to it
in the next section) we now analyze its properties.
The fact that the transition state  in Fig. \ref{fig:H2-Ag-elbows}
is found at small $Z$ and significantly
stretched $d_{\rm H-H}$ values
implies that the H--H bond is nearly broken when the molecule
has reached the top of the barrier.
In fact,  the detailed analysis
of the wave functions of the H$_2$--surface system at the geometry of
the barrier shows  that the differences between different metals
should be described in a covalent (or tight-binding) picture.
Earlier attempts, which applied a description in terms of Pauli 
repulsion,
and/or frontier orbitals of the unperturbed constituents do not
account properly for the character and strength of the interaction:
At the barrier we find that the interaction between the molecule and
the substrate is already significant and the electronic states ruling
the energetics are very different from those of the clean
surface.
\begin{figure}[tb]
\unitlength1cm
   \begin{picture}(0,9.0)
\centerline{
       \psfig{file=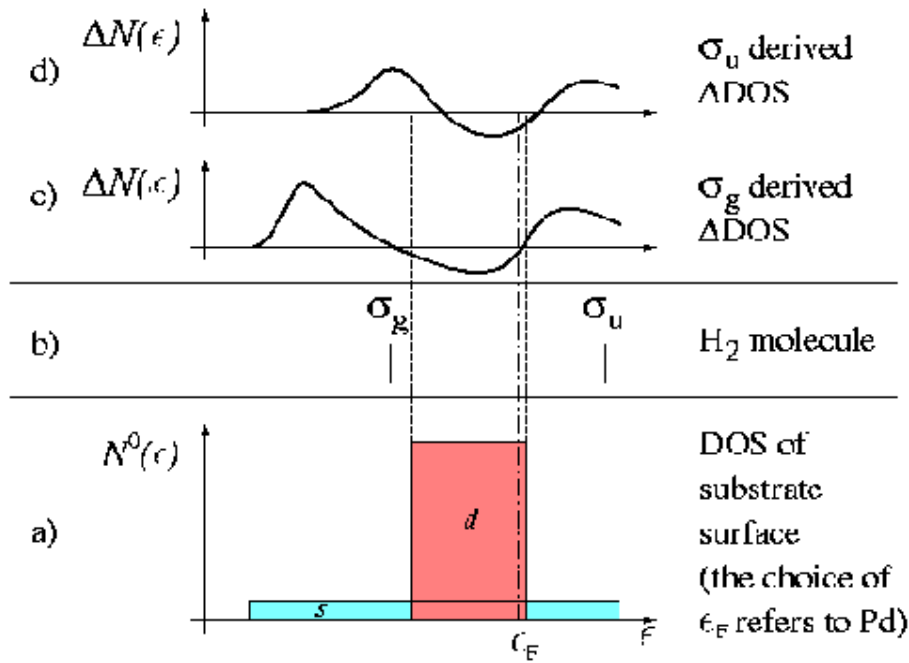,width=12.0cm}
}
   \end{picture}
\vspace{-0.3cm}
   \caption{\small
Schematic description of the interaction of H$_2$ at the transition
state toward  dissociative adsorption at transition-metal surfaces.
The bottom panel (a) shows the density of states for a transition
metal before adsorption,  and panel (b)
shows the energy levels of a free H$_2$ molecule:
the bonding state $[\phi_{\sigma_{\rm g}} = \phi_{1s}({\bf R_1}) +
\phi_{1s}({\bf R_2})]$ and the antibonding state
$[\phi_{\sigma_{\rm u}} = \phi_{1s}({\bf R_1}) - \phi_{1s}({\bf R_2})]$.
The $\sigma_{\rm g}$-level is filled with two electrons and the
$\sigma_{\rm u}$-level is empty.
The interaction between the H$_2$ $\sigma_{\rm g}$-level and
the substrate
$s$- and $d$-bands gives rise to a broadening and the formation of an
antibonding level at about  the upper edge of the $d$-band and a bonding
level below the $d$-band,  see panel (c). Panel (d) shows that
the interaction between the H$_2$ $\sigma_{\rm u}$-level with the
substrate
$s$- and $d$-bands gives rise to a broadening and the formation of
a bonding
level (at about  the lower edge of the $d$-band) and an antibonding level
(above the $d$-band).
}
\label{fig:H2-tight-binding}
\end{figure}
Figure \ref{fig:H2-tight-binding} summarizes the view developed by
Hammer et al. (1994, 1995) in their analysis of H$_2$ at Cu(111)
and H$_2$ at NiAl(110) [see also the earlier study by Hjelmberg et al.
(1979) for H$_2$ at jellium].
Because of obvious reasons, this figure looks similar to that in
Section  \ref{adsorbate-substrate-interaction}, but now the interacting
particle is an H$_2$ molecule and the geometry is that of the
transition state, and not of an adsorbate equilibrium configuration.
At the transition state the interaction of the molecule with the
surface has already produced a clear splitting into states which are
bonding
between the molecule and the substrate and which are antibonding.
Assuming a substrate from the middle of the transition-metal series (e.g.,
Ru or Rh) implies that the low energy resonances, which are
$\sigma_{\rm g}$ and $\sigma_{\rm u}$ derived, are filled with electrons.
These states are bonding with respect to the molecule-substrate
interaction and thus their filling implies an attraction of the molecule
to the surface. But the filling of the $\sigma_{\rm u}$ resonance
also implies a weakening of the H--H bond. Thus, when the substrate
Fermi-level is in the middle of the $d$-band, we understand that
molecules are strongly attracted to the surface and at the same time
the molecular bond is broken.

On the other hand, when the substrate Fermi level is well above the
$d$-band, as for a noble metal, also the states which are
antibonding with respect to the molecule-surface interaction get filled
(the high energy DOS in panels (c) and (d) of Fig.
\ref{fig:H2-tight-binding}).
This implies that the net interaction between the molecule and
the substrate
is repulsive. Thus, an energy barrier is built up which hinders the
dissociation.\\

%%\subsubsection{
\noindent
{\bf  5.9.2.2~~The dynamics of H$_2$ dissociation at
transition-metal surfaces}\\
%%\label{H2-dynamics}
We have seen that the dissociative adsorption of H$_2$ at Pd(001)
can proceed without an energy barrier. However, this holds only for few
pathways; in particular, it is necessary that the molecule reaches the
surface with an orientation parallel to the surface in order to
be able to
dissociate. Molecules arriving with an orientation of their axis
perpendicular to the surface will be reflected.  In order to understand
in more detail what goes on and to determine the probability 
of dissociation
it is necessary to
calculate the dynamics along the high-dimensional PES.
This has been done by Gross et al. (Gross et al., 1995, 1998).
They treat also the hydrogen nuclei as quantum particles, and
a comparison with a classical treatment of the dynamics showed in great
detail, which quantum effects are important (mainly zero-point 
vibrations,
and only little tunneling).

\begin{figure}[b]
\unitlength1cm
   \begin{picture}(0,6.0)
\centerline{
      \psfig{file=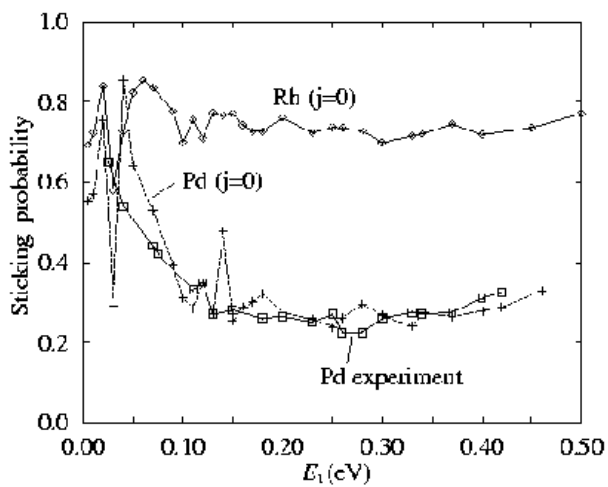,width=8.0cm}
}
   \end{picture}
\vspace{-0.3cm}
   \caption{\small Initial sticking probability versus kinetic energy for
an H$_2$ beam under normal incidence on a clean Pd(001)
and Rh(001) surface. The H$_2$ molecules are incident normal
to the surface and are in their rotational and vibrational
ground state. Theoretical results are from Eichler et
al. (1999a), and the experimental data are from Rendulic et al. (1989).
}
\label{fig:sticking}
\end{figure}

Figure \ref{fig:sticking} displays the sticking probability $S$ for
two different substrates.
The sticking probability is the probability that an incoming H$_2$
molecule dissociates and that the atoms then adsorb at the surface.
One could also say that $(1-S)$ is the probability
that an incoming H$_2$ molecule gets reflected back into the vacuum.
The calculated sticking curve exhibits many oscillations.
This is not noise, but
reflects the quantum nature of the dissociative adsorption
and the scattering event:
The reflected H$_2$ beams, which differ by reciprocal lattice
vectors of the surface, and the rotationally and vibrationally excited
beams are subject to quantum interferences. Thus, the
oscillations have a quantum mechanical origin, and when the
hydrogen nuclei were treated classically, they were absent.
Similar oscillations are well known for other quantum
mechanical scattering studies, such as the scattering of
electrons (LEED) or that of He. We trust that at some time
they will be also observed experimentally for the H$_2$ scattering at
metal surfaces, but clearly such experiments are very demanding.
More details on the nature of the quantum interference effects
of H$_2$/Pd scattering can be found in the paper by Gross
and Scheffler (1998).
When the theoretical results were broadened  with the
energy resolution of typical experiments, most of the oscillations
were averaged out.

One unexpected and surprising result of Fig. \ref{fig:sticking}
is that for low kinetic
energies ($E_i < 0.05$ eV) the sticking probabilities for the 
Pd and
Rh substrates are very similar (we recall that the PES of
Rh offers many pathways with vanishing energy barriers
toward dissociation but Pd only a few).
Despite their differences in electronic structure (Pd is more
noble than Rh) and the clear differences in the PESs,
both substrates give  an $S$ value as high as
75\% for low $E_i$. In fact, this low energy range  corresponds 
to the
typical thermal kinetic energies.
While for Rh the sticking probability always remains high,
it decreases for Pd to about 25\%.

The sticking probability for H$_2$ at Pd(001) starts at a high value
and then it goes down; this is found in the theory as well as in the
experimental data. Earlier, it had been suggested that this
behavior is due to the existence of a precursor adsorption, where
H$_2$ is trapped close to the surface from where it can undergo many
attempts to dissociate. Whereas this concept is valid for some systems,
the theoretical PES does not exhibit such a precursor state and thus,
the explanation for this behavior of $S(E_i)$ must have a different
origin. The analysis of the H$_2$ dynamics revealed that the effect is
best described by the word ``steering'', which
means the same as on the road. If one is going slowly,
one will make it well along a curvy street.
However, when one is going too fast
one is pulled out of the curve.
Thus, molecules which are approaching slowly will be able to follow
a pathway along curvy valleys and find (along the
high-dimensional PES) the way toward the point where they can
dissociate without an energy barrier. They will dissociate even
if their initial orientation is unfavorable (e.g., perpendicular to
the surface)
because they are steered toward a more favorable transition
state geometry.
On the other hand, fast molecules will not be able to make it 
around all
the curves. They will bump against an energy barrier and be reflected
back into the gas phase.

Thus, the behavior we see for the Pd substrate which
increases the sticking to nearly 75\% at low energies of the
H$_2$ beam, is  truly a dynamical effect.
In a static picture the high reactivity of Pd cannot be understood.
The efficiency of steering depends on the speed of the incoming molecule
and on the shape of the PES. Therefore, to evaluate the sticking
probability,
which we consider a good measure of the surface reactivity,
it is important to consider all degrees of freedom and
the dynamics of the nuclei. Obviously, as much as ``steering''
is important to
understand the high reactivity of Pd at low $E_i$, for other 
systems, which
on the grounds of the electronic structure alone
may be expected to exhibit a high reactivity, an ``anti-steering''
may occur, which drives approaching molecules not toward the best
transition state but against an energy barrier.

We add some words on the influence of rotations of the incoming
molecule. For rapidly rotating molecules steering is suppressed
because the molecules
will quickly rotate out of favorable orientations for dissociation
(so-called rotational hindering) (Darling and Holloway, 1994; 
Gross et al., 1996a, b, c).
But, there is also a steric effect (Gross et al., 1995):
molecules rotating in the so-called helicopter fashion
(i.e., parallel to the surface)
with $m=j$ dissociate more easily than molecules rotating
in the so-called cartwheel fashion (i.e., perpendicular to the
surface) with $m=0$, because the helicoptering molecules have
their axis already oriented preferentially parallel to
the surface which is the most favorable orientation for 
dissociation.  This orientation effect can be so strong 
that it even  over-compensates the rotational hindering.
Furthermore we note that, for example, the 
hindering effect of the rotational motion may be counterbalanced 
by the transfer of rotational energy to translational energy 
(see Eichler et al., 1999b, and Gross and Scheffler, 1996b, for
more details).

\subsection{The catalytic oxidation of CO}
\label{sec:catalytic}
\begin{figure}[b]
\unitlength1cm
   \begin{picture}(0,5.0)
\centerline{
    \psfig{file=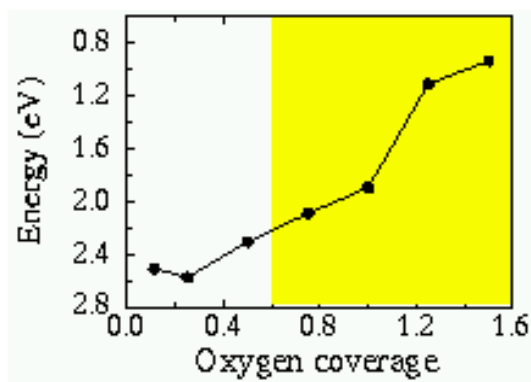,width=7.0cm}
}
   \end{picture}
\vspace{-0.3cm}
   \caption{\small
Average adsorption energy of O on Ru(0001) for various coverages, with
respect to 1/2 O$_2$. From Stampfl and Scheffler (1997).}
\label{oadsorp}
\end{figure}
Just as the dissociative adsorption of H$_{2}$ discussed above
in Section~\ref{sec:H2} may be regarded as the model
system for studying the dynamics of the dissociative adsorption
and associative desorption of small molecules,
the CO oxidation reaction may be regarded as the most simple
prototype model system
of a surface heterogeneous catalytic reaction --
a process involving molecular adsorption
(chemisorption of CO is non-dissociative under the conditions of
catalytic oxidation)
and dissociative (atomic) adsorption, surface diffusion, 
surface reaction, and desorption of products.
The oxidation of carbon monoxide
has been extensively
studied which is largely due to its
technological importance (e.g., in car exhaust catalytic converters
where the active components are transition metals such as
Pt, Pd, and Rh) but is also related to
its (relative)``simplicity'' (Engel and Ertl, 1979, 1982;
Biel\'{a}nski  and Haber, 1991; Peden, 1992, and references therein).
Therefore there exists a large data-base concerning its macroscopic
behavior and in these terms it is reasonably well understood.
On a {\em microscopic} level, however, an understanding
is still lacking.
Steps in this direction have recently been made via
first-principles calculations (Stampfl and Scheffler, 1997, 1999;
Alavi et al., 1998; Eichler and Hafner, 1999).
\begin{figure}[tb]
\unitlength1cm
   \begin{picture}(0,6.5)
\centerline{
       \psfig{file=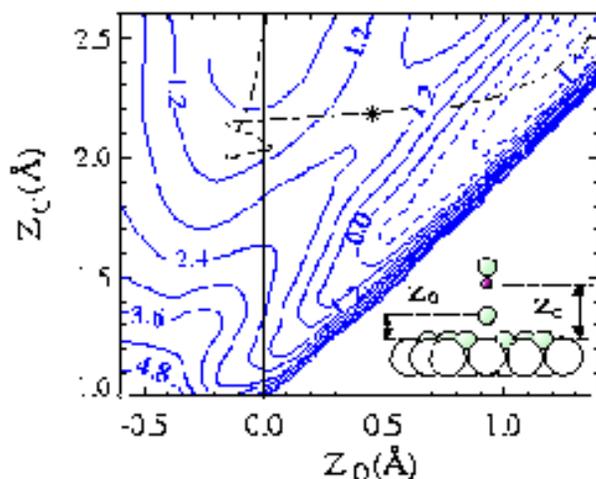,width=8.0cm}
}
   \end{picture}
\vspace{-0.5cm}
   \caption{\small Cut through the high-dimensional potential energy
surface (PES) as a function of the positions of the C atom,
Z$_{\rm C}$, and the O adatom, Z$_{\rm O}$ (see inset).
The molecular axis is constrained to be perpendicular to the surface.
Positive energies are shown as continuous lines, negative ones as
dashed lines.
The contour-line spacing is 0.6 eV.
The dot-dashed line indicates a possible reaction pathway. From 
Stampfl and Scheffler (1997). }
\label{pes-co2}
\end{figure}
The results of one of these studies is briefly discussed below.

We first like to mention that
the investigation of this type of
surface process is more complex than that described above for
H$_{2}$ dissociation; firstly significant substrate relaxations 
may be induced by CO and O adsorption, and the surface should not be treated
as rigid. Secondly, the nature of the problem is different; 
here a ``real'' surface chemical reaction takes place where a 
product is formed that desorbs from the surface.
So not only the coordinates of one adparticle need to be considered,
but also those of the reacting partner, i.e., the problem has an
even higher dimensionality.

Catalytic oxidation of CO at the ruthenium
transition metal surface has been reported to exhibit unusual
behavior compared to other transition metals when studied
at high pressures (Peden, 1992, and references therein). In particular,
the reaction rate over Ru for oxidizing conditions
(i.e., at CO/O$_{2}$ pressure ratios $ < 1$)
is the highest of the transition metals considered.
In contrast, under ultra-high vacuum conditions (UHV),
the rate is by far the lowest.
In addition, the kinetic data for the reaction over Ru
(e.g., the temperature and pressure
dependencies of the rate) deviate compared to the other metals and
highest reaction rates occur for high oxygen concentrations at the
surface. Interestingly, under these conditions, it was speculated 
that the
reaction mechanism may proceed via an Eley-Rideal
interaction (Peden et al., 1986, 1991), that is, a scattering 
reaction where a particle from
the gas-phase reacts with an adsorbed species without adsorbing
on the surface first. The usual mechanism by which reactions at 
surfaces take place
is via the Langmuir-Hinshelwood mechanism where both species are
adsorbed  on the surface prior to reaction.
\begin{figure}[tb]
\unitlength1cm
   \begin{picture}(0,6.0)
\centerline{
      \psfig{file=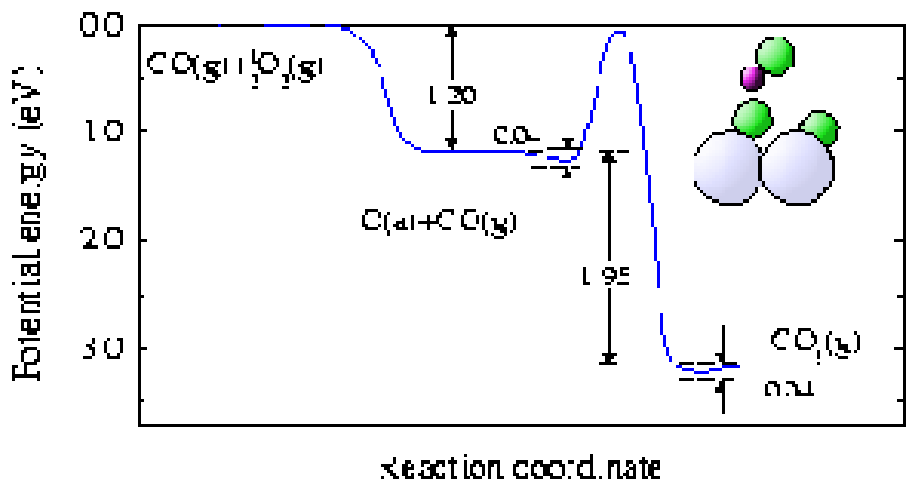,width=12.0cm}
}
   \end{picture}
\vspace{-0.3cm}
   \caption{\small Calculated energy diagram for a scattering reaction
of gas-phase CO (Eley-Rideal mechanism) with an adsorbed O atom
of the 1 ML phase on
Ru(0001). Note, 1.20 eV is the energy required to remove one O atom (relative to 1/2 O$_2$) from the 1 ML structure as calculated in a (2 $\times$ 2) surface unit cell -- or, equivalently, the energy gained by the system on adsorbing this atom in the vacant site, i.e.,  it is not the average adsorption energy at 1 ML as shown in Fig.~5.39.  The geometry of the
transition state is indicated in the inset. From Stampfl and Scheffler (1997).}
\label{en-dig-er}
\end{figure}

Under standard UHV conditions using O$_{2}$,
the saturation coverage has been reported to
be approximately half a monolayer.
Recent studies employing NO$_{2}$ or very high
exposures of O$_{2}$ have shown, however, that Ru(0001) can support
higher coverages; namely ordered structures
$(2 \times 2)$-3O     (Kostov et al., 1997; Kim et al., 1998;
Gsell et al., 1998) and
$(1 \times 1)$-O (Stampfl et al., 1996), as had
initially been predicted by DFT-GGA calculations (Stampfl and
Scheffler, 1996),
and that subsurface adsorption occurs after completion of the
monolayer structure at elevated temperatures ($\approx$600~K)
(Stampfl et al., 1996; Mitchell and Weinberg, 1996; B\"{o}ttcher and
Niehus, 1999).
Subsurface oxygen can apparently occur as well as the formation of
surface oxides\footnote{We note that the term ``surface
oxide'' is not well defined, and in particular for Ru, which can exist in
many oxidation states, it may be difficult, if not impossible,
to distinguish between a 2-3 layer thick ``surface oxide''
and an on-surface plus sub-surface adsorbate phase.}, and surfaces with different domains of high oxygen concentration but different stoichiometry are assumed to actuate the high catalytic reactivity of Ru.
Thus, the (apparent) oxygen saturation coverage noted above
for low (or room) temperature UHV conditions and typical
exposure by O$_2$ is solely due to kinetic hindering for O$_{2}$
dissociation.

The adsorption energy of O on Ru(0001) decreases notably with
increasing coverage; in particular, for concentrations
$\geq$ 1~ML the bond strength is atypically
weak compared to the lower coverage structures.
This can be seen from Fig.~\ref{oadsorp}.
A good catalyst should actuate dissociation of O$_{2}$
but at the same time should not bind the dissociated
entities too strongly, as then they have a good capability to 
diffuse and react.
Too strongly bound constituents would have little reason to 
react at all.
It is expected therefore that
these weaker adsorption energies for $\Theta \geq 1$
will lead to an enhanced reaction rate of CO$_2$ formation.
Indeed, recent experimental studies of CO oxidation
over Ru(0001) surfaces loaded with such high oxygen concentrations,
have found even higher rates (B\"{o}ttcher et al., 1997, 1999).

\begin{figure}[tb]
\unitlength1cm
   \begin{picture}(0,10.5)
\centerline{
       \psfig{file=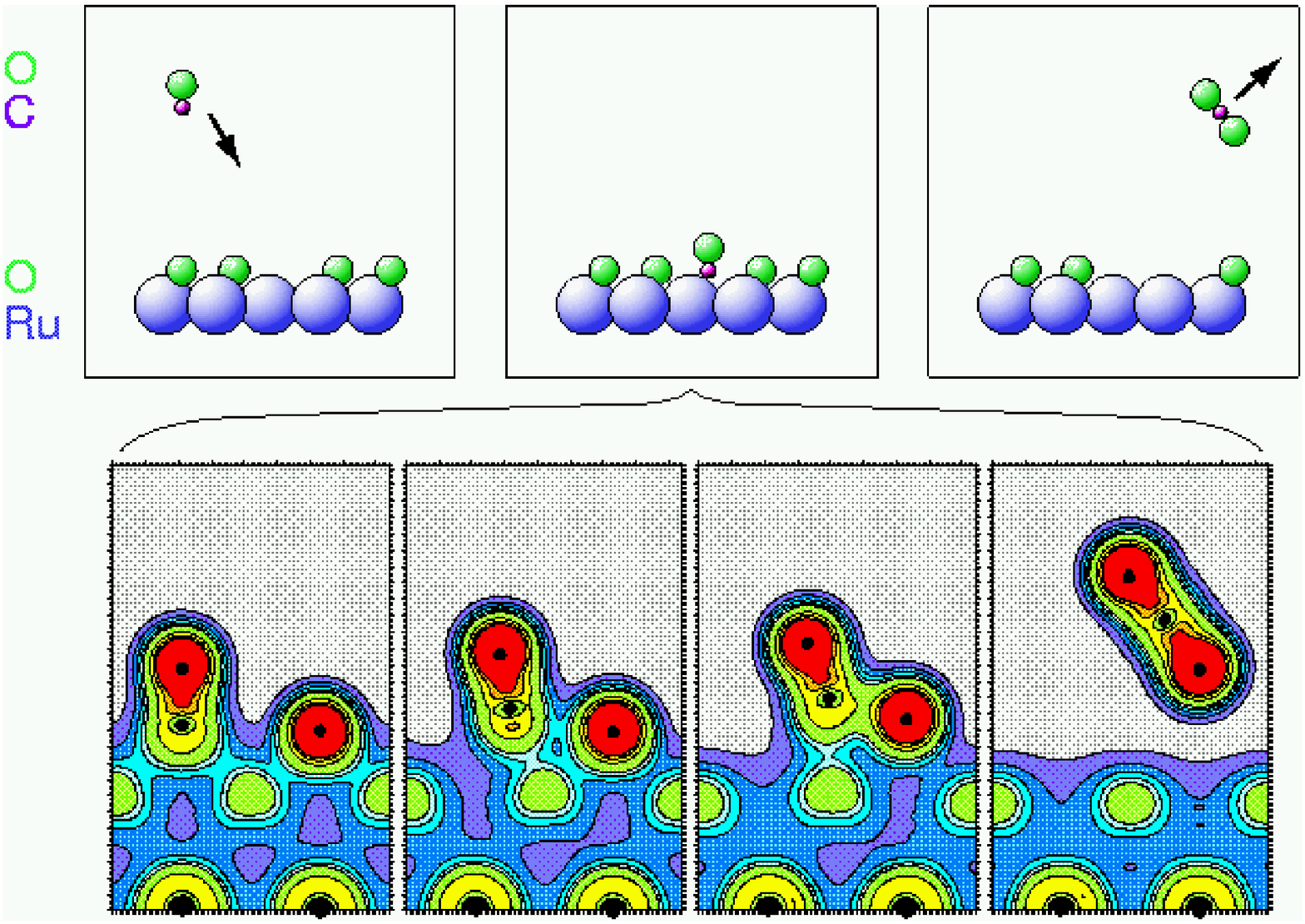,width=15.0cm}
}
   \end{picture}
\vspace{-0.3cm}
   \caption{\small Snapshots of the  Langmuir-Hinshelwood reaction
(top panel) of CO oxidation at Ru(0001).
The bottom panels display the electron density distribution
along a reaction path close to the transition states. The units
are bohr$^{-3}$, and the colors and contours are the same as in
Fig.~\ref{co+o}.}
\label{lh-chden}
\end{figure}

When a complete $(1 \times 1)$-O structure is present, the 
calculations show that CO cannot adsorb. However, a scattering
reaction of gas-phase CO with
adsorbed oxygen (i.e., an Eley-Rideal
mechanism) is possible, and the minimum energy
barrier was found to be about 1.1~eV with a corresponding
bent transition state geometry (Stampfl and Scheffler, 1997).
Figure~\ref{pes-co2} shows an appropriate cut through the
high-dimensional PES.
The minimum barrier corresponds to a tilted CO molecular axis of about
131$^{\circ}$ as depicted in the inset of Fig.~\ref{en-dig-er} 
which shows the corresponding energy diagram.
It can be seen that due to the surface reaction there would
be a significant energy gain of about 1.95~eV so that the
produced CO$_{2}$ molecules would be highly energetic.
Assuming this energy barrier would be the rate-limiting step
of CO$_{2}$ formation via this mechanism, then an estimate of the
reaction rate can be made using an Arrhenius-like equation.
The prefactor being taken as
the number of CO molecules hitting the surface per site per second
at a given temperature and pressure.
The rate obtained in this way was found to be significantly lower
than that measured experimentally (Peden and Goodman, 1986)
indicating that this mechanism alone cannot explain the enhanced 
CO$_{2}$ turnover frequency.
\begin{figure}[p]
\unitlength1cm
   \begin{picture}(0,5.0)
\centerline{
       \psfig{file=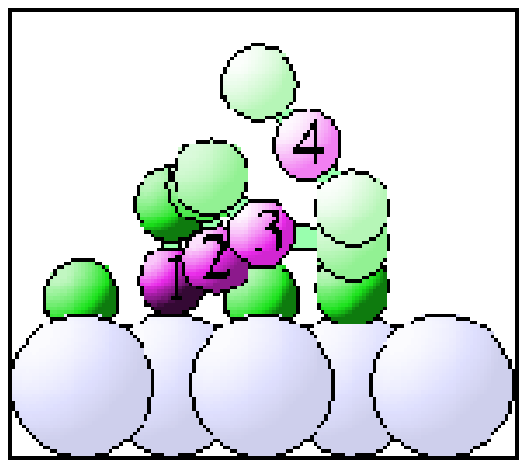,width=6.0cm}
}
   \end{picture}
\vspace{-0.3cm}
   \caption{\small Atomic positions along a reaction energy pathway to CO$_{2}$
formation.  The large, small, and
small (labeled) circles represent Ru, O, and C atoms, respectively. The
numbers indicate the sequence along the reaction path.
}
\label{react-path-geoms}
\end{figure}

To investigate other  possible reaction channels, it is
conceivable that there may be vacancies in the
$(1 \times 1)$-O adlayer (see Stampfl and Scheffler, 1997),
e.g., created by the above mentioned scattering reaction.
CO molecules may then adsorb at these vacant sites and react via
a Langmuir-Hinshelwood mechanism (see Fig.~\ref{lh-chden}).  In Section~\ref{sec:co-adsorption}
it was seen that there is an energy barrier for CO to adsorb in
such a vacancy; however, at the high pressures and elevated
temperatures employed in catalytic reactor experiments, it is 
expected that a barrier of such size can be readily overcome.

In view of the weaker CO-metal
bond strength compared to that of the O-metal bond strength at this
coverage, i.e., 0.85~eV compared to 2.12~eV\footnote{Here 2.12~eV is the average adsorption energy of O at coverage $\Theta = 0.75$ (cf. Fig. 5.39), i.e., corresponding to the (2 $\times$ 2)-3O structure.}
 (with respect to gas
phase 1/2 O$_{2}$), it may be expected that
the energetically favorable reaction pathway is via movement of
CO toward the O atom.
Indeed, among the various pathways investigated,
this turns out to be the case where the determined
energy barrier is about 1.5~eV.
Some selected geometries
along the reaction path are shown in Fig.~\ref{react-path-geoms}.
We refer to Stampfl and Scheffler (1999) for further details.
\begin{figure}[p]
\unitlength1cm
   \begin{picture}(0,7.0)
\centerline{
     \psfig{file=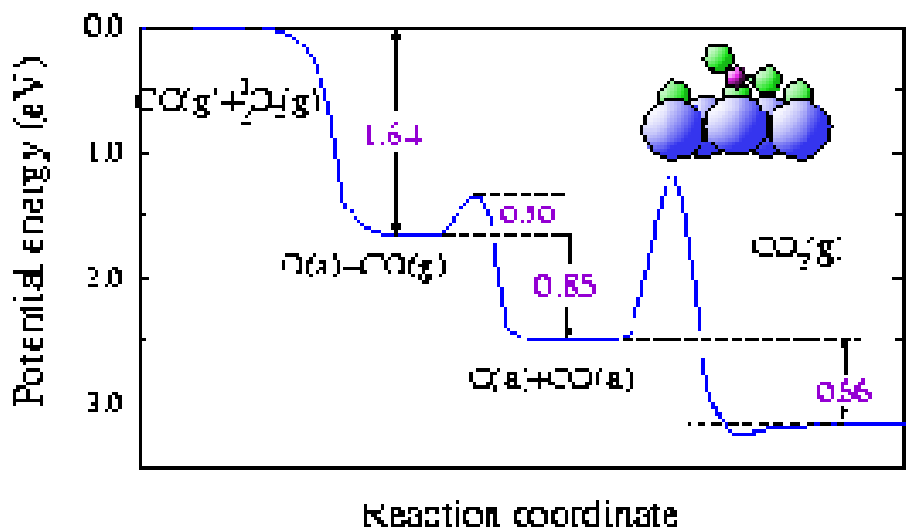,width=12.0cm}
}
   \end{picture}
\vspace{-0.3cm}
   \caption{\small Calculated energy diagram for a Langmuir-Hinshelwood
reaction mechanism at
Ru(0001) for high oxygen coverages on the surface. Note, 1.64 eV is the energy required to remove one O atom (relative to 1/2 O$_2$) from the (2 $\times$ 2)-3O structure -- or, equivalently, the energy gained by the system on adsorbing this atom in an hcp site of the (2 $\times$ 1) structure, yielding the (2 $\times$ 2)-3O structure.  The transition state
is indicated in the inset (from Stampfl and Scheffler, 1999).}

\label{en-dig-lh}
\end{figure}

It can be seen that at the transition state (also shown in the
inset of Fig.~\ref{en-dig-lh}), the C-O(a)
bond\footnote{
O(a) labels the oxygen atom which is adsorbed to the metal substrate}
is almost parallel to the surface.
The CO axis is bent away from O(a) yielding a bent
CO-O(a) complex with a bond angle of 125$^{\circ}$; similar
to that found for the Eley-Rideal mechanism. The  C-O(a) bond length is
1.59~\AA\, (about 29\% stretched compared to that in CO$_{2}$)
and the  CO bond length is 1.18~\AA\,.
At the transition state
CO and O begin to lift-off the
surface as they break their metal bonds in favor of developing
a C-O(a) bond.
This behavior can also be seen from the corresponding valence
electron density distribution shown in Fig.~\ref{lh-chden}.
The corresponding energy diagram is given in Fig.~\ref{en-dig-lh}.
In the work of Alavi et al. (1998) and Eichler and Hafner (1999) bent
transition state geometries were also identified for the CO 
oxidation reaction over the Pt(111) surface for the case of 
low  ($\Theta = 0.25$) oxygen coverages.
Alavi et al. (1998) attributed the main contribution to the activation
barrier to the weakening of the O-metal bond strength,
which supports the general understanding in this respect.

We end this section by noting that despite its high reactivity for CO
oxidation, elemental Ru will not be used in automotive catalysts because a
volatile Ru oxide exists which is  highly poisonous.
Nevertheless, trying to understand why Ru is so much more effective 
than other transition metals may help to design materials with similar 
properties as Ru. Furthermore, as noted above, there is an 
interesting and likely conceptionally important aspect of Ru 
studies: Ru can exist in many
oxidation states and the result (see Fig. \ref{oadsorp}) that its
surface region can be loaded with a high concentration of oxygen,
poses the question whether it is appropriate to call the reactive
surface an adsorbate system, or if it more appropriate to call this
a surface oxide (a RuO$_2$-like system in the present case). The highly reactive surface of O/Ru(0001) apparently is one with oxygen concentration much higher than $\Theta = 1$ and a coexistence of several domains of different stoichiometry.

\subsection{Summary outline of main points}
A deeper understanding of chemisorption, surface chemical reactions,
and heterogeneous catalysis is currently one of the main aims of surface
science (other important topics are, e.g., crystal growth, 
surface magnetism). In the context of this complex 
problem, the nature of the
chemisorption bond, the energetics of adsorption, and adsorbate
geometries and bond-lengths receive considerable attention. While
several simple rules have been found to be useful as in the area
of molecular chemistry, others should be applied with much caution.
Below we summarize some of the findings discussed in this chapter
in terms of a brief list:
\begin{itemize}
\item[1)] Density-functional theory has evolved into an important tool
for analyzing surface geometries. For example, the four-layer
surface alloy of Na on Al(111) was first predicted and
analyzed by DFT calculations (Stampfl and Scheffler, 1994c) and
subsequently confirmed by a
LEED intensity analysis (Burchhardt et al., 1995); in view of
the high number of structural parameters, it was indeed important
to start with the DFT analysis.
Similarly, the existence of a $(1 \times 1)$ ordered adlayer
of O on Ru(0001) and of the $(2 \times 2)$-3O adlayer was first predicted
by DFT calculations (Stampfl and Scheffler, 1996)
with detailed specification of the geometry.
And more examples exist.
\item[2)] In addition to providing accurate atomic geometries,
DFT calculations also
(and in particular) offer the potential of analysis of the underlying
mechanisms which determine if and how a certain geometry
can be attained and what the nature of the chemical bond is.
\item[3)] Whereas the  geometry is (typically) well described by
DFT-LDA and DFT-GGA calculations, the resulting {\em energies}
must be taken with some more caution. Adsorption energies
are typically not more accurate than about 0.2 eV per adatom.
And we do not expect that new exchange-correlation functionals
will  improve this to better than 0.1 eV per adatom in the near
future (the letter ``A'', as in LDA and GGA, will remain part of
exchange-correlation functionals employed in actual DFT calculations).
However, energy {\em differences} of chemically similar
bonding situations (in particular energies of small distortions
and phonons) are described with very well (possibly even  meV) accuracy.
\item[4)] The influence of thermally induced vibrations of the substrate
surface and adatoms is often ignored, even today. Still, it is 
quite clear
that some geometries (e.g., the unreconstructed surface of Au(111), or the
in-registry adsorption of Xe on fcc(111) surfaces) are stabilized by
vibrational energy and entropy.
We also mention that surface stress can be noticeably affected by the
surface thermal expansion.
Thus, sometimes it may be necessary to include the influence
of thermal expansion in theoretical studies (see, e.g.,
Cho and Scheffler, 1997; Xie et al., 1999).
\item[5)] Unlike a chemical reaction in molecular chemistry, 
adsorption on a surface involves very unequal partners. For
example, the substrate surface
gives rise to broadening of adsorbate levels, has an infinite 
number of electrons, and fixes the electron chemical potential.
\item[6)] Despite item 5), electronegativity differences between
adsorbate and substrate species appear to give a qualitatively
correct description of the nature of the chemical bond.
\item[7)] Item 6) may hold less true for higher coverages
when the adsorbate-adsorbate interaction becomes noticeable.
\item[8)] The correlation between local-coordination and bond-strength,
as noted for molecules by Pauling, appears to be (typically) fulfilled
for adsorbates. The energy per atom scales roughly proportional
to the square root of the coordination. For covalent systems,
some geometries may have a  more favorable energy than that of
this simple proportionality, e.g., when coordination numbers and/or 
bond-angles conform with the number of available valence electrons.\\
In this context we also re-emphasize that unexpected behavior of
atoms can occur at surfaces (i.e., the discussed substitutional 
adsorption and alloy formation). At this time knowledge about 
the energetics of the underlying processes, as, e.g., the formation 
of surface vacancies and the adsorption of adatoms at step sites,
is not sufficiently well developed that a prediction of the 
adsorbate site (in particular at low
coverage) is possible without performing a DFT calculation.
\item[9)] The validity of a correlation between  local
coordination and bond-length (in principle a consequence of the
local-coordination--bond-strength correlation), appears to be rarely
fulfilled for adsorbates. This is due to the fact that
geometries are quite constrained, as the atoms of the substrate
surface are well bonded to the rest of the substrate, and thus
the adsorbate bonds may not succeed to attain their optimum bond angle.
Thus, surface bonds my be subject to a ``frustrated'' hybridization of
the adatom orbitals or the substrate-surface orbitals.
\item[10)] A metal surface attempts to reach charge neutrality
on a rather short length scale. In fact, typically a perturbation is
even slightly over-screened on the length scale of the nearest-neighbor
distance and then the induced electron density is slowly decaying
in an oscillatory manner. This highly localized screening is achieved
by locally shifting valence electron density of states to 
higher or lower energies and thus changing the occupancy.
\item[11)] Reactivity concepts of molecular chemistry, which explore the
electron density of the {\em unperturbed} partners, are only of limited
value for dissociation at surfaces. It appears that the lowest
energy transition states are (often) so close to the surface where
the interaction is already so strong, that ``new states'' are formed, 
which differ considerably from the
clean surface states  and the  free adatom orbitals. The filling
of these ``new states'' (determined by the
substrate Fermi level) rules (at least partially) the reactivity.
\item[12)] In addition to item 11) it is important to include the
{\em dynamics} of the approaching molecules: It is not enough to know
just the energy of the lowest-energy transition state, but it is 
important
to know whether or not (or with what probability)
the transition state will be found by an approaching molecule. This
stresses
the importance of the high dimensionality of the potential-energy 
surface  on
which an approaching molecule travels and of {\em the statistics}.
\item[13)] Some often used concepts still need qualification. For 
example,
it is not clear how a term like that of a ``surface oxide'' 
or ``surface hydride''
describes a situation (or phase) which is different from an 
``adsorbate''. Clearly, this problem only arises for 
high-coverage adsorption ($\Theta > 1$).
\item[14)] We trust that the future will bring more studies
which are performed under finite pressure of a well defined atmosphere
(getting out of the vacuum) and in which the temperature is changed
systematically (warming up and cooling down). This will show which
of the previous studies were concerned with a thermal equilibrium
geometry and which dealt with a (possibly very special)
metastable state.\\[0.5cm]
\end{itemize}

\ni {\bf Acknowledgement}\\

We are grateful for the collaborations with many colleagues and friends,
the results of whom have been discussed in this chapter. In particular we
mention J{\"o}rg Bormet,  Axel Gross,  and  J{\"o}rg Neugbauer, but this 
list could easily be extended. We thank Kristen Fichthorn, Klaus Hermann,
and Karsten Reuter for helpful comments on the manuscript.\\[1cm]

%%\newpage

\noindent {\bf References}\\

\noindent 
Adams, D.L., 1996,
%           NEW PHENOMENA IN THE ADSORPTION OF ALKALI METALS ON
%           AL SURFACES. [Article]
            Appl. Phys. A {\bf 62}, 123.\\
Alavi, A., P. Hu, T. Deutsch, P.L. Silvestrelli and J. Hutter, 1998,
          Phys. Rev. Lett. {\bf 80}, 3650.\\
Aminpirooz, S., A. Schmalz, N. Pangher, J. Haase, M.M. Nielsen,
            D.R. Batchelor, E. B{\o}gh and D.L. Adams, 1992,
           Phys. Rev. B {\bf 46}, 15594.\\
Andersen, J.N., M. Qvarford, R. Nyholm, J.F. van Acker
           and E. Lundgren, 1992,
           Phys. Rev. Lett. {\bf 68}, 94.\\
Andersen, J.N., D. Hennig, E. Lundgren,
                       M. Methfessel, R. Nyholm and M. Scheffler, 1994,
                       Phys. Rev. B {\bf 50}, 17525.\\
Antoniewicz, P.R., 1978, Phys. Status Solidi (B) {\bf 86}, 645.\\
Bader, R.F.W., 1990, Atoms in Molecules. A Quantum Theory,
           International Series of Monographs on Chemistry, Vol. 22,
           Oxford University Press, Oxford.\\
Bader, R.F.W., 1994,
           Phys. Rev. B {\bf 49}, 13348.\\
Bagus, P.S., K. Hermann, W. M\"{u}ller, and C.J. Nelin, 1986,
                     Phys. Rev. Lett. {\bf 57}, 1496.\\
Barnes, C.J., 1994,
          in: The Chemical Physics of Solid
          Surfaces, Vol. 7, Phase Transitions and Adsorbate
          Restructuring at Metal Surfaces,
          eds. D.A. King and D.P. Woodruff.
          Elsevier, Amsterdam, p. 501.\\
Behm, R.J., 1989,
           in:  Physics and Chemistry of
           Alkali Metal Adsorption, eds. H.P. Bonzel,
           A.M. Bradshaw and G. Ertl. Elsevier, Amsterdam, p. 111.\\
Berndt, W., D. Weick, C. Stampfl, A.M. Bradshaw and M. Scheffler, 1995,
% Structural analysis of the two c(2$\times$2) phases of Na adsorbed
% on Al(001).
            Surf. Sci. {\bf 330}, 182.\\
Biela\'{n}ski, A. and J. Haber, 1991,
       Oxygen in Catalysis. 
       Dekker, New York, p. 472.\\
Blyholder, G., 1964,
           J. Phys. Chem. {\bf 68}, 2772.\\
Blyholder, G., 1975,
           J. Vac. Sci. Technol. {\bf 11}, 865.\\
Bockstedte, M., A. Kley, J. Neugebauer and M. Scheffler, 1997,
%%  Density-functional theory calculations for poly-atomic systems:
%%  Electronic structure, static and elastic properties and
%%  {\em ab initio} molecular dynamics.
           Comput. Phys. Commun. {\bf 107}, 187.\\
Bormet, J., J. Neugebauer and M. Scheffler, 1994a,
           Phys. Rev. B {\bf 49}, 17242.\\
Bormet, J., B. Wenzien, J. Neugebauer and M. Scheffler, 1994b,
                     Comput. Phys. Commun. {\bf 79}, 124.\\
Born, M. and R. Oppenheimer, 1927,
           Ann. Phys. {\bf 84}, 457.\\
Born, M. and K. Huang, 1954,
          in: Dynamical Theory of Crystal Lattices, eds. N.F. Mott
          and E. C. Bullard.
          Clarendon, Oxford, p. 420.\\
B\"{o}ttcher, A. and Niehus, 1999, J. Chem. Phys. {\bf 110}, 3186.\\
B\"{o}ttcher, A., Niehus, S. Schwegmann, H. Over and G. Ertl, 1997,
          J. Phys. Chem. {\bf 101}, 11185.\\
B\"{o}ttcher, A., M. Rogozia, H. Niehus, H. Over and G. Ertl, 1999, 
          J. Chem. Phys., accepted.\\
Bradshaw, A.M. and M. Scheffler, 1979, J. Vac. Sci. Technol. 
         {\bf 16}, 447.\\
Brivio, G.P. and M.I. Trioni, 1999,
          Rev. Mod. Phys. {\bf 71}, 231.\\
Bruch, L.W.,  M.W. Cole and E. Zaremba, 1997,
              Physical Adsorption: Forces and Phenomena,
              Clarendon, Oxford.\\
Burchhardt, J., M.M. Nielsen, D.L. Adams, E. Lundgren,
          J.N. Andersen, C. Stampfl, M. Scheffler, A. Schmalz,
          S. Aminpirooz and J. Haase, 1995,
                Phys. Rev. Lett. {\bf 74}, 1617.\\
Callaway, J., 1964, J. Math. Phys. {\bf 5}, 783.\\
Callaway, J., 1967, Phys. Rev. {\bf 154}, 515.\\
Campuzano, J.C., 1990,
                in: The Chemical Physics of Solid
                Surfaces and Heterogeneous Catalysis, Vol. 3a:
                Chemisorption Systems,
                eds. D.A. King and D.P. Woodruff.
                Elsevier, Amsterdam, p. 389.\\
Car, R. and M. Parrinello, 1985,
              Phys. Rev. Lett. {\bf 55}, 2471. \\
Cho, J.-H. and M. Scheffler, 1997,
%% Surface relaxation and ferromagnetism of Rh(001),
              Phys. Rev. Lett. {\bf 78}, 1299.\\
Christensen, O.B. and K.W. Jacobsen, 1992,
           Phys. Rev. B {\bf 45}, 6893.\\
Christensen, S.V., J. Nerlov, K. Nielsen, J. Burchhardt, M.M. Nielsen
           and D.L. Adams, 1996,
           Phys. Rev. Lett. {\bf 76}, 1892.\\
Darling, G.R. and S. Holloway, 1994,
            J. Chem. Phys. {\bf 101}, 3268.\\
Dreizler, R.M. and E.K.U.  Gross, 1990,   Density Functional Theory.
                    Springer, Berlin.\\
Echenique, P.M. and J.B. Pendry, 1975, J. Phys. C {\bf 8}, 2936.\\
Eichler, A., J. Hafner, A. Gross and M. Scheffler, 1999a,
              Phys. Rev. B {\bf 59}, 13297.\\
Eichler, A., J. Hafner, A. Gross and M. Scheffler, 1999b,
              Chem. Phys. Lett. {\bf 311}, 1.\\
Eichler A. and J. Hafner, 1999, Phys. Rev. B {\bf 59}, 5960.\\
Engel, T. and G. Ertl, 1979,
             J. Chem. Phys. {\bf 69}, 1267;
             Adv. Catal.  {\bf 28}, 1.\\
Engel, T. and G. Ertl, 1982,
             in:  The Chemical Physics of Solid Surfaces
             and Heterogeneous Catalysis, Vol. 4, Fundamental Studies of
             Heterogenous Catalysis,
             eds. D.A. King and D.P. Woodruff.
             Elsevier, Amsterdam, p. 73.\\
Fasel, R.,  P. Aebi, R.G. Agostino, L. Schlapbach
                   and J. Osterwalder, 1996,
              Phys. Rev. B {\bf 54}, 5893.\\
Feibelman, P.J., 1990, Phys. Rev. Lett. {\bf  65}, 729.\\
Finnis, M.W., R. Kaschner, C. Kruse, J. Furthm{\"u}ller
              and M. Scheffler, 1995,
%  The interaction of a point charge with a metal surface: theory and
%  calculations for (111), (001) and (110) aluminium surfaces.
           J. Phys.: Condens. Matter {\bf 7}, 2001.\\
Fiorentini, V., M. Methfessel and M. Scheffler, 1993,
           Phys. Rev. Lett. {\bf 71}, 1051; 1998, Phys. Rev. Lett. {\bf 81}, 2184.\\
Ganduglia-Pirovano, M.V., J. Kudrnovsk\'{y} and M. Scheffler, 1997,
                Phys. Rev. Lett. {\bf 78}, 1807.\\
Grimley, T.B., 1975, Prog. Surf. Membr. Sci. {\bf 9}, 71.\\
Gross, A., S. Wilke and M. Scheffler, 1995,
              Phys. Rev. Lett. {\bf 75}, 2718.\\
Gross, A.  and M. Scheffler, 1996a,
%%  Influence of molecular vibrations on dissociative adsorption.
              Chem. Phys. Lett. {\bf 256}, 417.\\
Gross, A.  and M. Scheffler, 1996b,
%%  Steering and ro-vibrational effects in the dissociative adsorption
%%  and associative desorption of H$_{2}$/Pd(001).
            Prog. Surf. Sci. {\bf 53}, 187.\\
Gross, A., S. Wilke and M. Scheffler, 1996c,
            Surf. Sci. {\bf 357/358}, 614. \\
Gross, A., M. Bockstedte  and M. Scheffler, 1997,
              Phys. Rev. Lett. {\bf 79}, 701.\\
Gross, A. and M. Scheffler, 1998,
              Phys. Rev. B  {\bf 57}, 2493.\\
Gross, A., 1998,
              Surf. Sci. Rep. {\bf 32}, 291.\\
Gsell, M., M. Stichler, P. Jacob and D. Menzel, 1998,
               Israel J. Chem. {\bf 38}, 339.\\
Gurney, R.W., 1935,
              Phys. Rev. {\bf 47}, 479.\\
Hammer, B.,  M. Scheffler, K.W. Jacobsen and J.K. N{\o}rskov, 1994,
            Phys. Rev. Lett. {\bf 73}, 1400.\\
%                   Multidimensional potential energy surface for H$_{2}$
%                   dissociation over Cu(111).
Hammer, B. and M. Scheffler, 1995,
%                  Local chemical reactivity of a metal alloy surface
                   Phys. Rev. Lett. {\bf 74}, 3487.\\
Hammer, B. and J.K. N{\o}rskov, 1997,
      %  We note that similar work had been published also by other
      %  groups (see, e.g., and references therein
      in: Chemisorption and Reactivity on Supported Clusters and Thin
      Films, eds. R.M. Lambert and G. Pacchioni.
      Kluwer, Dordrecht, p. 285.\\
Heine, V. and D. Marks, 1986,
       Surf. Sci. {\bf 115}, 65.\\
Hennig, D.,  M.V. Ganduglia-Pirovano and M. Scheffler, 1996,
                     Phys. Rev. B {\bf 53}, 10344.\\
Hermann, K. and P.S. Bagus, 1977,
                    Phys. Rev. B {\bf 16}, 4195.\\
Hermann, K., P.S. Bagus, and C.J. Nelin, 1987,
                     Phys. Rev. B {\bf 35}, 9467.\\
Hjelmberg, H., B.I. Lundqvist and J.K. N{\o}rskov, 1979,
             Phys. Scr. {\bf 20}, 192.\\
Hoffmann, P., C. von Muschwitz, K. Horn, K. Jacobi, A.M. Bradshaw,
          K. Kambe and M. Scheffler, 1979,
          Surf. Sci. {\bf 89}, 327.\\
Hoffmann, R., 1988,
            Rev. Mod. Phys. {\bf 60}, 601.\\
Horn, K., M. Scheffler and A.M. Bradshaw, 1978, 
          Phys. Rev. Lett. {\bf 41}, 822.\\
Hu, P., D.A. King, M.-H. Lee and M.C. Payne, 1995,
               Chem. Phys. Lett. {\bf 246}, 73.\\
Jacobi, K., C. von Muschwitz and K. Kambe, 1980, 
             Surf. Sci. {\bf 93}, 310.\\
Janak, J.F., 1978, Phys. Rev. B {\bf 18}, 7165.\\
Jennings, P.J., 1979, Surf. Sci. {\bf 88}, L25.\\
Kambe, K. and M. Scheffler, 1979, Surf. Sci. {\bf 89}, 262.\\
Kim, Y.D., S. Wendt, S. Schwegmann, H. Over and G. Ertl, 1998,
                   Surf. Sci. {\bf 418}, 267.\\
Kleinman, L., 1997, Phys. Rev. B {\bf 56}, 16029.\\
Kohn, W. and K.H. Lau, 1976, Solid State Commun. {\bf 18}, 553.\\
Kostov, K.L., H. Rauscher and D. Menzel, 1992,
             Surf. Sci. {\bf 278}, 62.\\
Kostov, K.L., M. Gsell, P. Jakob, T. Moritz,
           W. Widdra and D. Menzel, 1997,
           Surf. Sci. {\bf 394}, L138.\\
Kresse, G. and J. Furthm{\"u}ller, 1996,
           Phys. Rev. B {\bf 54}, 11169.\\
Kroes, G.J., E.J. Baerends and R.C. Mowrey, 1997, Phys. Rev. Lett. {\bf 78}, 3583; 1998,
                   Phys. Rev. Lett. {\bf 81}, 4781.\\
Kroes, G.J., 1999, Prog. Surf. Sci. {\bf 60}, 1.\\
Lang, N.D. and W. Kohn, 1971, Phys. Rev. B {\bf 3}, 1215.\\
Lang, N.D., 1971, Phys. Rev. B {\bf 4}, 4234.\\
Lang, N.D., 1973,  Solid State Phys. {\bf 28}, 225.\\
Lang, N.D. and A.R. Williams, 1977,
                     Phys. Rev. B {\bf 16}, 2408.\\
Lang, N.D. and A.R. Williams, 1978,
                     Phys. Rev. B {\bf 18}, 616.\\
Langmuir, I., 1932,
                J. Am. Chem. Soc. {\bf 54}, 2798.\\
Liebsch, A., 1978, Phys. Rev. B {\bf 17}, 1653.\\
Lindroos, M., H. Pfn\"{u}r, G. Held and D. Menzel, 1989,
            Surf. Sci. {\bf 222}, 451.\\
McRae, E.G., 1971, Surf. Sci. {\bf 25}, 491.\\
Methfessel, M., D. Hennig and M. Scheffler, 1992a,
%%   Trends of the surface relaxations, surface energies, and work
%%   functions of the $4d$ transition metals.
           Phys. Rev. B {\bf 46}, 4816.\\
Methfessel, M., D. Hennig and M. Scheffler, 1992b,
                Appl. Phys. A {\bf 55}, 442.\\
Methfessel, M., D. Hennig and M. Scheffler, 1995,
                Surf. Rev. Lett. {\bf 2}, 197.\\
Miller, A.R., 1946, Proc. Cambridge Philos. Soc. {\bf 42}, 492.\\
Mitchell, W.J. and W.H. Weinberg, 1996, J. Chem. Phys. {\bf 104}, 9127.\\
Muscat, J.P. and D.M. Newns, 1978, Prog. Surf. Sci. {\bf 9}, 1.\\
Muscat, J.P. and D.M. Newns, 1979, Phys. Rev. B {\bf 19}, 1270.\\
Narloch, B., G. Held and D. Menzel, 1994,
            Surf. Sci. {\bf 317}, 131.\\
Narloch, B., G. Held and D. Menzel, 1995,
            Surf. Sci. {\bf 340}, 159.\\
Naumovets, A.G., 1994,
                   in: The Chemical Physics of
                   Solid Surfaces, Vol. 7, Phase Transitions
                   and Adsorbate Restructuring at Metal Surfaces,
                   eds. D.A. King and D.P. Woodruff.
                   Elsevier, Amsterdam, p. 163.\\
Neugebauer, J. and M. Scheffler, 1992,
                     Phys. Rev. B {\bf 46}, 16067.\\
Neugebauer, J. and M. Scheffler, 1993,
                     Phys. Rev. Lett. {\bf 71}, 577.\\
Nilsson, A., N. Wassdahl, M. Weinelt, O. Karis, T. Weill,
       P. Bennich, J. Hasselstr\"{o}m, A. F\"{o}hlisch, J. St\"{o}hr
       and M. Samant, 1997,
                     Appl. Phys. A {\bf 65}, 147.\\
N{\o}rskov, J.K., 1990,
                 Rep. Prog. Phys. {\bf 53}, 1253.\\
Nouvertn\'e, F.,  U. May, M. Bamming, A. Rampe, U. Korte,
             G. G{\"u}ntherodt, R. Pentcheva and M. Scheffler, 1999,
             Phys. Rev. B {\bf 60},14382.\\
%   Atomic exchange processes and {\em bimodal} initial
%   growth of Co/Cu(001)}
Oppo, S., V. Fiorentini and M. Scheffler, 1993,
%                 Theory of adsorption and surfactant effect of Sb on
%                 Ag (111)
                  Phys. Rev. Lett. {\bf 71}, 2437.\\
Over, H., H. Bludau, M. Gierer, and G. Ertl, 1995,
                  Surf. Rev. Lett. {\bf 2}, 409.\\
Palmberg, P.W., 1971, Surf. Sci. {\bf 25}, 598.\\
Paul, J., 1987,
                 J. Vac. Sci. Technol. A {\bf 5}, 664.\\
Pauling, L., 1960,
                  The Nature of the Chemical Bond and the
                  Structure of Molecules and Crystals: An Introduction
                  to Modern Structural Chemistry.
                  Cornell University Press.\\
Payne, M.C., M.P. Teter, D.C. Allan, T.A. Arias 
                 and J.D. Joannopulos, 1992,
       Rev. Mod. Phys. {\bf 64}, 1045.\\
Payne, M.C., I.J. Robertson, D. Thomson and V. Heine, 1996,
      Philos. Mag. B {\bf 73}, 191.\\
%%\bibitem{peden3}
Peden, C.H.F.,  D.W. Goodman, M.D. Weisel and F.M. Hoffmann, 1991,
          Surf. Sci. {\bf 253}, 44.\\
%%\bibitem{peden2}
Peden, C.H.F. and D.W. Goodman, 1986,
          J. Phys. Chem. {\bf 90}, 1360.\\
Peden, C.H.F., 1992,
        in: Surface Science of Catalysis: In Situ Probes and Reaction
        Kinetics, eds. D.J. Dwyer and F.M. Hoffmann.
        Am. Chem. Soc., Washington DC.\\
Pedersen, M.O., I.A. Bonicke, E. Laegsgaard, I. Stensgaard, A. Ruban,
            J.K. N{\o}rskov and F. Besenbacher, 1997,
%  GROWTH OF CO ON CU(111) - SUBSURFACE GROWTH OF TRILAYER CO ISLANDS.
            Surf. Sci. {\bf 387}, 86. \\
Pentcheva, R. and M. Scheffler, 2000,
            Phys. Rev. B, in print.\\
Perdew, J.P. and M. Levy, 1997, Phys. Rev. B {\bf 56}, 16021.\\
Petersen, M., P. Ruggerone and M. Scheffler, 1996,
Phys. Rev. Lett. {\bf 76}, 995; 2000,
        Phys. Rev. B, submitted.\\
Pleth Nielsen, L.,  F. Besenbacher, I. Stensgaard, E. L{\ae}gsgaard,
          C. Engdahl, P. Stoltze, K.W. Jacobsen 
          and J.K. N{\o}rskov, 1993,
%%    Initial growth of Au on Ni(110): Surface alloying of
%%     non-miscible metals,
            Phys. Rev. Lett. {\bf 71}, 754.\\
Porteus, J.O., 1974,
                  Surf. Sci. {\bf 41}, 515.\\
Rader O., W. Gudat, C. Carbone, E. Vescovo, S. Bl\"{u}gel, R. Kl\"{a}sges,
        W. Eberhardt, M. Wuttig, J. Redinger and F.J. Himpsel, 1997,
%  ELECTRONIC STRUCTURE OF TWO-DIMENSIONAL MAGNETIC ALLOYS
%  - C(2X2) MN ON CU(001) AND NI(001).
          Phys. Rev. B {\bf 55}, 5404.\\
Rendulic, K.D., G. Anger and A. Winkler, 1989,
          Surf. Sci. {\bf 208}, 404.\\
Robertson, I.J., D.I. Thomson, V. Heine and M.C. Payne, 1994,
           J. Phys. Condens. Matter {\bf 6}, 9963. \\
Scheffler, M., K. Kambe and F. Forstmann, 1978,
           Solid State Commun. {\bf 25}, 93.\\
Scheffler, M., K. Horn, A.M. Bradshaw and K. Kambe, 1979,
           Surf. Sci. {\bf 80}, 69.\\
Scheffler, M. and A.M. Bradshaw, 1983,
%                     The electronic structure of adsorbed layers.
                      in: The Chemical Physics of Solid Surfaces and
                      Heterogeneous Catalysis, Vol. 2: Adsorption
                      at Solid Surfaces,
                      eds. D.A. King and D.P. Woodruff.
                      Elsevier, Amsterdam, p. 165.\\
Scheffler, M., Ch. Droste, A. Fleszar, F. M\'{a}ca, G. Wachutka
           and G. Barzel, 1991,
           Physica B {\bf 172}, 143.\\
Schiffer, A., P. Jakob and D. Menzel, 1997,
           Surf. Sci. {\bf 389}, 116.\\
Schmalz, A., S. Aminpirooz, L. Becker, J. Haase, J. Neugebauer,
           M. Scheffler, D.R. Batchelor, D.L. Adams and E. B{\o}gh, 1991,
           Phys. Rev. Lett. {\bf 67}, 2163.\\
Seyller T., M. Caragiu, R.D. Diehl, P. Kaukasoina and M. Lindroos, 1998,
%          M. OBSERVATION OF TOP-SITE ADSORPTION FOR XE ON CU(111).
           Chem. Phys. Lett. {\bf 291}, 567.\\
Skriver, H., 1985, Phys. Rev. B {\bf 31}, 1909.\\
Somorjai, G.A. and M.A. Van Hove,  1989,
           Prog. Surf. Sci. {\bf 30}, 201.\\
Spanjaard, D. and M.C. Desjonqu\`{e}res, 1990,
           in: Interaction of Atoms and Molecules with Surfaces,
           eds. V. Bortolani, N.H. March and N.P. Tosi. Plenum,
           New York, London, p. 255.\\
SRL, 1995, Surf. Rev. Lett. {\bf 2}, 315.
           A special issue of {\em Surface Review and Letters} devoted to
           alkali-metal adsorption,
           containing review articles by several groups.\\
Stampfl, C., M. Scheffler, H. Over, J. Burchhardt,
                     M. Nielsen, D.L. Adams and W. Moritz, 1992,
                     Phys. Rev. Lett. {\bf 69}, 1532.\\
Stampfl, C., M. Scheffler, H. Over, J. Burchhardt,
                     M. Nielsen, D.L. Adams and W. Moritz, 1994a,
%   LEED structural analysis of Al(111)-K(sqrt(3) x sqrt(3))R30:
%   Identification of stable and metastable adsorption sites,
                     Phys. Rev. B {\bf 49}, 4959.\\
Stampfl, C., J. Neugebauer and M. Scheffler, 1994b,
                    Surf. Sci. {\bf 307/309}, 8.\\
%                   Alkali-metal adsorption on Al(111) and Al(001)
Stampfl, C. and  M. Scheffler, 1994c,
%          Theoretical identification of a (2$\times$2) composite
%          double layer ordered surface alloy of Na on Al(111).
           Surf. Sci. {\bf 319}, L23.\\
Stampfl, C. and M. Scheffler, 1995,
%                   Theory of alkali-metal adsorption on close-packed
%                   metal surfaces
           Surf. Rev. Lett. {\bf 2}, 317.\\
Stampfl, C. and M. Scheffler, 1996,
             Phys. Rev. B {\bf 54}, 2868.\\
Stampfl, C., 1996, Surf. Rev.
             Lett. {\bf 3}, 1567.\\
Stampfl, C., S. Schwegmann, H. Over, M. Scheffler 
            and G. Ertl, 1996,
            Phys. Rev. Lett. {\bf 77}, 3371.\\
Stampfl, C. and M. Scheffler, 1997,
           Phys. Rev. Lett. {\bf 78}, 1500;
           J. Vac. Sci. Technol. A {\bf 15}, 1635;
           Surf. Sci. {\bf 377-379},  808.\\
Stampfl, C., K. Kambe, R. Fasel, P. Aebi and M. Scheffler, 1998,
%  Theoretical analysis of the electronic structure of
%  the stable and metastable c(2x2) phases of Na on Al(001): Comparison
%  with angle-resolved ultraviolet photoemission spectra.
           Phys. Rev. B {\bf 57}, 15251.\\
Stampfl, C. and M. Scheffler, 1998,
%       Coadsorption of CO and O on Ru(0001): A structural
%       analysis by density functional theory.
        Israel J. Chem. {\bf 38}, 409.\\
Stampfl, C. and M. Scheffler, 1999, Surf. Sci. {\bf 433-435}, 119.\\
Taylor, J.B. and I. Langmuir, 1933,
                   Phys. Rev. {\bf 44}, 423.\\
Topping, J., 1927, Proc. Roy. Soc. London Ser. A {\bf 114}, 67.\\
Wang, X.-G., W. Weiss, Sh.K. Shaikhutdinov, M. Ritter, M. Petersen,
F. Wagner, R. Schl{\"o}gl and M. Scheffler, 1998,
%    The hematite (Alpha-Fe2O3)(0001) surface: Evidence for domains of
%    distinct chemistry.
               Phys. Rev. Lett. {\bf 81}, 1038.\\
Wilke, S. and M. Scheffler, 1996,
%%  Potential-energy surface for H$_{2}$ dissociation over Pd\,(001).
               Phys. Rev. B {\bf 53}, 4926.\\
Wilke, S., M.H. Cohen and M. Scheffler, 1996,
              Phys. Rev. Lett. {\bf 77}, 1560.\\
Wenzien, B., J. Bormet, J. Neugebauer and M. Scheffler, 1993,
        Surf. Sci. {\bf 287/288}, 559.\\
Wenzien, B., J. Bormet and M. Scheffler, 1995,
         Comput. Phys.~Commun. {\bf 88}, 230.\\
Wimmer, E., C.L. Fu and A.J. Freeman, 1985,
         Phys. Rev. Lett. {\bf 55}, 2618.\\
Xie, J., S. de Gironcoli, S. Baroni and M. Scheffler, 1999,
%%  Temperature dependent surface relaxations of Ag(111).
        Phys. Rev. B {\bf 59}, 970.\\
Yang, L., G. Vielsack and M. Scheffler, 1994, unpublished.\\
Yu, B.D. and M. Scheffler, 1997,
           Phys. Rev. B {\bf 56}, R15569. \\

\end{document}